\pdfoutput = 1

\documentclass[12pt,a4paper]{article}

\usepackage{amssymb,epsf,epsfig,amsmath,color}
\usepackage{multirow}
\usepackage{accents}
\usepackage{braket}
\usepackage{float}
\usepackage[noadjust]{cite}

\newcommand{\lsim}{
\mathrel{\hbox{\rlap{\hbox{\lower4pt\hbox{$\sim$}}}\hbox{$<$}}}}
\newcommand{\gsim}{
\mathrel{\hbox{\rlap{\hbox{\lower4pt\hbox{$\sim$}}}\hbox{$>$}}}}

\begin{document}
\begin{titlepage}
\vspace*{-0.7truecm}
\begin{flushright}
Nikhef-2021-002

SI-HEP-2021-02

P3H-21-007
\end{flushright}

\vspace{1.6truecm}

\begin{center}
\boldmath
{\Large{\bf Mapping out the Space for New Physics with\\
\vspace*{0.3truecm}
Leptonic and Semileptonic $B_{(c)}$ Decays}
}
\unboldmath
\end{center}

\vspace{0.8truecm}

\begin{center}
{\bf Robert Fleischer\,${}^{a,b}$, Ruben Jaarsma\,${}^{a}$ and 
Gilberto~Tetlalmatzi-Xolocotzi\,${}^{c}$}

\vspace{0.5truecm}

${}^a${\sl Nikhef, Science Park 105, NL-1098 XG Amsterdam, Netherlands}

${}^b${\sl  Faculty of Science, Vrije Universiteit Amsterdam,\\
NL-1081 HV Amsterdam, Netherlands}

${}^c${\sl Theoretische Physik 1, Naturwissenschaftlich-Technische Fakult\"{a}t, Universit\"{a}t Siegen,Walter-Flex-Strasse 3, D-57068 Siegen, Germany.}

\end{center}

\vspace*{1.7cm}

\begin{abstract}
\noindent 
Decays of $B$ mesons with leptons in the final state offer an interesting laboratory to search for possible effects of physics from beyond the Standard Model. In view of puzzling patterns in experimental data, the violation of lepton flavour universality is an interesting option. We present a strategy, utilising ratios of leptonic and semileptonic $B$ decays, where the elements $|V_{ub}|$ 
and $|V_{cb}|$ of the Cabibbo--Kobayashi--Maskawa (CKM) matrix cancel, to constrain the short-distance coefficients of (pseudo)-scalar, vector and tensor operator contributions. The individual branching ratios allow us then to extract also the CKM matrix elements, even in the presence of new-physics contributions. Bounds on unmeasured leptonic and semileptonic decays offer important additional constraints. In our comprehensive analysis, we give also predictions for decays which have not yet been measured in a variety of scenarios. 

\end{abstract}

\vspace*{2.1truecm}

\vfill

\noindent
April 2021

\end{titlepage}

\thispagestyle{empty}
\vbox{}
\newpage

\setcounter{page}{1}

\section{Introduction}\label{sec:intro}
Decays of $B$ mesons caused by semileptonic $b\to u \ell \bar \nu_{\ell}$ and $b\to c \ell \bar \nu_{\ell}$ quark-level transitions provide valuable insights into the quark-flavour sector of the Standard Model (SM) that is described by the 
Cabibbo--Kobayashi--Maskawa (CKM) matrix. These channels play a key role to determine the CKM matrix elements $|V_{ub}|$ 
and $|V_{cb}|$ from experimental data, where we are facing extractions from inclusive and exclusive modes that are not fully consistent with one another \cite{Ricciardi:2021shl}. Decays of this kind have also received a lot of attention in view of experimental data indicating violations of lepton flavour universality, which is a key feature of the SM (see, e.g., Refs.~\cite{Bernlochner:2021vlv,Hiller:2018ijj,Fajfer:2019rjq} and references therein). Is it actually violated in Nature? Analyses of New Physics (NP) effects in leptonic and semileptonic modes mediated by these transitions have been performed in, e.g., Refs.~\cite{Akeroyd:2002cs,Khodjamirian:2011ub,Crivellin:2014zpa,Bernlochner:2014ova,Tanaka:2016ijq,Dutta:2016eml,Ivanov:2017hun,Jung:2018lfu,Hu:2018veh,Blanke:2018yud,Blanke:2019qrx,Ivanov:2020iad,Iguro:2020cpg,Becirevic:2020rzi,Cornella:2021sby,Bobeth:2021lya}.

For NP analyses to constrain the corresponding short-distance parameters, care is needed concerning the values of $|V_{ub}|$ and $|V_{cb}|$ as their determinations may be affected by contributions from physics beyond the Standard Model. In view of this feature, we have proposed a strategy for the exclusive $b\to u \ell \bar \nu_{\ell}$ modes in Ref.~\cite{Banelli:2018fnx}, introducing suitable ratios of leptonic and semileptonic modes of this decay class, where the CKM factor 
$|V_{ub}|$ cancels. Utilising these observables, we then determined allowed regions for the short-distance coefficients of (pseudo)-scalar operators, arising from NP effects. Using the individual branching ratios, we may finally extract $|V_{ub}|$, while simultaneously allowing for physics from beyond the SM. 

In this paper, we build upon this strategy, addressing two new points. The first concerns the analysis of the effects of new vector and tensor operators in the exclusive $b\to u \ell \bar \nu_{\ell}$ decays: the $B^-\to \ell \bar \nu_{\ell}$ as well as $B\to \pi \ell \bar \nu_{\ell}$ and $B\to \rho \ell \bar \nu_{\ell}$ channels,  thereby complementing our previous study where we analysed only (pseudo)-scalar operators. The second -- and main -- point is the development of a similar strategy for exclusive $B$ decays
originating from $b\to c \ell \bar \nu_{\ell}$ processes, $B\to D \ell \bar \nu_{\ell}$ and $B\to D^{(*)} \ell \bar \nu_{\ell}$ transitions as well as the leptonic $B^-_c\to \ell \bar \nu_{\ell}$ modes. Here we consider the NP effects of vector,  (pseudo)-scalar and tensor operators, and constrain the corresponding short-distance coefficients through the currently available experimental data. Moreover, we extract the CKM matrix element $|V_{cb}|$ with the help of the individual branching ratios, and make predictions for decays which have not yet been measured. 

The theoretical framework of our study is given by the corresponding low-energy effective Hamiltonians, allowing for NP operators. We perform also the proper renormalisation group evolution to connect the short-distance coefficients at the high-energy scale at the $1\,$TeV level with the low-energy scale at the $\mu=m_b$ level of the $b$-quark mass that is relevant for the analysis of experimental data for the $B$-meson decays under consideration.

The outline of this paper is as follows: in Section~\ref{sec:theory-framework}, we introduce the theoretical tools, including the general expressions for the exclusive decay rates, the renormalisation group evolution and the strategy for extracting the CKM matrix elements 
$|V_{ub}|$ and $|V_{cb}|$. In Section~\ref{sec:btou}, we focus on the exclusive $b\to u \ell \bar \nu_{\ell}$ modes, and explore the space for NP effects arising from vector and tensor operators, as well as update our previous analysis of the (pseudo)-scalar operators. In Section~\ref{sec:VubPredbu}, these results are applied to determine the allowed range for $|V_{ub}|$ and certain branching ratios that have not been measured. The exclusive $b\to c \ell \bar \nu_{\ell}$ quark-level decays are then discussed in Section~\ref{sec:btoc}, giving a comprehensive picture of the constraints on the short-distance coefficients of the corresponding (pseudo)-scalar, vector and tensor operators following from the current experimental data.  In Section~\ref{sec:Vcb}, we present a detailed discussion of the picture for the CKM factor $|V_{cb}|$, emerging from our NP analysis and the experimental information available. There we also consider the leptonic $B_c$ decays, which have so far not been observed experimentally. Finally, we summarise our conclusions in Section~\ref{sec:concl}. In two appendices, we collect details of the specific form factor parameterisations applied in our numerical analysis, and the various input parameters that are used in our calculations.

\section{Theoretical Framework}\label{sec:theory-framework}

The effective Hamiltonian relevant to our studies is
\begin{eqnarray}\label{eq:EffH}
{\cal H}^q_{\rm eff}= \frac{4 G_{\rm F}}{\sqrt{2}}V_{qb}\left[ 
\tilde{C}^{q, \ell}_{V_L} {\cal O}^{q, \ell}_{V_L}  +   
\tilde{C}^{q, \ell}_{V_R} {\cal O}^{q, \ell}_{V_R}  +
\tilde{C}^{q, \ell}_{S}{\cal O}_{S}^{q, \ell}
+ 
\tilde{C}^{q, \ell}_{P}{\cal O}_{P}^{q, \ell}
+ 
\tilde{C}^{q, \ell}_{T}{\cal O}_{T}^{q, \ell}  \right] + \mbox{h.c.},
\end{eqnarray}
where $q=u, c$ and $\ell = e, \mu, \tau$. The corresponding operators are
\begin{eqnarray}
\label{eq:operators}
{\cal O}^{q, \ell}_{V_L}=(\bar q \gamma^\mu  P_L b)(\bar \ell \gamma_\mu P_L \nu_{\ell}),
\quad
{\cal O}^{q, \ell}_{V_R}=(\bar q \gamma^\mu  P_R b)(\bar \ell \gamma_\mu P_L \nu_{\ell}),
\quad
{\cal O}^{q, \ell}_{S}=\frac{1}{2} (\bar q b)(\bar \ell   P_L \nu_{\ell}),\nonumber\\
{\cal O}^{q, \ell}_{P}=\frac{1}{2}  (\bar q \gamma_5  b)(\bar \ell P_L  \nu_{\ell}),
\quad
{\cal O}^{q, \ell}_{T}=(\bar{q}\sigma^{\mu\nu}P_L b)(\bar{\ell}\sigma_{\mu\nu}P_L \nu_{\ell}),
\quad\quad\quad\quad\quad\quad
\end{eqnarray}
where $P_{L,R} = \frac{1}{2}(1 \mp \gamma_5)$.
The Wilson coefficients can be decomposed in their SM and NP contributions as
\begin{eqnarray}
\tilde{C}^{q, \ell}_a&=&C^{({\rm SM})q, \ell}_a + C^{q, \ell}_a,
\end{eqnarray}
with $a=\{V_L, V_R, S, P, T\}$ and
\begin{eqnarray} \label{eq:WCSM}
C^{({\rm SM})q, \ell}_{V_L}=1, &\quad\quad& C^{({\rm SM})q, \ell}_{V_R}=C^{({\rm SM})q, \ell}_{S}=C^{({\rm SM})q, \ell}_{P}=C^{({\rm SM})q, \ell}_{T}=0.
\end{eqnarray}
Potential NP contributions enter through the $C_a^{q, \ell}$. In view of Eq.~(\ref{eq:WCSM}), we have $\tilde{C}_{a'}^{q, \ell} = C_{a'}^{q, \ell}$ for $a' = \{V_R, S, P, T\}$, i.e., they are fully characterized by their NP component. We will consider CP-conserving NP effects and hence only real Wilson coefficients.

As discussed in \cite{Alonso:2016oyd, Cirigliano:2009wk,Alonso:2015sja}, within the framework of the Standard Model Effective Field Theory (SMEFT), the coefficient $C^{q, \ell}_{V_L}$ 
depends in general on the leptonic flavour. In contrast, at
the lowest order in the  $v^2/\Lambda^2$ expansion, where $v$ is the vacuum expectation value of the Higgs boson and $\Lambda$ is the NP scale, the coefficient $C^{q, \ell}_{V_R}$ turns out to be lepton-flavour universal. 
Therefore, unless indicated otherwise, we will assume
\begin{eqnarray}\label{eq:universality}
C^{q, \ell}_{V_R}=C^q_{V_R}.   
\end{eqnarray}

To constrain the Wilson coefficients in Eq.~(\ref{eq:EffH}), we will consider different leptonic and semileptonic $B$ meson decays. In the presence of pseudoscalar and vector interactions, the leptonic modes obey the following expression:
\begin{eqnarray}\label{eq:leptonic}
\mathcal{B}(B_q^-\rightarrow \ell^- \bar{\nu}_{\ell})=\mathcal{B}(B_q^-\rightarrow \ell^- \bar{\nu}_{\ell})|_{\rm SM} 
\Bigl| 1 + C^{q, \ell}_{V_L} - C^{q}_{V_R} + \frac{M_{B_q}^2}{m_\ell (m_b+m_q)} C^{q, \ell}_P \Bigl|^2,
\end{eqnarray}
where
\begin{equation}\label{SM-Br}
{\mathcal B}(B_q^-\to\ell^-\bar\nu_\ell)|_{\rm SM}=\frac{G_{\rm F}^2}{8\pi}
|V_{qb}|^2M_{B_q}m_\ell^2\left(1-\frac{m_\ell^2}{M_{B_q}^2}\right)^2f_{B_q}^2\tau_{B_q}.
\end{equation}
Since the  tensor matrix element vanishes, $\Braket{0|\bar{q}\sigma^{\mu\nu} P_L b|B_q}=0$, leptonic $B_q$ decays
do not help us to constrain the coefficient $C^{q, \ell}_T$. They also receive no contribution from the scalar operator.

In the case of semileptonic $B$ decays involving a vector meson $\tilde{V}$ in the final state, we have the following expression \cite{Sakaki:2013bfa}:
\begin{eqnarray}\label{eq:dBrV}
&&\frac{d\mathcal{B}(\bar{B}\rightarrow \tilde{V} \ell^- \bar{\nu}_{\ell})}{dq^2}=
\frac{G^2_F \tau_B |V_{qb}|^2}{24 \pi^3 m^2_{B}}
\Biggl\{ \nonumber\\
&&(|1 + C^{q, \ell}_{V_L}|^2 + |C^q_{V_R}|^2)\Biggl[\frac{1}{4}\Bigl(1 + \frac{m^2_{\ell}}{2q^2}\Bigl)
\Bigl(H^{\tilde{V}~2}_{V,+} + H^{\tilde{V}~2}_{V,-} + H^{\tilde{V}~2}_{V,0}\Bigl) + 
\frac{3}{8}\frac{m^2_{\ell}}{q^2} H^{\tilde{V}~2}_{V, t}\Biggl]   \nonumber \\
&&-2\Re\Bigl[(1+C^{q, \ell}_{V_L})C^{q *}_{V_R}\Bigl]\Biggl[
\frac{1}{4}\Bigl(1 + \frac{m^2_{\ell}}{2q^2}\Bigl)\Bigl(H^{\tilde{V}~2}_{V,0} 
+ 2 H^{\tilde{V}}_{V,+} H^{\tilde{V}}_{V,-}\Bigl)  + \frac{3}{8}\frac{m^2_{\ell}}{q^2} H^{\tilde{V}~2}_{V, t}\Biggl]  \nonumber \\
&&+\frac{3}{8}|C^{q, \ell}_P|^2 H^{\tilde{V}~2}_S
+ 2 |C^{q, \ell}_T|^2 \Bigl(1 + \frac{2 m^2_{\ell}}{q^2}\Bigl)\Bigl(H^{\tilde{V}~2}_{T,+} + H^{\tilde{V}~2}_{T,-} + H^{\tilde{V}~2}_{T,0}\Bigl)  \nonumber \\
&&+\frac{3}{4}\Re\Bigl[(1+C^{q, \ell}_{V_L} - C^{q}_{V_R})C^{q, \ell*}_P\Bigl]\frac{m_{\ell}}{\sqrt{q^2}}
H^{\tilde{V}}_S H^{\tilde{V}}_{V, t}\nonumber \\
&&-3\Re\Bigl[(1+C^{q, \ell}_{V_L})C^{q, \ell *}_T\Bigl]\frac{m_{\ell}}{\sqrt{q^2}}
\Bigl(H^{\tilde{V}}_{T,0} H^{\tilde{V}}_{V,0} + H^{\tilde{V}}_{T,+} H^{\tilde{V}}_{V,+} -H^{\tilde{V}}_{T,-} H^{\tilde{V}}_{V,-}\Bigl)\nonumber \\
&&+3\Re\Bigl[C^{q }_{V_R}C^{q, \ell *}_{T}\Bigl]\frac{m_{\ell}}{\sqrt{q^2}}\Bigl(H^{\tilde{V}}_{T,0} H^{\tilde{V}}_{V,0} + H^{\tilde{V}}_{T,+} H^{\tilde{V}}_{V,-} -H^{\tilde{V}}_{T,-} H^{\tilde{V}}_{V,+}\Bigl)
\Biggl\}\frac{(q^2-m^2_{\ell})^2}{q^2} |\vec{p}_{\tilde{V}}|.\end{eqnarray}
In addition, for a semileptonic $B$ decay into a pseudoscalar meson $P$, we have \cite{Sakaki:2013bfa}
\begin{eqnarray} \label{eq:dBrP}
&&\frac{d\mathcal{B}(\bar B\rightarrow P \ell^- \bar{\nu_{\ell}})}{dq^2}=\frac{G^2_F \tau_{B}|V_{qb}|^2}
{24 \pi^3 M^2_{B}}\Biggl\{ \nonumber\\
&&|1 + C^{q, \ell}_{V_L} + C^q_{V_R}|^2
\Biggl[ 
\Bigl( 1+\frac{m^2_\ell}{2q^2} \Bigl)\frac{H^{P~2}_{V, 0}}{4} + \frac{3}{8}\frac{m^2_{\ell}}{q^2}H^{P~2}_{V, t}
\Biggl] + \frac{3}{8}|C^{q, \ell}_S|^2 H^{P~2}_S\nonumber\\
&&+ 2 |C^{q, \ell}_T|^2 \Bigl( 1+\frac{2m^2_\ell}{q^2}\Bigl) H^{P~2}_T + 
\frac{3}{4}\Re\Bigl[ C^{q, \ell *}_S ( 1 + C^{q, \ell}_{V_L} +  C^{q}_{V_R}  )\Bigl]\frac{m_{\ell}}{\sqrt{q^2}}
H^{P}_{S} H^{P}_{V,t}\nonumber\\
&&-3\Re\Bigl[( 1 + C^{q, \ell}_{V_L} +  C^{q}_{V_R}  )C^{q, \ell*}_T \Bigl]\frac{m_{\ell}}{\sqrt{q^2}}H^{P}_{T} H^{P}_{V,0}
\Biggl\}\frac{(q^2-m^2_{\ell})^2}{q^2} |\vec{p}_{P}|.
\end{eqnarray}
In Eqs.~(\ref{eq:dBrV})~and~(\ref{eq:dBrP}), $q^2$ represents the four momentum squared transferred to the leptonic system, composed of the $\ell$
and $\bar{\nu}_{\ell}$. In addition, the magnitude of the three momemtum of the meson in the final state has been denoted
as $|\vec{p}_{P}|$  and  $|\vec{p}_{\tilde{V}}|$. The expression of $|\vec{p}_{\tilde{V}}|$ in terms of $q^2$ is given by
\begin{equation}\label{eq:vecprho}
|\vec{p}_{\tilde{V}}|=\frac{\sqrt{\Bigl[(M_{B}-M_{\tilde{V}})^2-q^2\Bigl]\Bigl[(M_{B}+M_{\tilde{V}})^2-q^2\Bigl]}}{2M_{B}}.
\end{equation}
The equation for $|\vec{p}_{P}|$ is analogous to Eq.~(\ref{eq:vecprho}), with the straightforward replacement of $M_{\tilde{V}}$ by $M_P$.

Before continuing we would like to introduce some notation. In what follows,  we will use the symbol $\ell$ to denote either an electron, a muon or a tau lepton, on the other hand $\ell'$ will refer only to electrons and muons.

On the experimental side it is not always transparent what the branching ratios of the semileptonic modes represent. For the purpose of determining $|V_{qb}|$ under the assumption of the SM, it is advantageous to combine information from decays with electrons and muons in the final state.  Hence, they are often presented as $\mathcal{B}(\bar{B} \to \tilde{V} (P) \ell'^- \bar{\nu}_{\ell'})$, where $\ell'$ is an electron or a muon.   The exact definition of these branching ratios is often ambiguous. For example, they could be averages over both leptons, or involve just one of the two. In order to test the SM and search for NP, it would be instrumental to have measurements separately for electrons and muons. It would then also be pertinent that these analyses do not rely on theoretical assumptions such as a form factor parametrization, and are presented independently for the $B^-$ and $\bar{B}^0$ modes, which is currently not always the case. In this work, we will employ PDG averages whenever available, averaging their values for the $B^-$ and $\bar{B}^0$ channels, and interpret the experimental data for $\bar{B} \to \tilde{V} (P) \ell'^- \bar{\nu}_{\ell'}$ as averages over electronic and muonic modes.

Our aim is to search for potential NP effects, entering through the Wilson coefficients of the operators in Eq.~(\ref{eq:operators}), as well as to determine the $|V_{qb}|$ while allowing for new contributions. We follow the strategy introduced in Ref.~\cite{Banelli:2018fnx}, which is discussed in more detail in Section~\ref{subsec:strategy}.

Our analysis works under the general asumption of NP in $\tau$ leptons as well as electrons and muons. Therefore,
in principle, we should fit for three different coefficients $C^{e}_{X}$, $C^{\mu}_{X}$ 
and  $C^{\tau}_{X}$, where $X=P,S,V, T$. However, to simplify the analysis, we will consider scenarios where the coefficients for the electron and muon are correlated according to
\begin{eqnarray}
\label{eq:NPemucorr}
C^e_X&=& f^e_{\mu} C^{\mu}_X,
\end{eqnarray}
where $f^e_{\mu}$ is a constant. In general, there are three possibilities which can be addressed in a model-independent way: $C^e_X<C^{\mu}_X$, $C^e_X\sim C^{\mu}_X$ and $C^{\mu}_X < C^e_X$. For the purposes of illustration, we consider three possible scenarios:
\begin{eqnarray}
\label{eq:scenarios}	
\text{Scenario 1:}&& C^{e}_{X}=0.1 C^{\mu}_{X},\nonumber\\
\text{Scenario 2:}&& C^{e}_{X}= C^{\mu}_{X},\nonumber\\
\text{Scenario 3:}&& C^{e}_{X}= 10 C^{\mu}_{X}.
\end{eqnarray}

To evaluate the confidence regions allowed for our NP Wilson coefficients $C_X$ at the one-$\sigma$ level we consider the following equation:
\begin{eqnarray}
\Bigl|\mathcal{O}(C_X)_{\rm Theory}-\mathcal{O}_{\rm Experiment}\Bigl| < \sqrt{\sigma^2_{\rm Theory} + \sigma^2_{\rm Experiment}},
\end{eqnarray}
where $\mathcal{O}$ is a given physical observable with central value $\mathcal{O}_{\rm Experiment}$, and $\sigma_{\rm Theory}$ and $\sigma_{\rm Experiment}$ are the theoretical and experimental uncertainties respectively. The theoretical uncertainty is estimated based on
\begin{eqnarray}
\sigma^2_{\rm Theory}&=&\sum^{N}_{i=1}\Bigl(\frac{\partial \mathcal{O}(C_X)_{\rm Theory}}{\partial x_i}\Bigl)^2\Delta x^2_i,
\end{eqnarray}
where $x_i$ are the individual theoretical inputs on which $\mathcal{O}_{\rm Theory}$ depends. Notice that the theoretical uncertainty is in general not constant and depends on the particular value assigned to $C_X$.

\subsection{Introduction of New Physics Effects} \label{sec:RGE}

In this work, we will consider NP contributions
arising at a high energy scale of $\mu = \text{1~TeV}$. The evolution down to the bottom scale $\mu = m_b$ is given by the following matrix \cite{Iguro:2020keo}:
\begin{eqnarray}
\left(
\begin{array}{c}
C^{q, \ell}_{V_L}(m_b)\\
C^{q}_{V_R}(m_b)\\
C^{q, \ell}_S(m_b)\\
C^{q, \ell}_P(m_b)\\
C^{q, \ell}_T(m_b)\\
\end{array}
\right)
\simeq
\left(
\begin{array}{ccccc}
1 & 0 & 0 & 0 & 0\\
0 & 1 &0&0&0\\
0 & 0 & 1.71 & 0 & -0.27 \\
0 & 0 & 0 & 1.71 & 0.27 \\
0 & 0 & 0 & 0 & 0.84 \\
\end{array}
\right)
\!\!\left(
\begin{array}{c}
C^{q, \ell}_{V_L}(\text{1~TeV})\\
C^{q}_{V_R}(\text{1~TeV})\\
C^{q, \ell}_S(\text{1~TeV})\\
C^{q, \ell}_P(\text{1~TeV})\\
C^{q, \ell}_T(\text{1~TeV}).\\
\end{array}
\right).
\label{eq:RGEepsilon_b}
\end{eqnarray}
 Consequently, the short-distance coefficients at the bottom scale entering in the $B$-meson branching fractions are essentially functions of the coefficients at $\mu = \text{1~TeV}$. Notice that, as the result of the renormalization group evolution, in our basis the coefficient $C_T^{q, \ell}(\text{1~TeV})$ yields contributions to $C_S^{q, \ell}$ and $C_P^{q, \ell}$ at the scale $\mu = m_b$.

In the following discussion, we will determine allowed regions for the Wilson coefficients at the scale $\mu=\text{1~TeV}$, making the connection with the bottom scale implicitly through Eq.~(\ref{eq:RGEepsilon_b}). From this point onwards and  in order to simplify the notation, wherever we write $C^{q, \ell}_i$ (where $i \in \{V_L, V_R, S, P, T\}$), i.e., without specifying the scale $\mu$, we are referring to the coefficients at $\mu = \text{1~TeV}$. There are a few exceptions, for which we will clearly indicate in the text that we are considering the scale $\mu = m_b$.

\boldmath
\subsection{Strategy for the Determination of $|V_{ub}|$ and $|V_{cb}|$} \label{subsec:strategy}
\unboldmath

Typically, the $|V_{qb}|$ are determined assuming no NP contributions (see for instance Ref.~\cite{Amhis:2019ckw}). We have proposed a strategy to take into account potential NP effects arising from (pseudo)-scalar operators when determining $|V_{ub}|$ from exclusive $b \to u \ell \bar{\nu}_\ell$ modes in Ref.~\cite{Banelli:2018fnx}. In this section, we describe this strategy, generalizing it to $b \to q \ell \bar{\nu}_\ell$ and $|V_{qb}|$ for all operators in Eq.~(\ref{eq:operators}).

There are analyses that have evaluated $|V_{ub}|$ and $|V_{cb}|$ considering simultaneously 
NP contributions in scalar, vector and tensor operators, see for instance \cite{Crivellin:2014zpa,Bernlochner:2014ova,Jung:2018lfu,Iguro:2020cpg}. These studies are based on simultaneous fits to $|V_{qb}|$ and the different NP Wilson coefficients, and take advantage of the correlation
between these quantities established in branching fractions and differential distributions.

Our strategy follows a different approach, which allows us to extract NP effects independently of $|V_{qb}|$ for $q=u,c$. The key steps are as follows:
\begin{enumerate}
\item We first establish the possible regions for the different NP Wilson coefficients using ratios of branching fractions where 
$|V_{qb}|$ cancels out.
\item We then proceed with the evaluation of $|V_{qb}|$ by substituting the allowed NP regions established in the previous step inside
specific observables sensitive to the different NP Wilson coefficients and $|V_{qb}|$. We would like to stress that --- since the NP contributions
were already extracted in a $|V_{qb}|$ independent fashion --- observables such as branching fractions only depend on one unknown parameter, which is precisely $|V_{qb}|$. We can now finally extract this parameter, obtaining a value that takes potential NP contributions into account.
\end{enumerate}
This procedure can, in certain cases, be further refined to account for correlations between
the different observables and parameters considered during the analysis, as done in 
Ref.~\cite{Banelli:2018fnx}.

\boldmath
\section{The $b \to u$ Transitions} \label{sec:btou}
\unboldmath
The discussion in this section focuses on the determination of the $b \to u \ell \bar{\nu}_\ell$ short-distance coefficients, and complements the analysis in Ref.~\cite{Banelli:2018fnx}. There, the possibility of having NP effects from
scalar and pseudoscalar interactions entering in different leptonic and semileptonic decays was explored. Here, we will consider in addition the vector and tensor structures, taking into account the renormalization group evolution as introduced in Section~\ref{sec:RGE}. For completeness, we will include the (pseudo)-scalar operators as well, switching on their coefficients at $\mu = 1$ TeV, whereas the results in Ref.~\cite{Banelli:2018fnx} relate to $\mu = m_b$.

Our determinations will employ the following experimental measurements for $B^-$ leptonic decays \cite{Prim:2019gtj,Tanabashi:2018oca}:
\begin{eqnarray} \label{eq:brLepExp}
{\mathcal B}(B^-\to\mu^- \bar\nu_\mu)&=&(5.3 \pm 2.2)\times10^{-7},\nonumber\\
{\mathcal B}(B^-\to\tau^-\bar\nu_\tau)&=& (1.09\pm0.24)\times10^{-4}.
\end{eqnarray}
For the corresponding process involving electrons, we will consider the upper bound obtained by the Belle collaboration in 2007 \cite{Satoyama:2006xn}:
\begin{eqnarray}\label{eq:Belle-enu-limit}
{\mathcal B}(B^- \to e^- \bar\nu_e) < 9.8 \times 10^{-7} \, \mbox{(90\% C.L.)}.
\end{eqnarray}

Within the SM, we have
\begin{eqnarray}
{\mathcal B}(B^-\to\tau^-\bar\nu_\tau)|_{\rm SM} &=& (8.58 \pm 0.71) \times 10^{-5}, \label{eq:leptutauSM} \\
{\mathcal B}(B^-\to\mu^-\bar\nu_\mu)|_{\rm SM} &=& (3.86 \pm 0.32) \times 10^{-7}, \label{eq:leptumuSM} \\
{\mathcal B}(B^-\to e^-\bar\nu_e)|_{\rm SM} &=& (9.03 \pm 0.75) \times 10^{-12},\label{eq:leptueSM}
\end{eqnarray}
where we have used \cite{Amhis:2019ckw}
\begin{eqnarray}
\label{eq:Vub}	
\left| V_{ub} \right|&=& (3.67 \pm 0.15) \times 10^{-3}
\end{eqnarray}
and \cite{Aoki:2019cca}
\begin{eqnarray}
f_{B^-}&=&0.1900 \pm 0.0013~\rm{GeV}.	
\end{eqnarray}

The value of $|V_{ub}|$ in Eq.~(\ref{eq:Vub}) is determined by HFLAV from $B \to \pi \ell' \nu$ data, where $\ell'$ is an electron or a muon, with form factors from lattice QCD (LQCD) and light-cone sum-rule (LCSR) calculations, assuming no NP in the light leptonic generations. This should be kept in mind when considering our evaluations for the branching fractions in the SM given in Eqs.~(\ref{eq:leptutauSM}),~(\ref{eq:leptumuSM})~and~(\ref{eq:leptueSM}). As argued in  
Ref.~\cite{Banelli:2018fnx}, decoupling the determination of $|V_{ub}|$ from
potential NP effects in $b\rightarrow u \ell \bar{\nu}_{\ell}$ transitions requires a
more careful approach which is developed further in this work. As a result, the values in Eqs.~(\ref{eq:leptutauSM}),~(\ref{eq:leptumuSM})~and~(\ref{eq:leptueSM}) serve an illustrative purpose and should not be seen as predictions. Moreover, we would like to point out that there is a long-standing tension between exclusive, as in Eq.~(\ref{eq:Vub}), and inclusive determinations of $|V_{ub}|$. For the latter, HFLAV gives the value \cite{Amhis:2019ckw}:
\begin{equation} \label{eq:VubInclHFLAV}
    |V_{ub}|_{\rm incl.} = (4.32 \pm 0.12_{-0.13}^{+0.12}) \times 10^{-3}.
\end{equation}

For the present analysis, we have also taken into account the semileptonic decay processes 
involving the $\rho$ vector meson in the final state. 
From the results reported by the Belle collaboration \cite{Sibidanov:2013rkk}, we obtain
\begin{eqnarray}\label{eq:Exprho}
\Braket{{\mathcal B}(\bar B^0\rightarrow \rho^+ \ell'^- \bar{\nu}_{\ell'})}_{[\ell'=~ e, \mu],~q^2\leq 12~\rm{GeV}^2}&=&(1.90 \pm 0.20)\times 10^{-4},\nonumber\\
2\Braket{{\mathcal B}( B^-\rightarrow \rho^0 \ell'^- \bar{\nu}_{\ell'})}_{[\ell'=~ e, \mu],~q^2\leq 12~\rm{GeV}^2}&=&(2.03 \pm 0.16) \times 10^{-4}.
\end{eqnarray}
Here we have taken into account only the $q^2$ region satisfying $q^2\leq12~\rm{GeV}^2$, as indicated by the subindex. The reason for this is that, on the theory side, we make use of form factors calculated using LCSRs, which are applicable at low values of $q^2$ \cite{Straub:2015ica}. Details on the form factors are provided in Appendix~\ref{sec:FFBpr}. We interpret the experimental results in Eq.~(\ref{eq:Exprho}) as an average 
over electrons and muons, as discussed in Sec.~\ref{sec:theory-framework}. Using the isospin symmetry,
we combine the results in Eq.~(\ref{eq:Exprho}) to obtain
\begin{eqnarray}\label{eq:Brhoaverage}
\Braket{{\mathcal B}(\bar B\rightarrow \rho \ell'^- \bar{\nu}_{\ell'})}_{[\ell'=~ e, \mu],~q^2\leq 12~\rm{GeV}^2}&=&(1.98 \pm 0.12) \times 10^{-4}.
\end{eqnarray}

For semileptonic decays including a pseudo-scalar meson in the final state, we consider the $\bar{B}\rightarrow \pi \ell'^- \bar{\nu}_{\ell'}$
processes, where $\ell'=e,~\mu$. On the experimental side, we have \cite{Tanabashi:2018oca}
\begin{eqnarray}\label{eq:Exppi}
\Braket{{\mathcal B}(\bar B^0\rightarrow \pi^+ \ell'^- \bar{\nu}_{\ell'})}_{[\ell'=~ e, \mu]}&=&(1.50 \pm 0.06)\times 10^{-4},\nonumber\\
2\Braket{{\mathcal B}( B^-\rightarrow \pi^0 \ell'^- \bar{\nu}_{\ell'})}_{[\ell'=~ e, \mu]}&=&(1.56 \pm 0.05) \times 10^{-4}.
\end{eqnarray}
Making use of the isospin symmetry, we get
\begin{eqnarray}\label{eq:piexp}
\Braket{{\mathcal B}(\bar{B}\rightarrow \pi \ell'^- \bar{\nu}_{\ell'})}_{[\ell'=~ e, \mu]}&=&(1.53 \pm 0.04)\times 10^{-4}.
\end{eqnarray}
In addition, we consider the following upper bound:
\begin{equation} \label{eq:BpitaunuBound}
{\mathcal B}(\bar{B}^0 \rightarrow \pi^+ \tau^- \bar{\nu}_{\tau}) < 2.5 \times 10^{-4} \, \mbox{(90\% C.L.)},
\end{equation}
as obtained by the Belle Collaboration \cite{Hamer:2015jsa}. In Appendix~\ref{sec:FFBpr} we provide details on the $B \to \pi$ form factors.

When studying potential NP contributions to these decays, we have to take into account that they are actually used to determine $|V_{ub}|$ (see, e.g., Ref.~\cite{Amhis:2019ckw}). Therefore, the value of $|V_{ub}|$ may be affected by NP, and we have to construct observables independent of $|V_{ub}|$ for consistency. To that end, we consider ratios of branching fractions to evaluate the possible values of the NP Wilson coefficients.

The $|V_{ub}|$-independent leptonic ratios are 
\begin{equation}\label{eq:Ru}
R^{\ell_1}_{\ell_2}\equiv 
\frac{m^2_{\ell_2}}{m^2_{\ell_1}}\left(\frac{M_{B^-}^2-m_{\ell_2}^2}{M_{B^-}^2-m_{\ell_1}^2}\right)^2
\frac{{\mathcal B}(B^-\to\ell_1^-\bar\nu_{\ell_1})}{{\mathcal B}(B^-\to \ell_2^-\bar\nu_{\ell_2})},
\end{equation}
where $\ell_1$, $\ell_2$ are any of the leptons $e, \mu, \tau$. Additionally, we include the following combinations of leptonic and semileptonic branching fractions:
\begin{eqnarray}\label{eq:leptonicoversemileptonicrho-theo}
\mathcal{R}^{\ell}_{\Braket{e, \mu}; \rho ~ [q^2\leq 12]~\rm{GeV}^2}&\equiv&\mathcal{B}(B^-\rightarrow \ell^- \bar{\nu}_{\ell})/
\Braket{{\mathcal B}(\bar B \rightarrow \rho \ell'^- \bar{\nu}_{\ell'})}_{[\ell'=~ e, \mu],~q^2\leq 12~\rm{GeV}^2},
\end{eqnarray}
\begin{eqnarray}\label{eq:leptonicoversemileptonicpi}
\mathcal{R}^{\ell}_{\Braket{e, \mu}; \pi}&\equiv&\mathcal{B}(B^-\rightarrow \ell^- \bar{\nu}_{\ell})/
\Braket{{\mathcal B}(\bar B \rightarrow \pi \ell'^- \bar{\nu}_{\ell'})}_{[\ell'=~ e, \mu]},
\end{eqnarray}
with $\ell=e, \mu, \tau$. Finally, we use also the ratio of semileptonic processes
\begin{eqnarray}
\mathcal{R}^{\Braket{e, \mu};\rho~[q^2 \leq 12]~\rm{GeV}^2}_{\Braket{e, \mu};\pi}
&\equiv&
\Braket{ \mathcal{B}(\bar{B}\rightarrow \rho \ell^{\prime -} \bar{\nu}_{\ell'})}_{[\ell'=~ e, \mu],~q^2\leq 12~\rm{GeV}^2}/
\Braket{\mathcal{B}(\bar{B}\rightarrow \pi \ell^{\prime -} \bar{\nu}_{\ell'})}_{[\ell'=e, \mu]}.
\label{eq:semirhosemipi}
\end{eqnarray}

With the currently available branching ratios in Eqs.~(\ref{eq:brLepExp}),~(\ref{eq:Brhoaverage})~and~(\ref{eq:piexp}), we have the following ratios at our disposal:
\begin{equation} \label{eq:btouObs}
    R_\mu^\tau, \  \mathcal{R}^{\tau}_{\Braket{e, \mu}; \rho ~ [q^2\leq 12]~\rm{GeV}^2}, \  \mathcal{R}^{\mu}_{\Braket{e, \mu}; \rho ~ [q^2\leq 12]~\rm{GeV}^2}, \  \mathcal{R}^{\tau}_{\Braket{e, \mu}; \pi}, \  \mathcal{R}^{\mu}_{\Braket{e, \mu}; \pi}, \  \mathcal{R}^{\Braket{e, \mu};\rho~[q^2 \leq 12]~\rm{GeV}^2}_{\Braket{e, \mu};\pi},
\end{equation}
where we use the notation defined in Eqs.~(\ref{eq:Ru})--(\ref{eq:semirhosemipi}). Given the number of independent branching fractions, at most three of these observables are independent. However, they are sensitive to different short-distance coefficients. In the following sections, we will consider each of the structures $i \in \{P, S, V_L, V_R, T\}$ independently, constraining the coefficients $C_i^{u, \mu}$ and $C_i^{u, \tau}$ through three appropriate ratios of branching fractions. For the $C_i^{u, e}$ coefficient, we will consider the three scenarios from Eq.~(\ref{eq:scenarios}).

There are two exceptions to the approach discussed above: The Wilson coefficient $C_{V_R}$ is lepton-flavour universal (see Eq.~(\ref{eq:universality})), and in case of the $C_S$ short-distance coefficient, we have only sensitivity to $C_S^{u, e}(\text{1~TeV})$ and $C_S^{u, \mu}(\text{1~TeV})$ through $\Braket{{\mathcal B}(\bar{B}\rightarrow \pi \ell^{\prime -} \bar{\nu}_{\ell'})}_{[\ell'=~ e, \mu]}$, so we will constrain these coefficients without assuming any relation between them.

\subsection{Constraints on (Pseudo)-Scalar Wilson Coefficients}
We will start our analysis of the $b\rightarrow u$ transitions by considering new contributions to the Wilson coefficients of the (pseudo)-scalar operators 
${\cal O}^{u, \ell}_{P}$ and ${\cal O}^{u, \ell}_{S}$ introduced in Eq.~(\ref{eq:operators}). We have considered these operators before in Ref.~\cite{Banelli:2018fnx}. The main difference with respect to our previous analysis is that here we will include RGE effects, considering the short-distance coefficients at the scale $\mu = \text{1~TeV}$. From Eq.~(\ref{eq:RGEepsilon_b}), we can see that this involves a rescaling of the coefficients at $\mu = m_b$. In addition, we use more recent data on $\mathcal{B}(B^- \to \mu^- \bar{\nu}_\mu)$ \cite{Prim:2019gtj}.

\boldmath
\subsubsection{New Physics Entering through $C_P^{u, \ell}$} \label{subsubsec:CPul}
\unboldmath
The short-distance coefficient $C_P^{u, \ell}$ contributes to all branching ratios considered here with the exception of $\Braket{{\mathcal B}(\bar{B}\rightarrow \pi \ell^{\prime -} \bar{\nu}_{\ell'})}_{[\ell'=~ e, \mu]}$. Still we may consider ratios involving this decay mode, as it may be still be used to cancel $|V_{ub}|$. Actually, all six observables in Eq.~(\ref{eq:btouObs}) are sensitive to the $C_P^{u, \ell}$ coefficient, but, as stated before, they are not all independent. We will consider the following three ratios:
\begin{equation} \label{eq:cPobsList}
    R_\mu^\tau, \qquad \mathcal{R}^{\mu}_{\Braket{e, \mu}; \rho ~ [q^2\leq 12]~\rm{GeV}^2}, \qquad \mathcal{R}^{\Braket{e, \mu}; \rho ~ [q^2\leq 12]~\rm{GeV}^2}_{\Braket{e, \mu}; \pi}.
\end{equation}
We have verified, in the scenario where $C_P^{u, e} = C_P^{u, \mu}$, that a different subset of observables would not have a meaningful impact on the results. Interestingly, of the ratios in Eq.~(\ref{eq:cPobsList}), the latter two do not involve a $\tau$ in the final state, so they constrain only $C_P^{u, \mu}$, whereas $R_\mu^\tau$ is sensitive to both $C_P^{u, \mu}$ and $C_P^{u, \tau}$.

\begin{center}
\begin{figure}[H]
\begin{center}
\includegraphics[width=0.48\textwidth]{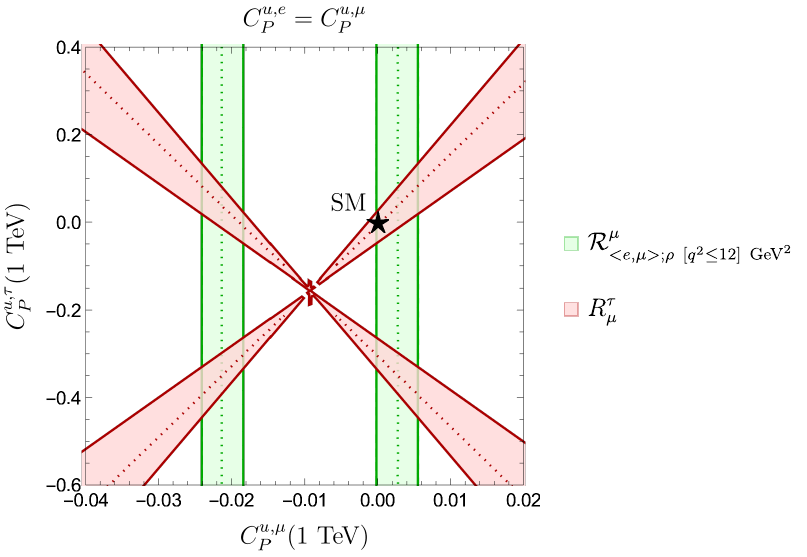} \\
\includegraphics[width=0.48\textwidth]{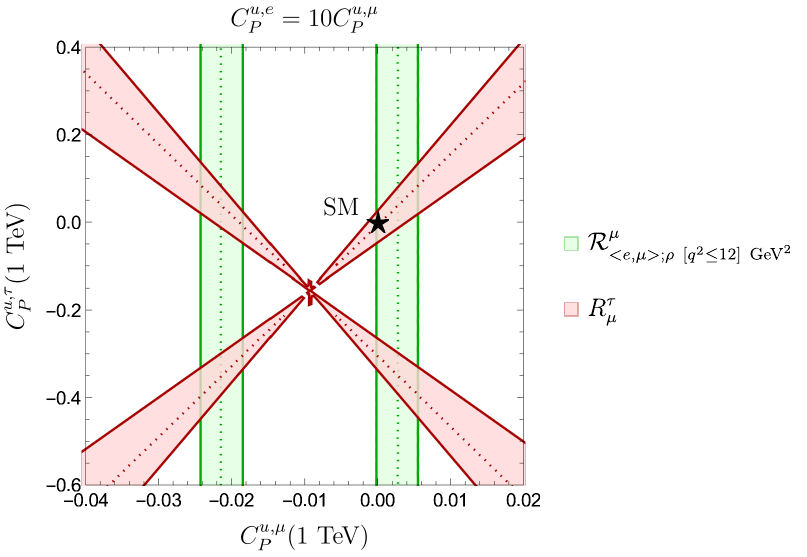}
\includegraphics[width=0.48\textwidth]{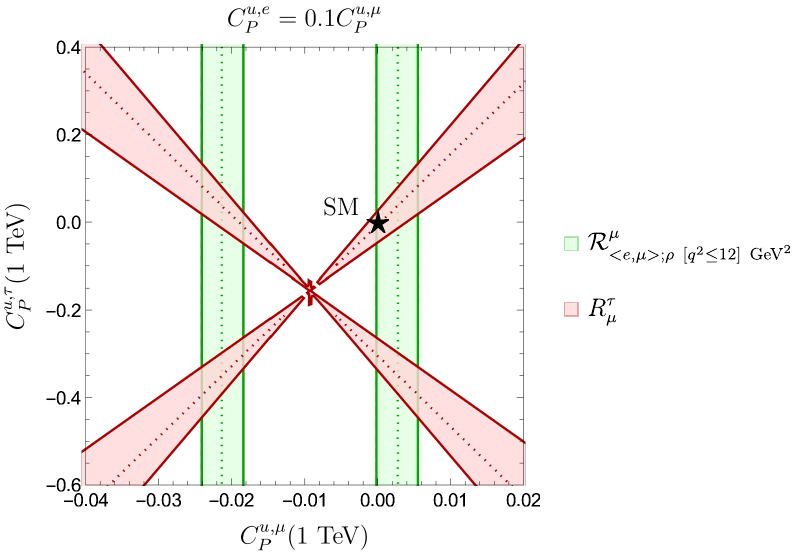}
\caption{Constraints on $C_P^{u, \mu}$ and $C_P^{u, \tau}$ under different assumptions about the relation between $C_P^{u, e}$ and $C_P^{u, \mu}$, as indicated in the plots.}
\label{fig:btouCP}
\end{center}
\end{figure}
\end{center}

As it turns out, there is no agreement between theory and experiment for the observable $\mathcal{R}^{\Braket{e, \mu}; \rho ~ [q^2\leq 12]~\rm{GeV}^2}_{\Braket{e, \mu}; \pi}$ in any of the three scenarios for $C_P^{u, e}$. We will address this in the next paragraph. First, we present in Fig.~\ref{fig:btouCP} the constraints on $C_P^{u, \mu}$ and $C_P^{u, \tau}$ from the other two observables listed in Eq.~(\ref{eq:cPobsList}). We note that the results are very insensitive to the assumptions about $C_P^{u, e}$. In each scenario, we obtain four distinct regions where both constraints overlap, one of which includes the SM.

Studying the ratio $\mathcal{R}^{\Braket{e, \mu}; \rho ~ [q^2\leq 12]~\rm{GeV}^2}_{\Braket{e, \mu}; \pi}$ further, we find that the theoretical value and experimental data only barely disagree. In fact, small variations in the uncertainty may have a significant impact in Fig.~\ref{fig:btouCP}, yielding a contour that agrees with all four regions defined by the overlap of $R_\mu^\tau$ and $\mathcal{R}^{\mu}_{\Braket{e, \mu}; \rho ~ [q^2\leq 12]~\rm{GeV}^2}$. Specifically, we find that, on the basis of this observable, we cannot exclude these solutions at the $1~\sigma$ level.

\boldmath
\subsubsection{New Physics Entering through $C_S^{u, \ell}$}
\unboldmath
The only observable sensitive to $C_S^{u, \ell}$ is $\Braket{{\mathcal B}(\bar{B}\rightarrow \pi \ell^{\prime -} \bar{\nu}_{\ell'})}_{[\ell'=~ e, \mu]}$. Consequently, we cannot constrain $C_S^{u, \tau}$ and we will hence consider $C_S^{u, e}$ and $C_S^{u, \mu}$ only. We will still determine the bounds from all three ratios involving the $\Braket{{\mathcal B}(\bar{B}\rightarrow \pi \ell^{\prime -} \bar{\nu}_{\ell'})}_{[\ell'=~ e, \mu]}$ branching fraction; the distinction is that the cancellation of $|V_{ub}|$ is obtained from different modes in each ratio. Specifically, we take
\begin{equation} \label{eq:btouCSObs}
    \mathcal{R}^{\tau}_{\Braket{e, \mu}; \pi}, \qquad  \mathcal{R}^{\mu}_{\Braket{e, \mu}; \pi}, \qquad  \mathcal{R}^{\Braket{e, \mu};\rho~[q^2 \leq 12]~\rm{GeV}^2}_{\Braket{e, \mu};\pi}.
\end{equation}

\begin{center}
\begin{figure}[H]
\begin{center}
\includegraphics[width=0.30\textwidth]{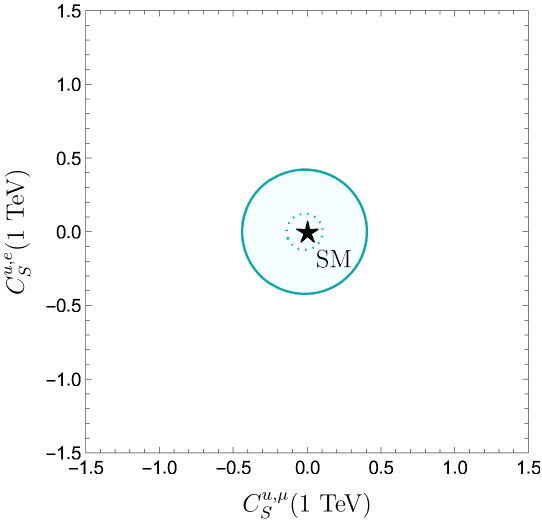}
\includegraphics[width=0.30\textwidth]{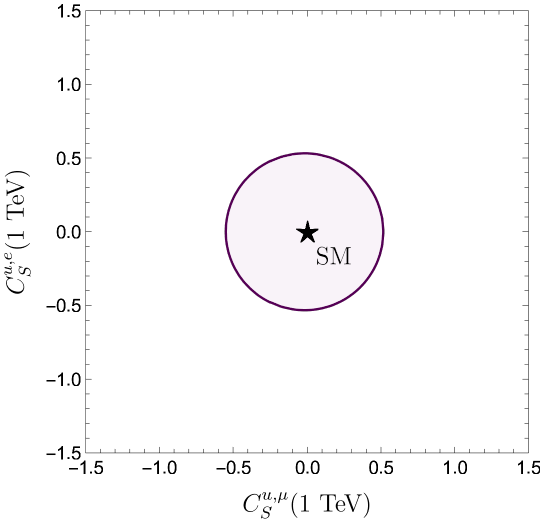}
\includegraphics[width=0.30\textwidth]{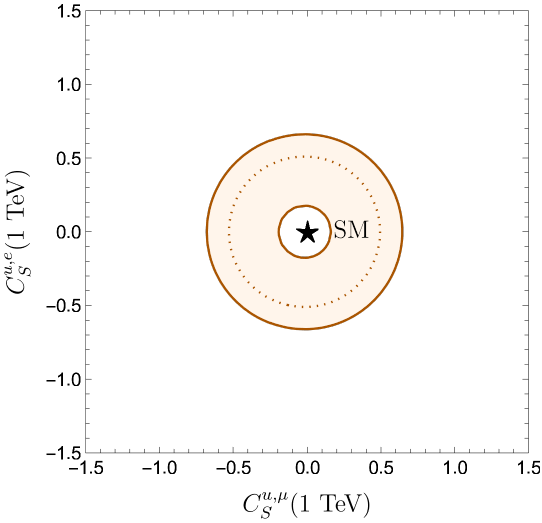}
\includegraphics[width=0.60\textwidth]{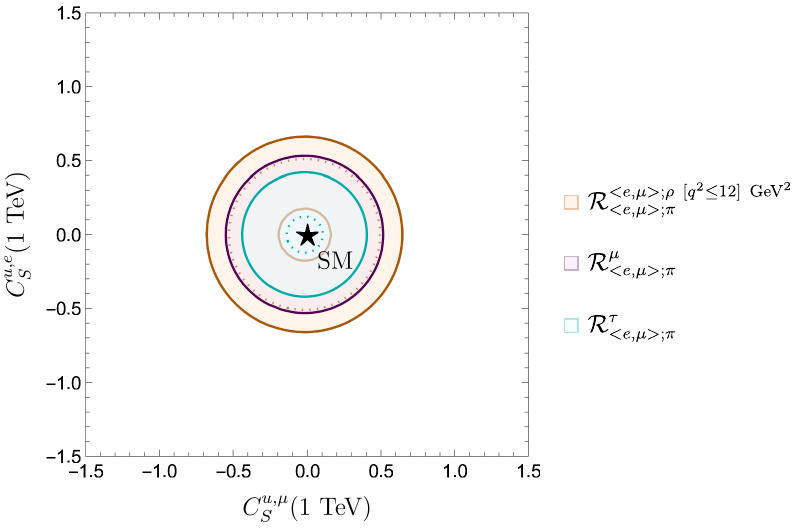}
\caption{Constraints in the $C_S^{u, \mu}$--$C_S^{u, e}$ plane following from the ratios of branching fractions listed in the text. The top row shows the allowed regions from each individual observable. On the bottom, we show them all together.}
\label{fig:btouCS}
\end{center}
\end{figure}
\end{center}

The constraints in the $C_S^{u, \mu}$--$C_S^{u, e}$ plane corresponding to the observables in Eq.~(\ref{eq:btouCSObs}) are shown in Fig.~\ref{fig:btouCS}. Since they all overlap, we show them for clarity separately on the top row, and all together on the bottom. In case of the $\mathcal{R}^{\mu}_{\Braket{e, \mu}; \pi}$ ratio, we note that there is no solution for the central value; a region only appears once the uncertainties are taken into account. Furthermore, the ratio $\mathcal{R}^{\Braket{e, \mu};\rho~[q^2 \leq 12]~\rm{GeV}^2}_{\Braket{e, \mu};\pi}$ yields a doughnut-shaped constraint with a hole around the SM point. However, this is slightly over $1~\sigma$, i.e., not a significant deviation. 

\subsection{Constraints on Vector Wilson Coefficients}
\label{subsec:Vector}
We continue by considering the vector operators $\mathcal{O}_{V_L}^{u, \ell}$ and $\mathcal{O}_{V_R}^{u, \ell}$. First we will allow for NP to enter through $C_{V_L}^{u, \ell}$, where $\ell$ is any of the charged leptons $e$, $\mu$ and $\tau$. Next, we will consider a non-vanishing $C_{V_R}^u$, which is lepton-flavour universal, as we discussed in Section~\ref{sec:theory-framework}.

\boldmath
\subsubsection{New Physics Entering through $C^{u, \ell}_{V_L}$}
\unboldmath
Allowing NP to enter only through $C^{u, \ell}_{V_L}$, the expressions in Eqs.~(\ref{eq:leptonic}), (\ref{eq:dBrV}) and (\ref{eq:dBrP}) take the following form:
\begin{eqnarray}\label{eq:leptonicsimp}
\mathcal{B}(B^-\rightarrow \ell^- \bar{\nu}_{\ell})=\left.\mathcal{B}(B^-\rightarrow \ell^- \bar{\nu}_{\ell})\right|_{\rm SM} 
\left| 1 + C^{u, \ell}_{V_L} \right|^2,
\end{eqnarray}
\begin{eqnarray}\label{eq:dBrVisimp}
&&\frac{d\mathcal{B}(\bar{B}\rightarrow \rho \ell^- \bar{\nu}_{\ell})}{dq^2} =
\left.\frac{d\mathcal{B}(\bar{B}\rightarrow \rho \ell^- \bar{\nu}_{\ell})}{dq^2}\right|_{\rm SM} \left|1 + C^{u, \ell}_{V_L}\right|^2
\end{eqnarray}
and
\begin{eqnarray} \label{eq:dBrPisimp}
&&\frac{d\mathcal{B}(\bar B\rightarrow \pi \ell^- \bar{\nu_{\ell}})}{dq^2} = \left.\frac{d\mathcal{B}(\bar B\rightarrow \pi \ell^- \bar{\nu_{\ell}})}{dq^2}\right|_{\rm SM} \left|1 + C^{u, \ell}_{V_L}\right|^2.
\end{eqnarray}
Given that all these expressions have exactly the same dependence with respect to $C^{u, \ell}_{V_L}$, any ratio of these quantities involving the same lepton flavour
will lead to the cancellation of the NP contributions. Consequently, to obtain constraints on the vector short-distance contributions,
we have to include ratios of leptonic and semileptonic branching ratios with different leptonic content in numerator and denominator. 
Hence, from the set of observables in the list in
Eq.~(\ref{eq:btouObs}) the only relevant ratios are
\begin{eqnarray}\label{eq:leptonicoversemileptonicrhotau}
R^{\tau}_{\mu}, \qquad \mathcal{R}^{\tau}_{\Braket{e, \mu}; \rho ~ [q^2\leq 12]~\rm{GeV}^2},\qquad \mathcal{R}^{\tau}_{\Braket{e, \mu}; \pi}.
\end{eqnarray}
Depending on the correlation between $C_{V_L}^{u, e}$ and $C_{V_L}^{u, \mu}$, some of the other observables in Eq.~(\ref{eq:btouObs}) may be considered as well. However, as indicated before, they are not independent. In addition, the observables in Eq.~(\ref{eq:leptonicoversemileptonicrhotau}), with the semileptonic branching fractions in $\mathcal{R}^{\tau}_{\Braket{e, \mu}; \rho ~ [q^2\leq 12]~\rm{GeV}^2}$ and $\mathcal{R}^{\tau}_{\Braket{e, \mu}; \pi}$ replaced by ${\mathcal B}(\bar B\rightarrow \rho \mu^- \bar{\nu}_{\mu})$ and ${\mathcal B}(\bar{B}\rightarrow \pi \mu^- \bar{\nu}_{\mu})$, respectively, would be the obvious candidates to constrain $C_{V_L}^{u, \mu}$ and $C_{V_L}^{u, \tau}$ in the future, should these measurements become available.

\begin{center}
\begin{figure}[H]
\begin{center}
\includegraphics[width=0.6\textwidth]{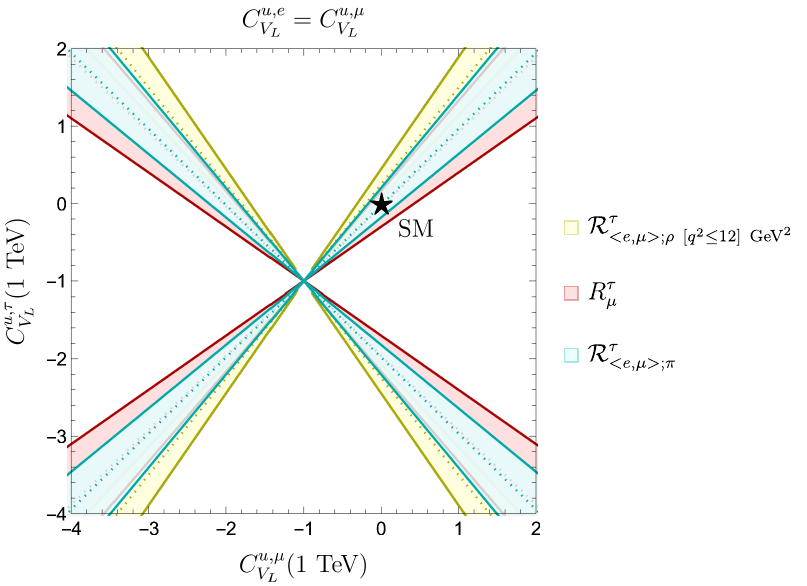}
\caption{Constraints in the $C_{V_L}^{u, \mu}$--$C_{V_L}^{u, \tau}$ plane in the scenario $C_{V_L}^{u, e} = C_{V_L}^{u, \mu}$.}
\label{fig:CVLmu-CVLtau}
\end{center}
\end{figure}
\end{center}

Let us start by making the assumption
\begin{eqnarray} \label{eq:AssumpemuVL}
C^{u, e}_{V_L}=C^{u, \mu}_{V_L},
\end{eqnarray}
allowing us to constrain the independent coefficients $C^{u, \mu}_{V_L}$ and $C^{u, \tau}_{V_L}$. The corresponding results are shown
in Fig.~\ref{fig:CVLmu-CVLtau}, where the cross-shaped coloured areas illustrate the NP values for the Wilson coefficients that are allowed by the observables in Eq.~(\ref{eq:leptonicoversemileptonicrhotau}). One of the four branches agrees with the SM, though the constraint from $\mathcal{R}^{\tau}_{\Braket{e, \mu}; \rho ~ [q^2\leq 12]~\rm{GeV}^2}$ is slightly more than $1~\sigma$ away.

Unfortunately, it is not possible to use the regions in Fig.~\ref{fig:CVLmu-CVLtau} to determine $|V_{ub}|$, or to make predictions for branching ratios that have not yet been measured, such as $\mathcal{B}(B^-\rightarrow e^-\bar{\nu}_{e})$, since they are not further constrained. Although at some point the contours following from the three ratios may no longer overlap, other effects will start to play a role. Since $|V_{ub}|$ cancels in the ratios in Eq.~(\ref{eq:leptonicoversemileptonicrhotau}), the region in Fig.~\ref{fig:CVLmu-CVLtau} may correspond to values of $|V_{ub}|$ that are completely unrealistic when compared to other CKM constraints.

\begin{center}
\begin{figure}[H]
\begin{center}
\includegraphics[width=0.6\textwidth]{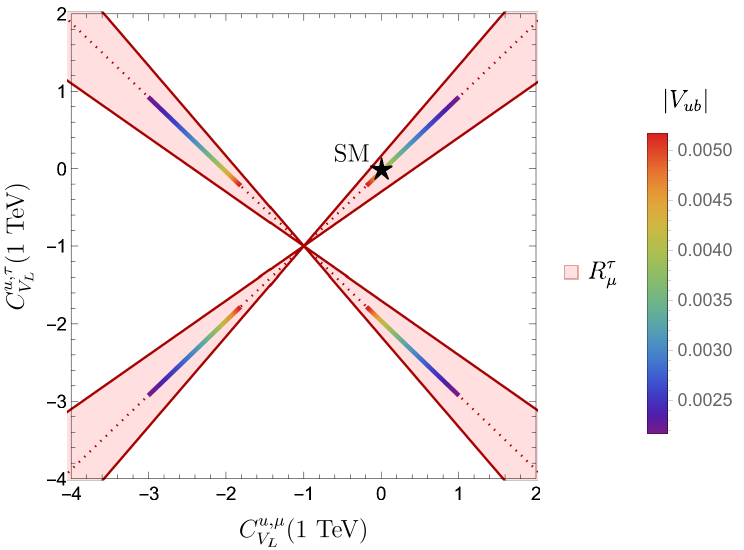}
\caption{Allowed region from $R_\mu^\tau$ in the $C_{V_L}^{u, \mu}$--$C_{V_L}^{u, \tau}$ plane, with the value of $|V_{ub}|$ indicated along the central values of the contours.}
\label{fig:leptonicVLmutauVubColour}
\end{center}
\end{figure}
\end{center}

Let us illustrate this feature by considering the leptonic decays. In Fig.~\ref{fig:leptonicVLmutauVubColour}, we show again the allowed region in the $C_{V_L}^{u, \mu}$--$C_{V_L}^{u, \tau}$ plane following from $R_\mu^\tau$. Moving along the dotted line, which indicates the central value, we obtain a relation between $C_{V_L}^{u, \mu}$ and $C_{V_L}^{u, \tau}$. This information allows us to determine the corresponding correlation with $|V_{ub}|$ from any of the branching fractions. In this case, we employ the $C_{V_L}^\tau$ short-distance coefficient to extract $|V_{ub}|$ from $\mathcal{B}(B^- \to \tau^- \bar{\nu}_\tau)$. The variation of $|V_{ub}|$, moving along the central value of $R_\mu^\tau$, is added to Fig.~\ref{fig:leptonicVLmutauVubColour}. Here, we have restricted ourselves to a range of ten times the uncertainty around the central value in Eq.~(\ref{eq:Vub}). We note that the value of $|V_{ub}|$ may differ significantly from that in Eq.~(\ref{eq:Vub}) for the range of short-distance coefficients considered in Figs.~\ref{fig:CVLmu-CVLtau}~and~\ref{fig:leptonicVLmutauVubColour}.

\begin{center}
\begin{figure}[H]
\begin{center}
\includegraphics[width=0.48\textwidth]{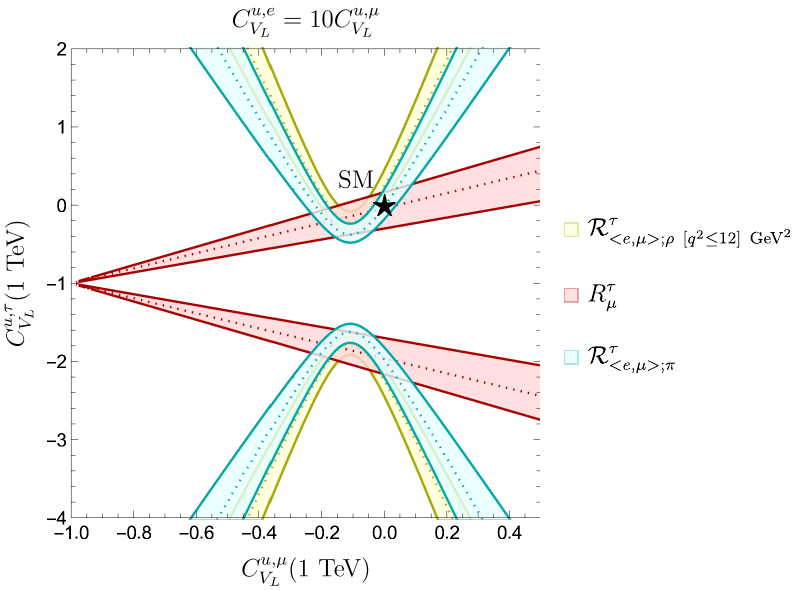}
\includegraphics[width=0.48\textwidth]{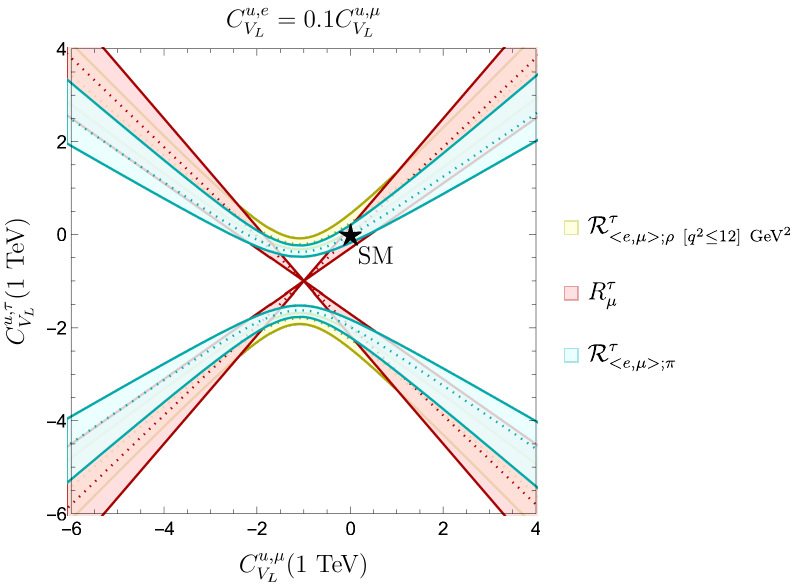}
\caption{Allowed regions in the $C_{V_L}^{u, \mu}$--$C_{V_L}^{u, \tau}$ plane following from the observables defined in the text. The left panel corresponds to the assumption $C_{V_L}^{u, e} = 10 C_{V_L}^{u, \mu}$, whereas the right panel shows $C_{V_L}^{u, e} = 0.1 C_{V_L}^{u, \mu}$.}
\label{fig:btouCVLnoUni}
\end{center}
\end{figure}
\end{center}

So far we have considered universality between $C_{V_L}^{u, e}$ and $C_{V_L}^{u, \mu}$. Let us continue with the other two scenarios from Eq.~(\ref{eq:scenarios}). The results can be found in Fig.~\ref{fig:btouCVLnoUni}. The regions agree with the SM, and in the case of $C_{V_L}^{u, e} = 10 C_{V_L}^{u, \mu}$, the regions are actually quite constrained in comparison with the other two scenarios considered here. In both plots in Fig.~\ref{fig:btouCVLnoUni}, we find agreement with the SM for one region, though at a bit more than $1~\sigma$ for $\mathcal{R}^{\tau}_{\Braket{e, \mu}; \rho ~ [q^2\leq 12]~\rm{GeV}^2}$, as was the case when making the assumption $C_{V_L}^{u, e} = C_{V_L}^{u, \mu}$.

\boldmath
\subsubsection{New Physics Entering through $C^{u}_{V_R}$} \label{sec:CVRonly}
\unboldmath
As indicated in Section~\ref{sec:theory-framework}, the short-distance coefficient $C_{V_R}^{q}$ is lepton-flavour universal. Switching on only this coefficient, we will determine the constraints from several observables. From Eq.~(\ref{eq:leptonic}), we find
\begin{equation}
    \mathcal{B}(B^- \to \ell^- \bar{\nu}_\ell) = \left.\mathcal{B}(B^- \to \ell^- \bar{\nu}_\ell)\right|_{\rm SM} \left|1 - C_{V_R}^u \right|^2.
\end{equation}

Considering that all the leptonic branching ratios have the same dependence on $C_{V_R}^u$,  taking ratios of purely leptonic processes cannot help us to constrain $C_{V_R}^u$ since its contribution will cancel. Nevertheless, we note that $R_\mu^\tau$ should agree with its SM prediction of $R_\mu^\tau = 1$. From the experimental data, we obtain
\begin{equation}
    R_\mu^\tau = 0.92 \pm 0.43,
\end{equation}
so this constraint is satisfied.

\begin{center}
\begin{figure}[H]
\begin{center}
\includegraphics[width=0.48\textwidth]{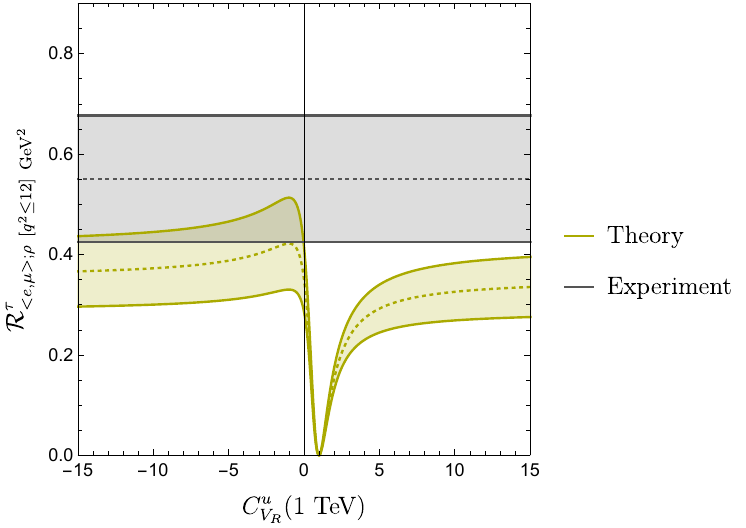}
\includegraphics[width=0.48\textwidth]{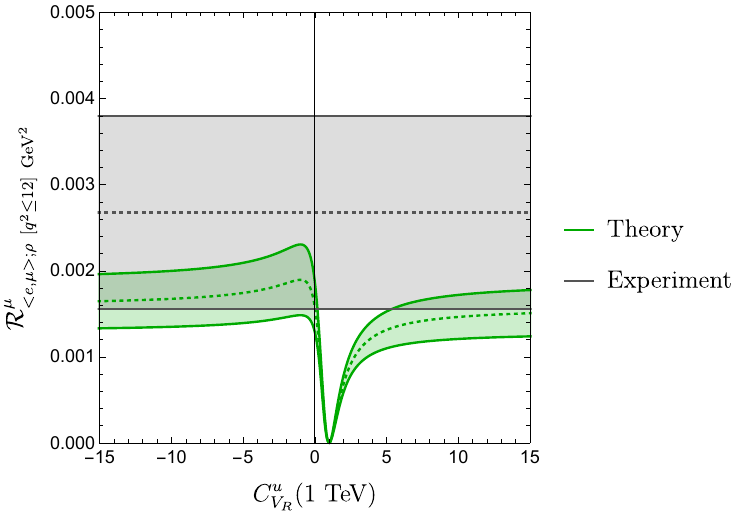}
\includegraphics[width=0.48\textwidth]{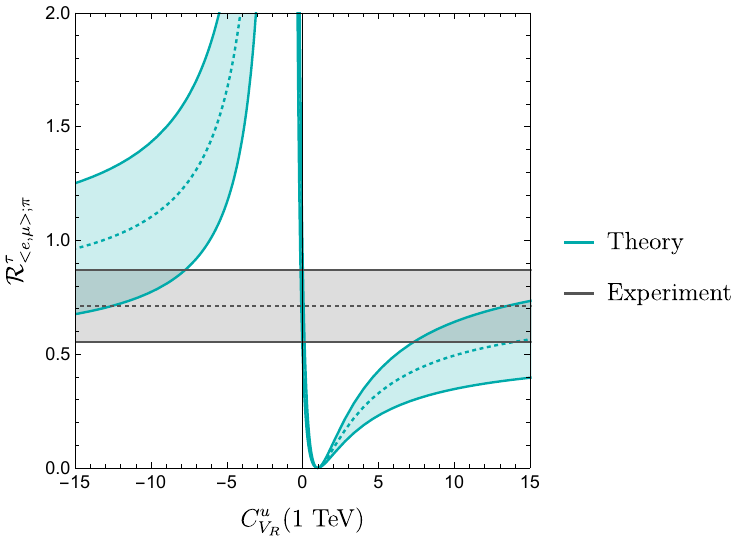}
\includegraphics[width=0.48\textwidth]{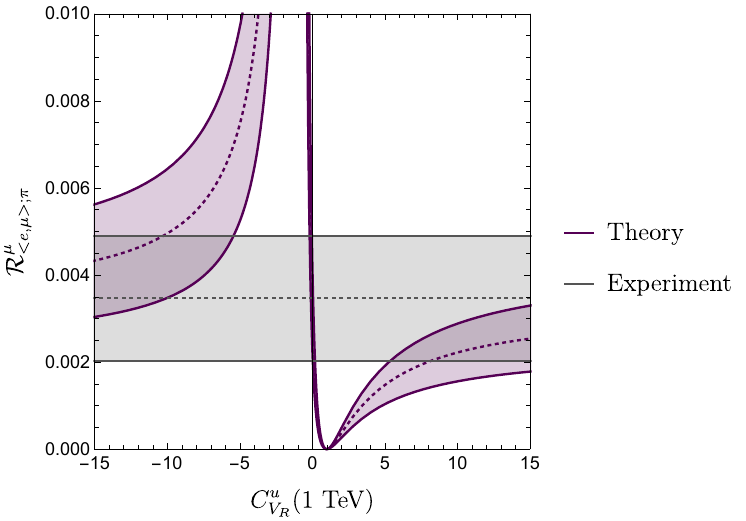}
\includegraphics[width=0.48\textwidth]{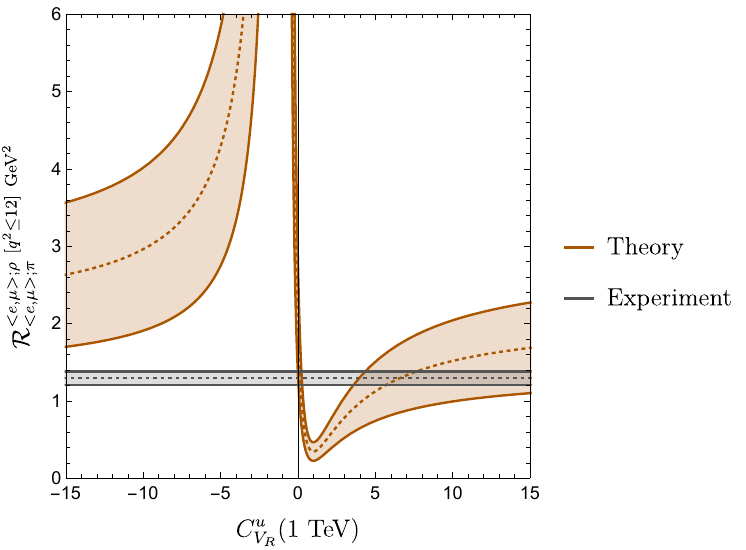}
\caption{Dependence of the five observables listed in the text on $C_{V_R}^u$, and comparison with the experimental regions.}
\label{fig:btouCVRuni}
\end{center}
\end{figure}
\end{center}

Let us now consider the other observables from the list in Eq.~(\ref{eq:btouObs}) that are also sensitive to $C_{V_R}^u$:
\begin{equation}
    \mathcal{R}^{\tau}_{\Braket{e, \mu}; \rho ~ [q^2\leq 12]~\rm{GeV}^2}, \  \mathcal{R}^{\mu}_{\Braket{e, \mu}; \rho ~ [q^2\leq 12]~\rm{GeV}^2}, \  \mathcal{R}^{\tau}_{\Braket{e, \mu}; \pi}, \  \mathcal{R}^{\mu}_{\Braket{e, \mu}; \pi}, \  \mathcal{R}^{\Braket{e, \mu};\rho~[q^2 \leq 12]~\rm{GeV}^2}_{\Braket{e, \mu};\pi}.
\end{equation}
In Fig.~\ref{fig:btouCVRuni}, we plot the dependence of these ratios on $C_{V_R}^u$, along with the region that corresponds to the experimental data. Comparing these contours in the range $-15 \leq C_{V_R}^u \leq 15$, we find that each observable yields agreement with the SM value $C_{V_R}^u = 0$ at the (1--2) $\sigma$ level. In addition, each ratio allows for one or more intervals of NP values for $C_{V_R}^u$, though not all of them overlap at $1~\sigma$.

\subsection{Constraints on Tensor Wilson Coefficients}
As a final step in our $b \to u$ analysis, we consider the tensor operator. From   Eq.~(\ref{eq:RGEepsilon_b}) we can see that taking a non-zero vale for $C_T$ at the scale $\mu = \text{1~TeV}$ yields contributions at the low energy scale $\mu = m_b$ not only in $C_T$, but also in the coefficients $C_P$ and $C_S$. In  consequence, even though the branching ratios of the leptonic modes do not depend on $C_T^{u, \ell}(m_b)$, they are sensitive to $C_T^{u, \ell}(\text{1~TeV})$ through $C_P^{u, \ell}(m_b)$. Hence, all six ratios in Eq.~(\ref{eq:btouObs}) can be employed to constrain the $C_T^{u, \ell}$ coefficients at the high-energy scale. Since they are not all independent, we will consider again three ratios of branching fractions,
\begin{equation} \label{eq:btouCTObs}
    R_\mu^\tau, \qquad \mathcal{R}^{\mu}_{\Braket{e, \mu}; \rho ~ [q^2\leq 12]~\rm{GeV}^2}, \qquad \mathcal{R}^{\Braket{e, \mu}; \rho ~ [q^2\leq 12]~\rm{GeV}^2}_{\Braket{e, \mu}; \pi},
\end{equation}
to constrain the coefficients $C_T^{u, \mu}$ and $C_T^{u, \tau}$ at $\mu = \text{1~TeV}$. We verified, in the scenario where $C_T^{u, e} = C_T^{u, \mu}$, that a different choice of observables would not meaningfully affect the results.

\begin{center}
\begin{figure}[H]
\begin{center}
\includegraphics[width=0.48\textwidth]{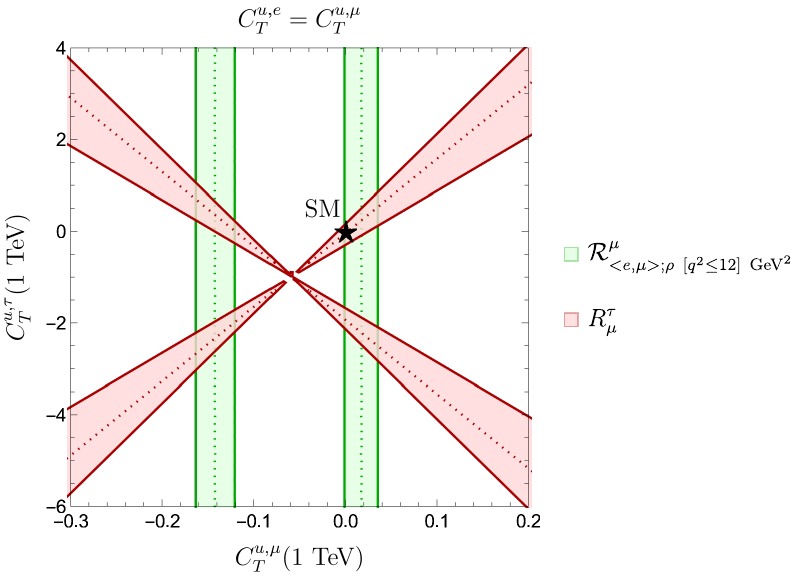} \\
\includegraphics[width=0.48\textwidth]{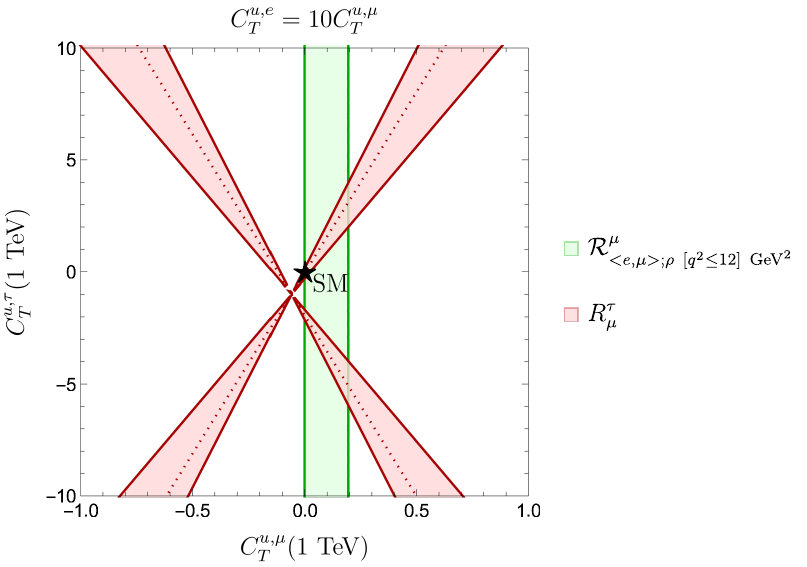}
\includegraphics[width=0.48\textwidth]{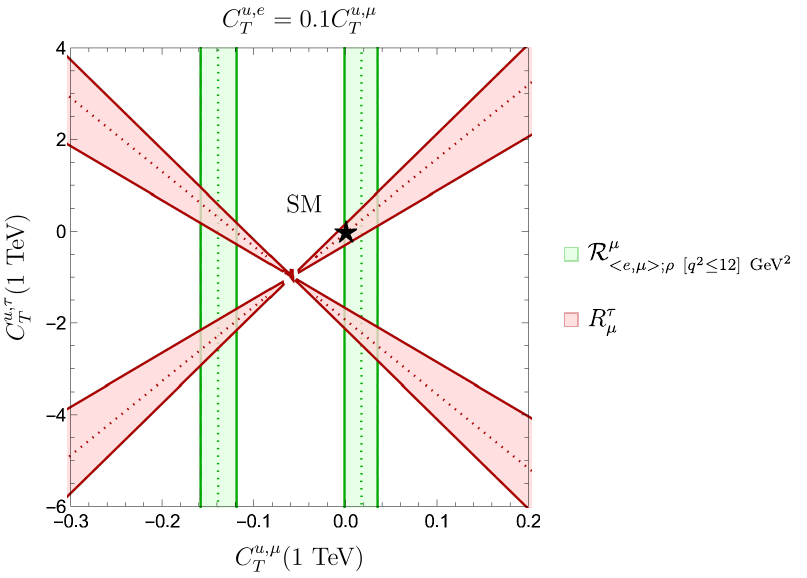}
\caption{Constraints in the $C_T^{u, \mu}$--$C_T^{u, \tau}$ plane for the observables listed in the text. The three plots correspond to different assumptions about the relation between $C_T^{u, e}$ and $C_T^{u, \mu}$, as indicated.}
\label{fig:btouCT}
\end{center}
\end{figure}
\end{center}

We find, similar to the pseudoscalar case, that there is no agreement between theory and experiment for $\mathcal{R}^{\Braket{e, \mu}; \rho ~ [q^2\leq 12]~\rm{GeV}^2}_{\Braket{e, \mu}; \pi}$. The contours from the remaining two ratios are shown in Fig.~\ref{fig:btouCT}. Each plot corresponds to a scenario from Eq.~(\ref{eq:scenarios}). Note that when we assume $C_T^{u, e} = C_T^{u, \mu}$ and $C_T^{u, e} = 0.1 C_T^{u, \mu}$, we obtain four regions where all contours overlap, one of which includes the SM. On the other hand, in the scenario where $C_T^{u, e} = 10 C_T^{u, \mu}$, the ratio $\mathcal{R}^{\mu}_{\Braket{e, \mu}; \rho ~ [q^2\leq 12]~\rm{GeV}^2}$ does not yield a solution for the central value. Still, a region appears once the uncertainties are taken into account, yielding two solutions from the combination of both observables, one of which agrees with the SM.

As for the ratio $\mathcal{R}^{\Braket{e, \mu}; \rho ~ [q^2\leq 12]~\rm{GeV}^2}_{\Braket{e, \mu}; \pi}$, the constraint is slightly stronger than in the case of $C_P^{u, \ell}$. One the one hand, in the scenarios where $C_T^{u, e} = C_T^{u, \mu}$ and $C_T^{u, e} = 0.1 C_T^{u, \mu}$, we find that the constraint from this observable is quite unstable, as was the case when we considered the $C_P^{u, \ell}$ coefficients. In particular, at the $1~\sigma$ level, it is actually still in agreement with all four regions following from the overlap of the contours from $R_\mu^\tau$ and $\mathcal{R}^{\mu}_{\Braket{e, \mu}; \rho ~ [q^2\leq 12]~\rm{GeV}^2}$. On the other hand, while the $\mathcal{R}^{\Braket{e, \mu}; \rho ~ [q^2\leq 12]~\rm{GeV}^2}_{\Braket{e, \mu}; \pi}$ ratio agrees with the SM at the $1~\sigma$ level in the scenario where $C_T^{u, e} = 10 C_T^{u,\mu}$, it does constrain the NP parameter space in this scenario.

\boldmath
\section{Determination of $|V_{ub}|$ and Predictions} \label{sec:VubPredbu}
\unboldmath
Now that we have determined allowed regions for the different short-distance coefficients in several scenarios, we may use these results to determine $|V_{ub}|$ and make predictions for branching ratios that have not yet been measured. We can consider any of the measured branching ratios to determine $|V_{ub}|$ while accounting simultaneously for NP. Then, we may use the results for $|V_{ub}|$ and the Wilson coefficients together to predict the branching fractions $\mathcal{B}(B^- \to e^- \bar{\nu}_e)$, ${\mathcal B}(\bar B\rightarrow \rho \tau^- \bar{\nu}_{\tau})_{q^2\leq 12~\rm{GeV}^2}$ and ${\mathcal B}(\bar{B}^0 \rightarrow \pi^+ \tau^- \bar{\nu}_{\tau})$. Since experimental limits exist for two of these branching ratios, $\mathcal{B}(B^- \to e^- \bar{\nu}_e)$ and ${\mathcal B}(\bar{B}^0 \rightarrow \pi^+ \tau^- \bar{\nu}_{\tau})$, we may find that (part of) our predicted range for these modes is already excluded. In that case, the bounds will be used to constrain the allowed parameter space even further.

The purpose of this section is to demonstrate our strategy for the determination of $|V_{ub}|$, as well as the constraining power and potential enhancements of branching ratios such as $\mathcal{B}(B^- \to e^- \bar{\nu}_e)$, which have not yet been measured. Therefore, we have chosen to limit ourselves to a selection of scenarios, which are:
\begin{itemize}
    \item NP through the pseudoscalar coefficients with the assumption $C_P^{u, e} = C_P^{u, \mu}$,
    \item NP through the left-handed vector coefficients with the assumption $C_{V_L}^{u, e} = 10 C_{V_L}^{u, \mu}$,
    \item NP through the tensor coefficients with the assumption $C_T^{u, e} = 0.1 C_T^{u, \mu}$.
\end{itemize}
We will use the branching fraction $\Braket{{\mathcal B}(\bar B\rightarrow \rho \ell^{\prime -} \bar{\nu}_{\ell'})}_{[\ell'=~ e, \mu],~q^2\leq 12~\rm{GeV}^2}$ for the determination of $|V_{ub}|$.

\subsection{The Pseudoscalar Coefficients}

The first step in our strategy is to obtain bounds on the short-distance coefficients $C_P^{u, \mu}$ and $C_P^{u, \tau}$ from the overlap of the contours from $R_\mu^\tau$ and $\mathcal{R}^{\mu}_{\Braket{e, \mu}; \rho ~ [q^2\leq 12]~\rm{GeV}^2}$ in the first plot in Fig.~\ref{fig:btouCP}. As indicated in Sec.~\ref{subsubsec:CPul}, the ratio $\mathcal{R}^{\Braket{e, \mu}; \rho ~ [q^2\leq 12]~\rm{GeV}^2}_{\Braket{e, \mu}; \pi}$ just about excludes the four solutions from the former two observables. However, as was discussed there, this constraint is quite unstable, so we will still apply our strategy using $R_\mu^\tau$ and $\mathcal{R}^{\mu}_{\Braket{e, \mu}; \rho ~ [q^2\leq 12]~\rm{GeV}^2}$. The resulting allowed ranges are given in Table~\ref{tab:CPuSDCVub}, where the regions are numbered $1$ to $4$ clockwise, starting at the SM-like solution in the top right corner of the first plot in Fig.~\ref{fig:btouCP}.

\begin{table}
    \centering
    \begin{tabular}{|c|c|c|c|}
        \hline
        \multicolumn{4}{|c|}{Scenario: $C_P^{u, e} = C_P^{u, \mu}$} \\
        \hline
        Region & $C_P^{u, \mu} (\rm{1~TeV})$ & $C_P^{u, \tau} (\rm{1~TeV})$ & $|V_{ub}|$ \\
        \hline
        $1$ & $[-0.0001, 0.0055]$ & $[-0.05, 0.13]$ & $[0.0030, 0.0037]$ \\
        \hline
        $2$ & $[-0.0001, 0.0055]$ & $[-0.44, -0.27]$ & $[0.0030, 0.0037]$ \\
        \hline
        $3^*$ & $[-0.024, -0.018]$ & $[-0.44, -0.27]$ & $[0.0030, 0.0037]$ \\
        \hline
        $4^*$ & $[-0.024, -0.018]$ & $[-0.05, 0.13]$ & $[0.0030, 0.0037]$ \\
        \hline
    \end{tabular}
    \caption{Bounds on the short-distance coefficients from the constraints on $C_P^{u, \mu}$ and $C_P^{u, \tau}$, along with the allowed range for $|V_{ub}|$ from $\Braket{{\mathcal B}(\bar B\rightarrow \rho \ell^{\prime -} \bar{\nu}_{\ell'})}_{[\ell'=~ e, \mu],~q^2\leq 12~\rm{GeV}^2}$, in the scenario where $C_P^{u, e} = C_P^{u, \mu}$. The regions marked by the asterisk are fully excluded by the experimental upper bound on $\mathcal{B}(B^- \to e^- \bar{\nu}_e)$.}
    \label{tab:CPuSDCVub}
\end{table}

The next step is then to determine $|V_{ub}|$ from $\Braket{{\mathcal B}(\bar B\rightarrow \rho \ell^{\prime -} \bar{\nu}_{\ell'})}_{[\ell'=~ e, \mu],~q^2\leq 12~\rm{GeV}^2}$. Here we scan the ranges for $C_P^{u, \mu}$ in Table~\ref{tab:CPuSDCVub} together with values for $|V_{ub}|$ to find agreement with the measurement in Eq.~(\ref{eq:Brhoaverage}). The resulting allowed ranges for $|V_{ub}|$ are added in the fourth column of Table~\ref{tab:CPuSDCVub}. Since the branching ratio $\Braket{{\mathcal B}(\bar B\rightarrow \rho \ell^{\prime -} \bar{\nu}_{\ell'})}_{[\ell'=~ e, \mu],~q^2\leq 12~\rm{GeV}^2}$ is very stable on such a small scale for $C_P^{u, \mu}$, we find the same result from all four regions. The range contains the HFLAV result for $|V_{ub}|$ from exclusive decays in Eq.~(\ref{eq:Vub}) at the upper end, but not $|V_{ub}|_{\rm incl.}$ as given in Eq.~(\ref{eq:VubInclHFLAV}).

Now that the short-distance coefficients and $|V_{ub}|$ have been determined, we can apply them to predict branching ratios that have not yet been measured. Of the three modes that were introduced for this purpose in the introduction of this section, two are sensitive to $C_P^{u, \ell}$: $\mathcal{B}(B^- \to e^- \bar{\nu}_e)$ and ${\mathcal B}(\bar B\rightarrow \rho \tau^- \bar{\nu}_{\tau})_{q^2\leq 12~\rm{GeV}^2}$. It turns out that the branching ratio prediction of $\mathcal{B}(B^- \to e^- \bar{\nu}_e)$ is larger than the experimental bound in Eq.~(\ref{eq:Belle-enu-limit}) for both regions $3$ and $4$, thereby excluding these solutions. The allowed ranges for regions 1 and 2 are given in Table~\ref{tab:CPuPred}. As we found in Ref.~\cite{Banelli:2018fnx}, as well as in Ref.~\cite{Fleischer:2017ltw} for neutral leptonic $B$-meson decays, the $C_P$ coefficient potentially enhances the branching ratio of the $B^- \to e^- \bar{\nu}_e$ decay by several orders of magnitude because it lifts the helicity suppression.

\begin{table}
    \centering
    \begin{tabular}{|c|c|c|}
        \hline
        \multicolumn{3}{|c|}{Scenario: $C_P^{u, e} = C_P^{u, \mu}$} \\
        \hline
        Region & $\mathcal{B}(B^- \to e^- \bar{\nu}_e)$ & ${\mathcal B}(\bar B\rightarrow \rho \tau^- \bar{\nu}_{\tau})_{q^2\leq 12~\rm{GeV}^2}$ \\
        \hline
        $1$ & $[0, 1.4 \times 10^{-7}]$ & $[5.2 \times 10^{-5}, 1.2 \times 10^{-4}]$ \\
        \hline
        $2$ & $[0, 1.4 \times 10^{-7}]$ & $[4.3 \times 10^{-5}, 9.3 \times 10^{-5}]$ \\
        \hline
    \end{tabular}
    \caption{Predictions for branching ratios that have not yet been measured corresponding to the allowed ranges of $|V_{ub}|$ and $C_P^{u, \mu}$ or $C_P^{u, \tau}$, in the scenario where $C_P^{u, e} = C_P^{u, \mu}$.}
    \label{tab:CPuPred}
\end{table}

\subsection{The Left-Handed Vector Coefficients}

Let us now apply the same strategy to the Wilson coefficient $C_{V_L}^{u, \ell}$, specifically the scenario where we assume $C_{V_L}^{u, e} = 10 C_{V_L}^{u, \mu}$. The relevant contours are given in the left panel of Fig.~\ref{fig:btouCVLnoUni}. There are a few key differences with respect to the $C_P^{u, \ell}$ scenario:
\begin{itemize}
    \item The regions are now defined by the overlap of three different contours.
    \item Within the uncertainties, two solutions merge into one region, thereby giving two instead of four regions in total.
    \item We can give predictions for all three branching ratios $\mathcal{B}(B^- \to e^- \bar{\nu}_e)$, ${\mathcal B}(\bar B\rightarrow \rho \tau^- \bar{\nu}_{\tau})_{q^2\leq 12~\rm{GeV}^2}$ and ${\mathcal B}(\bar{B}^0 \rightarrow \pi^+ \tau^- \bar{\nu}_{\tau})$, since they are all sensitive to the left-handed vector coefficient.
\end{itemize}
In this instance, we find that neither the upper bound on $\mathcal{B}(B^- \to e^- \bar{\nu}_e)$ nor the one on ${\mathcal B}(\bar{B}^0 \rightarrow \pi^+ \tau^- \bar{\nu}_{\tau})$ gives any further constraints.

The results for the short-distance coefficients and $|V_{ub}|$ are given in Table~\ref{tab:CVLuSDCVub}. Region 1 is the SM-like solution on the top in the left panel of Fig.~\ref{fig:btouCVLnoUni}, and region 2 is the other solution on the bottom. We find quite a large range for $|V_{ub}|$, which includes the HFLAV value from exclusive modes in Eq.~(\ref{eq:Vub}), as well as the inclusive result in Eq.~(\ref{eq:VubInclHFLAV}). The large size of the range is dominated by the NP effects, i.e., the variation of $|V_{ub}|$ with respect to $C_{V_L}^{u, \mu}$. This is in contrast to the $C_P^u$ scenario, where $|V_{ub}|$ was very stable over the range of the short-distance coefficient. On the other hand, this means that improving the determination of $C_{V_L}^{u, \mu}$ has the potential to drastically reduce the allowed range for $|V_{ub}|$ as well.

\begin{table}
    \centering
    \begin{tabular}{|c|c|c|c|}
        \hline
        \multicolumn{4}{|c|}{Scenario: $C_{V_L}^{u, e} = 10 C_{V_L}^{u, \mu}$} \\
        \hline
        Region & $C_{V_L}^{u, \mu} (\rm{1~TeV})$ & $C_{V_L}^{u, \tau} (\rm{1~TeV})$ & $|V_{ub}|$ \\
        \hline
        $1$ & $[-0.19, 0.02]$ & $[-0.32, 0.19]$ & $[0.0028, 0.0058]$ \\
        \hline
        $2$ & $[-0.19, 0.02]$ & $[-2.2, -1.7]$ & $[0.0028, 0.0058]$ \\
        \hline
    \end{tabular}
    \caption{Allowed ranges for the left-handed vector coefficients and $|V_{ub}|$ in the scenario where $C_{V_L}^{u, e} = 10 C_{V_L}^{u, \mu}$.}
    \label{tab:CVLuSDCVub}
\end{table}

The predictions for the branching ratios are given in Table~\ref{tab:CVLuPred}. In contrast to the results for $C_P$, the helicity supression of the leptonic decays is not lifted. Consequently, the potential enhancement of $\mathcal{B}(B^- \to e^- \bar{\nu}_e)$ is not as dramatic as when NP would enter through the pseudoscalar coefficient.

\begin{table}
    \centering
    \begin{tabular}{|c|c|c|c|}
        \hline
        \multicolumn{4}{|c|}{Scenario: $C_{V_L}^{u, e} = 10 C_{V_L}^{u, \mu}$} \\
        \hline
        Region & $\mathcal{B}(B^- \to e^- \bar{\nu}_e)$ & ${\mathcal B}(\bar B\rightarrow \rho \tau^- \bar{\nu}_{\tau})_{q^2\leq 12~\rm{GeV}^2}$ & ${\mathcal B}(\bar{B}^0 \rightarrow \pi^+ \tau^- \bar{\nu}_{\tau})$ \\
        \hline
        $1$ & $[0, 1.0 \times 10^{-11}]$ & $[6.1 \times 10^{-5}, 1.6 \times 10^{-4}]$ & $[5.3 \times 10^{-5}, 1.3 \times 10^{-4}]$ \\
        \hline
        $2$ & $[0, 1.0 \times 10^{-11}]$ & $[6.1 \times 10^{-5}, 1.6 \times 10^{-4}]$ & $[5.3 \times 10^{-5}, 1.3 \times 10^{-4}]$ \\
        \hline
    \end{tabular}
    \caption{Predictions for branching ratios that have not yet been measured in the scenario where NP enters through new contributions to the left-handed vector coefficient, making the assumption that $C_{V_L}^{u, e} = 10 C_{V_L}^{u, \mu}$.}
    \label{tab:CVLuPred}
\end{table}

\subsection{The Tensor Coefficients}

Finally, we consider the Wilson coefficient $C_T^{u, \ell}$ in the scenario where $C_T^{u, e} = 0.1 C_T^{u, \mu}$. This is a very interesting situation for a few reasons. First of all, per Eq.~(\ref{eq:RGEepsilon_b}), switching on $C_T^{u, \ell}(\text{1~TeV})$ gives contributions in $C_{S, P}^{u, \ell}(m_b)$ as well as in $C_T^{u, \ell}(m_b)$. Secondly, the upper bound on ${\mathcal B}(\bar{B}^0 \rightarrow \pi^+ \tau^- \bar{\nu}_{\tau})$ in Eq.~(\ref{eq:BpitaunuBound}) severely restricts the allowed regions, but does not fully exclude any of the four solutions.

In Table~\ref{tab:CTuSDCVub}, we give the allowed ranges of the short-distance coefficients and $|V_{ub}|$ that are in agreement with the upper bound on ${\mathcal B}(\bar{B}^0 \rightarrow \pi^+ \tau^- \bar{\nu}_{\tau})$. The corresponding predictions for the branching ratios are given in Table~\ref{tab:CTuPred}. We have numbered the regions again in a clockwise fashion, starting with 1 for the SM-like solution in the top right corner of the relevant plot in Fig.~\ref{fig:btouCT}.

\begin{table}
    \centering
    \begin{tabular}{|c|c|c|c|}
        \hline
        \multicolumn{4}{|c|}{Scenario: $C_T^{u, e} = 0.1 C_T^{u, \mu}$} \\
        \hline
        Region & $C_T^{u, \mu} (\rm{1~TeV})$ & $C_T^{u, \tau} (\rm{1~TeV})$ & $|V_{ub}|$ \\
        \hline
        $1$ & $[-0.001, 0.035]$ & $[-0.29, 0.84]$ & $[0.0030, 0.0037]$ \\
        \hline
        $2$ & $[-0.0011, 0.0032]$ & $[-1.72, -1.67]$ & $[0.0030, 0.0031]$ \\
        \hline
        $3$ & $[-0.13, -0.12]$ & $[-1.8, -1.7]$ & $[0.0029, 0.0031]$ \\
        \hline
        $4$ & $[-0.16, -0.12]$ & $[-0.26, 0.95]$ & $[0.0029, 0.0035]$ \\
        \hline
    \end{tabular}
    \caption{Allowed ranges for the tensor short-distance coefficients and $|V_{ub}|$ in the scenario where $C_T^{u, e} = 0.1 C_T^{u, \mu}$.}
    \label{tab:CTuSDCVub}
\end{table}

\begin{table}
    \centering
    \begin{tabular}{|c|c|c|c|}
        \hline
        \multicolumn{4}{|c|}{Scenario: $C_T^{u, e} = 0.1 C_T^{u, \mu}$} \\
        \hline
        Region & $\mathcal{B}(B^- \to e^- \bar{\nu}_e)$ & ${\mathcal B}(\bar B\rightarrow \rho \tau^- \bar{\nu}_{\tau})_{q^2\leq 12~\rm{GeV}^2}$ & ${\mathcal B}(\bar{B}^0 \rightarrow \pi^+ \tau^- \bar{\nu}_{\tau})$ \\
        \hline
        $1$ & $[3.6 \times 10^{-12}, 1.6 \times 10^{-9}]$ & $[5.1 \times 10^{-5}, 6.2 \times 10^{-4}]$ & $[3.4 \times 10^{-5}, 2.5 \times 10^{-4}]$ \\
        \hline
        $2$ & $[2.3 \times 10^{-12}, 2.8 \times 10^{-11}]$ & $[1.7 \times 10^{-3}, 2.4 \times 10^{-3}]$ & $[2.3 \times 10^{-4}, 2.5 \times 10^{-4}]$ \\
        \hline
        $3$ & $[9.3 \times 10^{-9}, 1.1 \times 10^{-8}]$ & $[1.7 \times 10^{-3}, 2.4 \times 10^{-3}]$ & $[2.3 \times 10^{-4}, 2.5 \times 10^{-4}]$ \\
        \hline
        $4$ & $[9.5 \times 10^{-9}, 2.3 \times 10^{-8}]$ & $[4.6 \times 10^{-5}, 6.8 \times 10^{-4}]$ & $[3.2 \times 10^{-5}, 2.5 \times 10^{-4}]$ \\
        \hline
    \end{tabular}
    \caption{Allowed ranges for branching ratios that have not yet been measured corresponding to the allowed ranges for $|V_{ub}|$ and $C_T^{u, \mu}$ or $C_T^{u, \tau}$, in the scenario where $C_T^{u, e} = 0.1 C_T^{u, \mu}$.}
    \label{tab:CTuPred}
\end{table}

Although we reach the upper bound of ${\mathcal B}(\bar{B}^0 \rightarrow \pi^+ \tau^- \bar{\nu}_{\tau})$ in all four regions, it mostly restricts solution 2 and 3. For these regions, the ranges of $|V_{ub}|$ are also quite small, and do not contain the HFLAV value from exclusive decays in Eq.~(\ref{eq:Vub}), whereas the ranges found in regions 1 and 4 do. None of them reach the inclusive result in Eq.~(\ref{eq:VubInclHFLAV}). In all four regions, the variation of $|V_{ub}|$ with respect to the short-distance coefficient has only a minor impact on the range that we obtain. As for the predictions of branching ratios that have not yet been measured, we find again potentially spectacular enhancements for $\mathcal{B}(B^- \to e^- \bar{\nu}_e)$, though the maximum values are somewhat smaller than in the case of $C_P^{u, \ell}$. It is not surprising to find such values, because the $C_T^{u, \ell}(\text{1~TeV})$ yields a contribution to $C_P^{u, \ell}(m_b)$ as stated before, thereby also lifting the helicity supression of the leptonic modes.

\begin{center}
\begin{figure}[H]
\begin{center}
\includegraphics[width=0.6\textwidth]{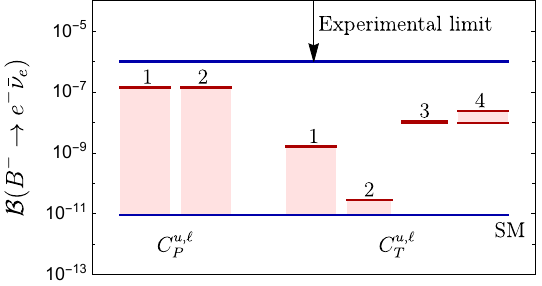}
\caption{Summary of the potential enhancements of $\mathcal{B}(B^- \to e^- \bar{\nu}_e)$ correspondig to the two allowed regions for the $C_P^{u, \ell}$ and the four solutions for $C_T^{u, \ell}$, along with the SM value and experimental limit.}
\label{fig:propPlotBen}
\end{center}
\end{figure}
\end{center}

Due to the potential enhancement of $\mathcal{B}(B^- \to e^- \bar{\nu}_e)$ by several orders of magnitude with respect to the SM value, we have decided to summarize these results in Fig.~\ref{fig:propPlotBen}. There we show the range of values larger than the SM that may be obtained for $\mathcal{B}(B^- \to e^- \bar{\nu}_e)$ in the two allowed regions for $C_P^{u, \ell}$ and the four regions for $C_T^{u, \ell}$. We also indicate the SM value from Eq.~(\ref{eq:leptueSM}) and the experimental limit from Eq.~(\ref{eq:Belle-enu-limit}). The results from the left-handed vector coefficient are not included because on the scale of this plot they would be essentially at the SM level.

\boldmath
\section{The $b \to c$ Transitions}
\label{sec:btoc}
\unboldmath
To continue our discussion, we focus on semileptonic processes induced by the $b \to c$ quark transition. More specifically,
we will concentrate on the semileptonic decays of $B$-mesons $B\rightarrow D \ell \bar{\nu}_{\ell}$ and $B\rightarrow D^{*} \ell \bar{\nu}_{\ell}$ with $\ell=e, \mu, \tau$. Since the $D$ and $D^*$ are pesudoscalar and vector meson states, respectively, this will allow us to probe different Lorenz structures involving $b\rightarrow c$ processes.

For the semileptonic decays involving the $B\rightarrow D$ transition, 
the following experimental averages calculated from 
independent measurements reported by the BaBar, Belle, CLEO and ALEPH collaborations \cite{Tanabashi:2018oca} are available:
\begin{eqnarray}
\label{eq:ExpD1}
\braket{{\mathcal B}(\bar B^0\rightarrow D^- \ell'^+ \bar{\nu}_{\ell'})}&=&(2.35\pm 0.09)\times 10^{-2},\nonumber\\
\braket{{\mathcal B}(B^-\rightarrow \bar{D}^0 \ell'^- \bar{\nu}_{\ell'})}&=&(2.35\pm 0.09)\times 10^{-2}.
\end{eqnarray}
For channels with a $\tau$ lepton in the final state, we have considered the processes
\begin{eqnarray}
\label{eq:ExpD2}
{\mathcal B}(\bar B^0\rightarrow D^+ \tau^- \bar{\nu}_{\tau})&=&(1.08\pm 0.23)\times 10^{-2},\nonumber\\
{\mathcal B}(B^-\rightarrow \bar{D}^0 \tau^- \bar{\nu}_{\tau})&=&(7.7\pm 2.5)\times 10^{-3},
\end{eqnarray}
where the first and the second ratios were obtained by the BaBar \cite{Aubert:2007dsa} and the Belle \cite{Bozek:2010xy} experiments, respectively. 

On the other hand, for the $B\rightarrow D^{*} \ell' \bar{\nu}_{\ell'}$ decays we have the following
averages which combine partial measurements by Belle, BaBar, DELPHI, 
CLEO, OPAL, ARGUS and ALEPH \cite{Tanabashi:2018oca}:
\begin{eqnarray}
\label{eq:ExpDstar1}
\braket{{\mathcal B}(\bar B^0\rightarrow D^{*+} \ell'^- \bar{\nu}_{\ell'})} &=&(5.05\pm 0.14)\times 10^{-2},\nonumber\\
\braket{{\mathcal B}( B^-\rightarrow \bar{D}^{*0} \ell'^- \bar{\nu}_{\ell'})}&=&(5.66\pm 0.22)\times 10^{-2}.
\end{eqnarray}

For the analogous processes involving a $\tau$ in the final state, we have the following experimental results available \cite{Tanabashi:2018oca}:
\begin{eqnarray}
\label{eq:ExpDstar2}
{\mathcal B}(\bar B^0\rightarrow D^{*+} \tau^- \bar{\nu}_{\ell})&=&(1.57\pm 0.09)\times 10^{-2},\nonumber\\
{\mathcal B}( B^-\rightarrow \bar{D}^{*0} \tau^- \bar{\nu}_{\ell})&=&(1.88\pm 0.20)\times 10^{-2},
\end{eqnarray}
which are based on measurements performed by the Belle and BaBar Collaborations. We can average both determinations in Eq.~(\ref{eq:ExpDstar2}), obtaining
\begin{eqnarray}
\label{eq:BDTauAv}
\braket{{\mathcal B}(\bar B\rightarrow D^{*} \tau^- \bar{\nu}_{\tau})}
&=&(1.62\pm 0.08)\times 10^{-2}.
\end{eqnarray}

To exemplify the application of our strategy for the extraction of the CKM element $|V_{cb}|$ in the presence of possible NP effects, we need observables where
$|V_{cb}|$ itself is not present. One possibility is to construct new 
quantities based on ratios of the branching fractions in 
Eqs.~(\ref{eq:ExpD1}), (\ref{eq:ExpD2}), (\ref{eq:ExpDstar1})
and (\ref{eq:ExpDstar2}). Fortunately, ratios of this nature
are already available directly from experimental determinations.

Indeed, the ratios
\begin{eqnarray}
\label{eq:RDandRDstar}	
\mathcal{R}(D)=\frac{\mathcal{B}(\bar{B}\rightarrow D \tau^-\bar{\nu}_{\tau})}{\mathcal{B}(\bar{B}\rightarrow D \ell'^-\bar{\nu}_{\ell'})},
\quad\quad
\mathcal{R}(D^*)=\frac{\mathcal{B}(\bar{B}\rightarrow D^* \tau^-\bar{\nu}_{\tau})}{\mathcal{B}(\bar{B}\rightarrow D^* \ell'^-\bar{\nu}_{\ell'})}
\end{eqnarray}
satisfy the requirement of $|V_{cb}|$ independence. Furthermore, they 
posses the  feature of showing a combined tension with the
SM determination at the $3~\sigma$ level, as reported in Ref.~\cite{Amhis:2019ckw}.
The most recent averages include the latest measurement reported by
Belle \cite{Abdesselam:2019dgh,Belle:2019rba}, and reads as follows \cite{Amhis:2019ckw}:
\begin{eqnarray}
\mathcal{R}(D)&=&0.340\pm 0.027\pm 0.013=0.340\pm 0.030,\nonumber\\
\mathcal{R}(D^*)&=&0.295\pm 0.011 \pm 0.008=0.295\pm 0.014,
\label{eq:Updated_Experimental}
\end{eqnarray}
being in tension with the SM at $3.08~\sigma$, as reported in
\cite{Amhis:2019ckw}. We would like to emphasize that $\mathcal{R}(D)$ and 
$\mathcal{R}(D^*)$ are defined in terms of physical quantities 
which depend on the CKM factor $|V_{cb}|$. However,  
this factor actually cancels out in the ratios which define 
$\mathcal{R}(D)$ and $\mathcal{R}(D^*)$ themselves. This is precisely the
desirable feature we need to decouple the determination of NP effects in the different Wilson coefficients and the independent extraction 
of $|V_{cb}|$.

Our SM evaluations of the observables  $\mathcal{R}(D)$ and 
$\mathcal{R}(D^*)$ are
\begin{eqnarray}
\mathcal{R}(D)|_{\rm SM}=0.300 \pm 0.006, &\quad\quad&
\mathcal{R}(D^*)|_{\rm SM}=0.253 \pm 0.005,
\end{eqnarray}
where we have performed the corresponding theoretical evaluations 
using the form factors for the $B\rightarrow D$ and $B\rightarrow D^*$ transitions obtained from QCD sum-rule (QCDSR) and LQCD calculations. The technical details of the corresponding paramterizations are summarized in Appendix \ref{sec:FFBD}.

In addition, different $|V_{cb}|$-independent polarization observables can, in principle,
serve our purposes as well \cite{Blanke:2018yud}. Nevertheless, for most of them,
their implementation suffers from large experimental uncertainties or from 
the lack of experimental information at all. 

In this analysis, we consider the following polarization ratio which allows us to obtain  useful constraints on NP coefficients associated with semileptonic processes involving a $D^*$ meson in the final state:
\begin{eqnarray}
\label{eq:FL}
F_{L}(D^*)&=&
\frac{\Gamma(B\rightarrow D^*_{L} \tau \bar{\nu}_{\tau})}
	{\Gamma(B\rightarrow D^* \tau \bar{\nu}_{\tau})}.
\end{eqnarray}
Within the SM, we find
\begin{eqnarray}
F_{L}(D^*)|_{\rm SM}&=&0.458\pm 0.004,	
\end{eqnarray}
which is compatible with the current experimental result
\cite{Abdesselam:2019wbt}
\begin{eqnarray}
F_{L}(D^*)&=&0.60\pm 0.08\pm 0.04,
\end{eqnarray}
at the $1.6~\sigma$ level, in agreement with other studies reported
in the literature \cite{Huang:2018nnq}.

Finally, the Belle Collaboration \cite{Abdesselam:2018nnh} has performed an experimental test which quantifies the flavour universality between electrons and muons through the semileptonic ratio
\begin{eqnarray}
\label{eq:Rlight}
R^{e}_{\mu}(D^*)=\frac{\mathcal{B}(B^0\rightarrow D^{*-}e^+\nu_{e})}{\mathcal{B}(B^0\rightarrow D^{*-}\mu^+\nu_{\mu})}.
\end{eqnarray}
The corresponding experimental result reads \cite{Abdesselam:2018nnh}
\begin{eqnarray}
\label{eq:RemuExp}
R^{e}_{\mu}(D^*)=1.01\pm 0.01\pm 0.03,
\end{eqnarray}
which can be compared with our SM evaluation
\begin{eqnarray}
R^{e}_{\mu}(D^*)|_{\rm SM}&=&=1.0045(1).
\end{eqnarray}
A striking feature of $R^{e}_{\mu}(D^*)$ is its small uncertainty on both the theoretical and the experimental side. This quantity will play an important role in restricting NP in the light generations of leptons.

For completeness, we present a set of semi-numerical formulae which
display the dependence of the different observables in our study
on the presence of the different NP contributions.
To simplify the presentation, we show two cases 
separately depending on whether NP enters either through the $\tau$ leptons 
or through the light generations, i.e., muons and electrons. In addition, we don't give the uncertainty of the numerical coefficients, which arises from, e.g., the form factors. However, it is important to mention that we do take these uncertainties into account in our analysis. Finally, we would like 
to stress that, in order to compare with other sources available in the literature, we present results where the Wilson coefficients are evaluated
at the scale $\mu=m_b$. The connection with the $1$~$\rm{TeV}$ value can be established straightforwardly using the renormalization group evolution given in Eq.~(\ref{eq:RGEepsilon_b}).

In the case of NP entering through the $\tau$ leptons, we have
\begin{eqnarray}\label{eq:RDstarTau}
\mathcal{R}(D^{*})/\mathcal{R}(D^{*})^{\rm SM}\Bigl|_{\tau}&=&|1+C^{c,\tau}_{V_L}|^2 +|C^{c,\tau}_{V_R}|^2
 -1.80 \Re[(1+C^{c,\tau}_{V_L}) C^{c,\tau *}_{V_R}] \nonumber \\
&&+ 0.11\Re[(1 + C^{c,\tau}_{V_L} - C^{c, \tau}_{V_R}) C^{c,\tau *}_P ]
+ 0.034 |C^{c,\tau}_P|^2 \nonumber \\
&&-5.02 \Re[(1+C^{c,\tau}_{V_L}) C^{c,\tau *}_T] 
+ 15.94 |C^{c,\tau}_T|^2 + 6.60\Re[C^{c,\tau}_{V_R} C^{c, \tau *}_T],
\nonumber\\
\mathcal{R}(D)/\mathcal{R}(D)^{\rm SM}\Bigl|_{\tau}&=&
|1+C^{c,\tau}_{V_L}+C^{c,\tau}_{V_R}|^2+
	1.46\Re[(1+ C^{c,\tau}_{V_L} + C^{c,\tau}_{V_R}) 
	C^{c,\tau*}_S ] \nonumber \\
	&&+ 0.98|C^{c,\tau}_S|^2
+1.14\Re[(1+C^{c,\tau}_{V_L}+C^{c,\tau}_{V_R})C^{c,\tau *}_T]
+0.91|C^{c,\tau}_T|^2,\nonumber\\
F_{L}(D^{*})&=&\Biggl(\mathcal{R}(D^*)^{\rm SM}/\mathcal{R}(D^*)\Bigl|_{\tau}\Biggl)\Biggl[0.46|1+C^{c,\tau}_{V_L}-C^{c,\tau}_{V_R}|^2 \nonumber \\
&&+ 0.11\Re[(1+C^{c,\tau}_{V_L} -C^{c,\tau}_{V_R})C^{c,\tau *}_P] +0.034 |C^{c,\tau}_P|^2 \nonumber \\
	&&-1.95\Re[(1+C^{c,\tau}_{V_L}-C^{c,\tau}_{V_R})C^{c,\tau *}_T]
	+3.08|C^{c,\tau}_T|^2\Biggl].
\end{eqnarray}	
On the other hand, should NP affect the light generations, the effects on
$R_{D^{*}}$ are described as
\begin{eqnarray}
\label{eq:RDstarLight}
\mathcal{R}(D^{*})/\mathcal{R}(D^{*})^{\rm SM}\Bigl|_{e,\mu}&=&
\frac{1}{G^{D*}_{\mu} + G^{D*}_{e}},		
\end{eqnarray}
with
\begin{eqnarray}
G^{D*}_{\mu}&=&0.499|1+C^{c,\mu}_{V_L}|^2 + 0.499 |C^{c,\mu}_{V_R}|^2 
	- 0.874\Re[(1+C^{c,\mu}_{V_L})C^{c,\mu *}_{V_R}] \nonumber \\
	&&+ 0.009\Re[(1+C^{c,\mu}_{V_L}-C^{c,\mu}_{V_R})C^{c,\mu *}_P] 
	+ 0.025|C^{c,\mu}_{P}|^2 -0.221\Re[(1+C^{c,\mu}_{V_L})C^{c,\mu *}_T] \nonumber \\
	  &&+ 7.710|C^{c,\mu}_T|^2 + 0.356\Re[C^{c,\mu}_{V_R} C^{c,\mu *}_T],
\nonumber\\
G^{D*}_{e}&=&0.501|1+C^{c,e}_{V_L}|^2 + 0.501|C^{c,e}_{V_R}|^2 
	- 0.878\Re[(1+C^{c,e}_{V_L})C^{c,e *}_{V_R}] \nonumber \\
	&&+ 4.50\times 10^{-5}\Re[(1+C^{c,e}_{V_L}-C^{c,e}_{V_R})C^{c,e *}_P] 
	+ 0.026|C^{c,e}_{P}|^2 \nonumber \\
	&&- 0.001\Re[(1+C^{c,e}_{V_L})C^{c,e *}_T]
	  + 7.743|C^{c,e}_T|^2 + 1.75\times 10^{-3}\Re[C^{c,e}_{V_R} C^{c,e *}_T]. \label{eq:GDast}
\end{eqnarray}
For $\mathcal{R}(D)$, the corresponding contributions to the light
generations are given by
\begin{eqnarray}
\mathcal{R}(D)/\mathcal{R}(D)^{\rm SM}\Bigl|_{e, \mu}&=&
\frac{1}{G^{D}_{\mu} + G^{D}_{e}}
\end{eqnarray}
with
\begin{eqnarray}
G^{D}_{\mu}&=&0.50 |1 + C^{c,\mu}_{V_L} + C^{c,\mu}_{V_R}|^2
+ 0.07\Re[(1+C^{c,\mu}_{V_L} + C^{c,\mu}_{V_R})C^{c,\mu *}_{S}]
+0.52|C^{c,\mu}_S|^2\nonumber\\
&&+0.10\Re[(1+C^{c,\mu}_{V_L}+C^{c,\mu}_{V_R})C^{c,\mu *}_{T}]+0.37
|C^{c,\mu}_{T}|^2,\nonumber\\
G^{D}_{e}&=&0.50|1+ C^{c,e}_{V_L} + C^{c,e}_{V_R}|^2 
	+ 3.6\times 10^{-4}\Re[(1+C^{c,e}_{V_L} + 
	C^{c,e}_{V_R})C^{c, e *}_S]+0.53|C^{c, e}_S|^2\nonumber\\
	&&+5.0\times 10^{-4}\Re[(1+C^{c,e}_{V_L} + 
	   C^{c,e}_{V_R})C^{c,e*}_{T}] + 0.37 |C^{c,e}_T|^2. \label{eq:GD}
\end{eqnarray}
For the light semileptonic ratio $R^e_{\mu}(D^*)$ introduced in Eq.~(\ref{eq:Rlight}), we have
\begin{eqnarray}
R^{e}_{\mu}(D^*)/R^{e~\rm{SM}}_{\mu}(D^*)&=&
\frac{\tilde{G}^{D^*}_e}{\tilde{G}^{D^*}_{\mu}},
\end{eqnarray}
with 
\begin{eqnarray}
\tilde{G}^{D^*}_{e}&=&|1+C^{c,e}_{V_L}|^2+|C^{c,e}_{V_R}|^2+
1.753\Re[(1+C^{c,e}_{V_L})C^{c,e *}_{V_R}] \nonumber \\
&&+8.981\times 10^{-5}\Re[(1+C^{c, e}_{V_L}-C^{c, e}_{V_R})C^{c, e *}_P]
+0.051|C^{c, e}_{P}|^2 \nonumber \\
&&+ 2.164\times 10^{-3}\Re[(1+C^{c, e}_{V_L})C^{c, e *}_{T}]+3.504\times 10^{-3}\Re[C^{c,e}_{V_R}C^{c, e*}_{T}]+15.455|C^{c, e}_{T}|^2,\nonumber\\
\tilde{G}^{D^*}_{\mu}&=&|1+C^{c, \mu}_{V_L}|^2+|C^{c, \mu}_{V_R}|^2+ 1.753\Re[(1+C^{c, \mu}_{V_L})C^{c, \mu *}_{V_R}] \nonumber \\
&&+ 0.018\Re[(1+C^{c, \mu}_{V_L}-C^{c, \mu}_{V_R})C^{c, \mu *}_P]
+0.051|C^{c, \mu}_{P}|^2+0.443\Re[(1+C^{c, \mu}_{V_L})C^{c, \mu *}_{T}] \nonumber \\
&&+ 0.714\Re[C^{c, \mu}_{V_R}C^{c, \mu *}_{T}]+15.458|C^{c, \mu}_{T}|^2. \label{eq:GtilDast}
\end{eqnarray}
In order to avoid any potential confusion, let us emphasize again that the short-distance coefficients in Eqs.~(\ref{eq:RDstarTau}),~(\ref{eq:GDast}),~(\ref{eq:GD})~and~(\ref{eq:GtilDast}) correspond to the scale $\mu = m_b$.

\subsection{Constraints Pseudo-scalar Wilson Coefficients}
\label{subsec:pseudoscalar}

\begin{center}
\begin{figure}[H]
\begin{center}
\includegraphics[width=0.6\textwidth]{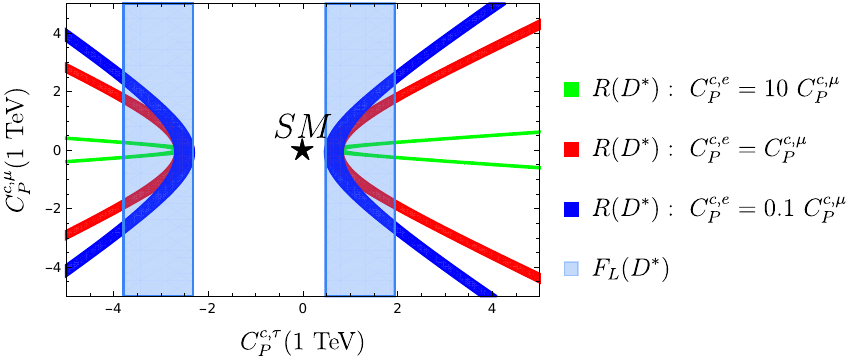}
\caption{Constraints in the $C^{c, \tau}_P$--$C^{c, \mu}_P$
plane considering different correlations 
between $C^{c, e}_{P}$ and $C^{c, \mu}_P$.}
\label{fig:CPRDstar}
\end{center}
\end{figure}
\end{center}

In order to constrain the possible values of the pseudo-scalar Wilson coefficients for $\tau$
and light leptons $e, \mu$,  we make a combined scan which includes the observables $\mathcal{R}(D^{*})$ and $F_L(D^{*})$ introduced in Eq.~(\ref{eq:RDandRDstar})  and Eq.~(\ref{eq:FL}), respectively. The corresponding results are presented in Fig.~\ref{fig:CPRDstar} from which we can read off the intervals allowed for $C^{c, \mu}_P$ and $C^{c, \tau}_P$ at $1$~$\rm{TeV}$. We summarize these results in the $2$nd and the $4$th columns
of Table~\ref{tab:CPvalues}. Here we see how the regions for $C^{c, \tau}_P$ fall into any of the two intervals $(-3.78,-2.32)$ or $(0.48,1.94)$ regardless of the assumed NP scenario correlating $C^{c, e}_{P}$ with $C^{c, \mu}_P$. 

\begin{table}
\begin{center}
\begin{tabular}{ | c | c | c | c | c |}
\hline
Scenario&\multicolumn{2}{|c|}{$C^{c,\mu}_P(1~\rm{TeV})$} & \multicolumn{2}{|c|}{$C^{c,\tau}_{P}(1~\rm{TeV})$}\\
\hline
& $\mathcal{R}(D^*)$ 
& 
& $\mathcal{R}(D^*)$ 
& $\mathcal{R}(D^*)$, $F_L(D^*)$ \\
&and  
& $R^e_{\mu}(D^*)$
&and 
&and  
\\
&$F_L(D^*)$ 
&
&
 $F_L(D^*)$
& $R^e_{\mu}(D^*)$\\
\hline
&$[-0.27,0.27]$&& $[-3.78, -2.32]$&$[-2.76,-2.32]$\\
$C^{c,e}_P=10C^{c,\mu}_P$& &$[-0.05,0.05]$& &\\
&$[-0.27,0.27]$&& $[0.48, 1.94]$&$[0.48,0.90]$\\
\hline
&$[-2.0, 1.90]$&& $[-3.78,-2.32]$&$[-3.52, -2.32]$\\
$C^{c,e}_P=C^{c,\mu}_P$&&$[-1.46,1.04]$&&\\
&$[-2.00, 1.89]$&&$[0.48,1.94]$&$[0.48, 1.66]$\\
\hline	
&$[-2.86, 2.65]$&& $[-3.78, -2.32]$& $[-2.73, -2.32]$\\
$C^{c,e}_P=0.1 C^{c,\mu}_P$&&$[-0.55,0.34]$&&\\
&$[-2.84, 2.64]$&&$[0.48,1.94]$&$[0.48,0.89]$\\
\hline
\end{tabular}
\end{center}
\caption{Bounds for $C^{c,\mu}_P$ and $C^{c,\tau}_P$ in
different scenarios correlating $C^{c,e}_P$ and $C^{c,\mu}_P$.}
\label{tab:CPvalues}
\end{table}

\begin{center}
\begin{figure}[H]
\begin{center}
\includegraphics[width=0.6\textwidth]{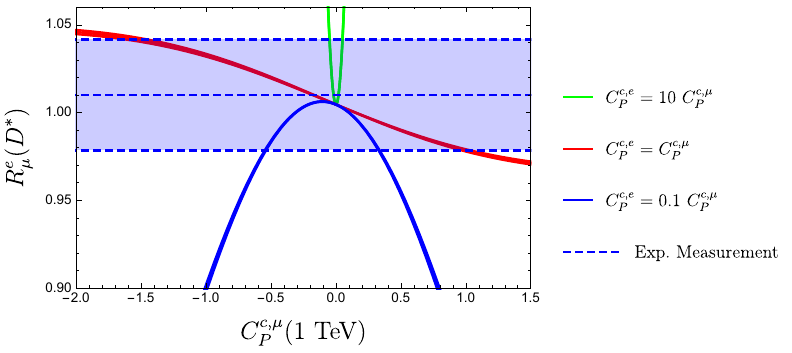}
\caption{Constraints on the $C^{c, \mu}_P$ and $C^{c, e}_P$
Wilson coefficients considering different correlations between them. The observable taken into account is $R^e_{\mu}(D^*)$.}
\label{fig:CPLight}
\end{center}
\end{figure}
\end{center}

In the case of the light generations, the NP intervals depend on the scenario under consideration, see the $2$nd column in Table~\ref{tab:CPvalues}, and can be relatively sizeable, reaching values as large as $|C^{c, \mu}_P|\sim 3$. We can, however, refine these bounds by including the light universality ratio $R^e_{\mu}(D^*)$. Unlike $\mathcal{R}(D^*)$  and $F_L(D^*)$, the observable $R^e_{\mu}(D^*)$ is only sensitive to potential NP in light generations and has the potential of further constraining $C^{c, \mu}_P$ by  one order of magnitude in most of the cases. We show the effect of 
$R^e_{\mu}(D^*)$
in Figure~\ref{fig:CPLight}, and present our numerical results in the $3$rd column of Table~\ref{tab:CPvalues}. Interestingly, since $C^{c, \mu}_P$ and $C^{c, \tau}_P$ are correlated by  $\mathcal{R}(D^*)$  and $F_L(D^*)$, we can return to these ratios with our new findings for $C^{c, \mu}_P$ to improve the available regions for  $C^{c, \tau}_P$. We present the corresponding results in the $5$th column of Table~\ref{tab:CPvalues}.

We have seen  how the observable $R^{e}_{\mu}(D^*)$, introduced in Eq.~(\ref{eq:Rlight}), plays an important role in constraining the NP contributions to the light generations of leptons. This results from its high experimental precision, as given in Eq.~(\ref{eq:RemuExp}). An interesting question is whether this observable can impose any restriction on the correlation factor $f^e_{\mu}$  which relates the short distance contributions for $e$ and  $\mu$ presented in  Eq.~(\ref{eq:NPemucorr}). To
answer this, we study the relationship between $f^e_{\mu}$ and $|C^{c, \mu}_P(m_b)|$, finding
\begin{eqnarray}
f^e_{\mu}&=&\pm\frac{1}{C^{c, \mu}_P}\sqrt{\frac{1}{0.051}\Biggl[\frac{R^e_{\mu}(D^*)}{R^{e, SM}_{\mu}(D^*)}\Biggl(1 +  0.051|C^{c, \mu}_P|^2\Biggl)-1 \Biggl]},
\label{eq:CPRemu}
\end{eqnarray}
where the  Wilson coefficients are evaluated at $\mu=m_b$.

\begin{center}
\begin{figure}[H]
\begin{center}
\includegraphics[width=0.6\textwidth]{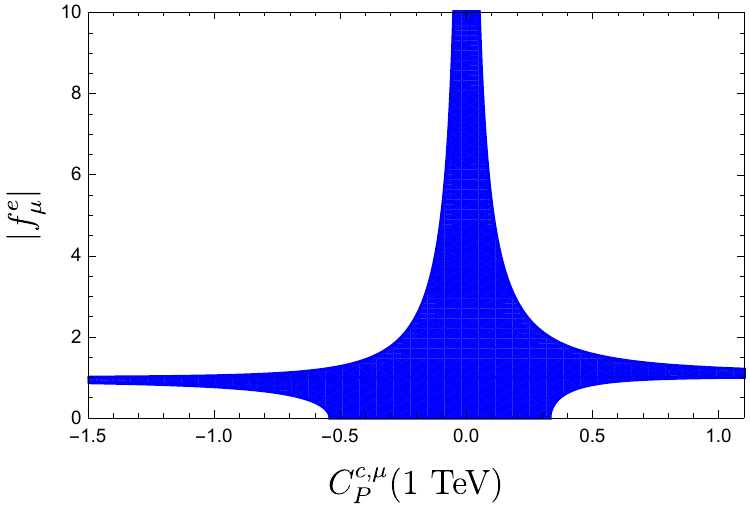}
\caption{Correlation following from Eq.~(\ref{eq:CPRemu}) between $C^{c, \mu}_P$ and the NP enhancement factor $f^e_{\mu}$, the latter relating the former with $C^{c, e}_{P}$ as defined through Eq.~(\ref{eq:NPemucorr}).}
\label{fig:CPfemu}
\end{center}
\end{figure}
\end{center}

A careful analysis of the expression inside the square root shows that as long as the quantity $R^e_{\mu}(D^*)/R^{e, SM}_{\mu}(D^*)$ is compatible with one, there is no upper bound on  $|f^e_{\mu}|$. If we take $R^e_{\mu}(D^*)$ equal to the experimental value given in Eq.~(\ref{eq:RemuExp}) we see that, within the current theoretical and experimental uncertainties, we have $R^e_{\mu}(D^*)=R^{e, SM}_{\mu}(D^*)$. Hence,  we conclude that, in spite of the precision in theory and experiment available for $R^e_{\mu}(D^*)$, it is not possible to restrict the values for $f^e_{\mu}$. For completeness, we present graphically the relationship between $f^e_{\mu}$ and $C^{c, \mu}_P$ in Fig.~\ref{fig:CPfemu}.

\subsection{Constraints on Scalar Wilson Coefficients}
\label{subsec:scalar}

\begin{center}
\begin{figure}[H]
\begin{center}
\includegraphics[width=0.35\textwidth]{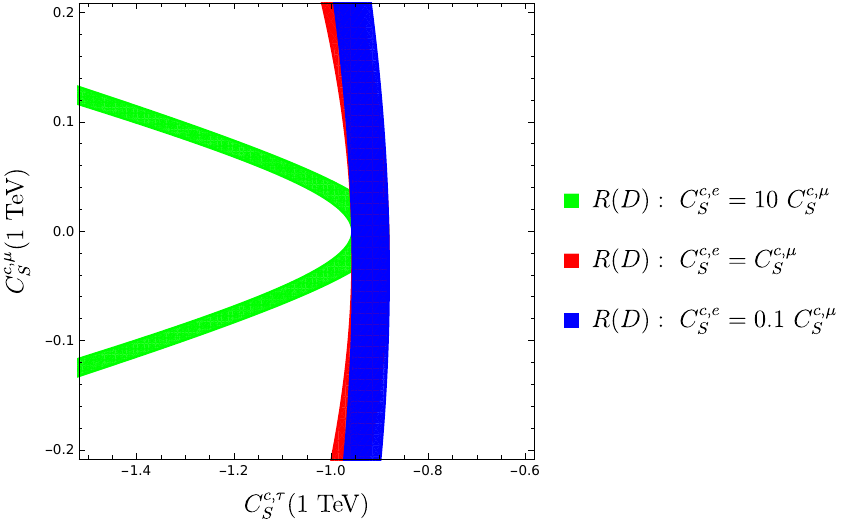}
\includegraphics[width=0.35\textwidth]{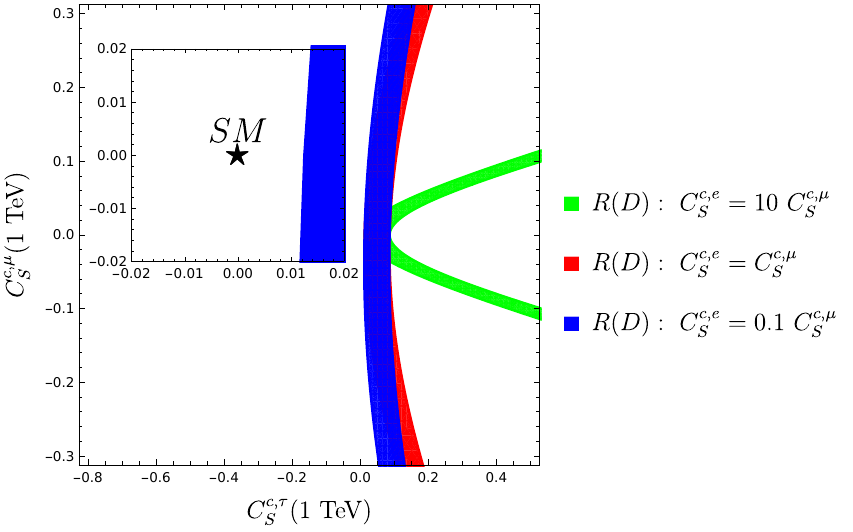}
\caption{Constraints in the $C^{\tau}_S$--$C^{\mu}_S$ plane considering different correlations 
between $C^e_S$ and $C^{\mu}_S$.}
\label{fig:CS}
\end{center}
\end{figure}
\end{center}

The scalar Wilson coefficients for semileptonic $b\rightarrow c$ transitions  enter in our analysis only 
through the observable $\mathcal{R}(D)$. Even though $\mathcal{R}(D)$ correlates $C^{c, \tau}_S$ with $C^{c, \mu}_S$, we find that this observable does not lead to strong bounds on these coefficients. We illustrate the interplay between $C^{c, \tau}_S$ and $C^{c, \mu}_S$ as imposed by $\mathcal{R}(D)$ in Fig.~\ref{fig:CS} for the different assumed scenarios between $C^{c, e}_S$ and $C^{c, \mu}_S$.

\subsection{Constraints on Vector Wilson Coefficients}
\label{subsec:vector}

\begin{center}
\begin{figure}[H]
\begin{center}
\includegraphics[width=0.40\textwidth]{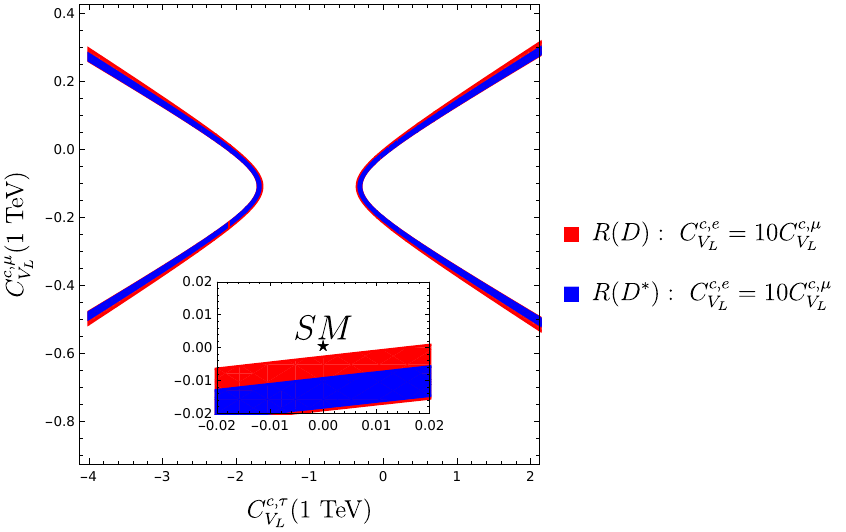}
\includegraphics[width=0.40\textwidth]{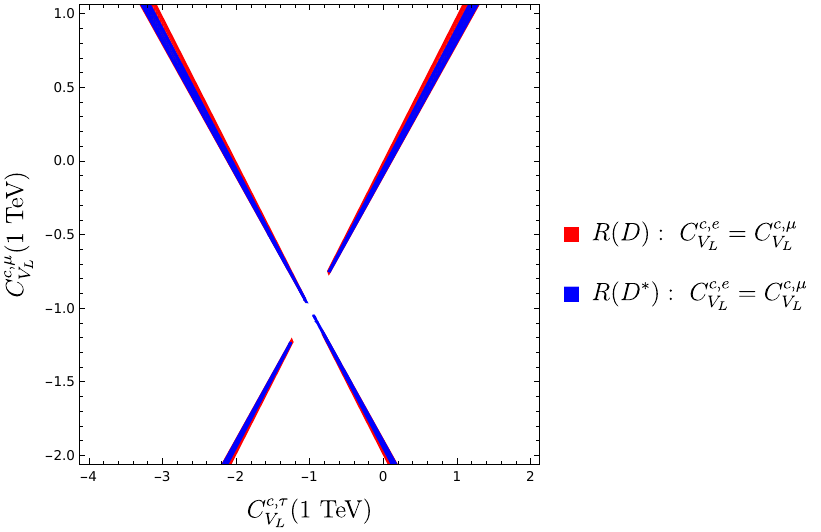}
\includegraphics[width=0.40\textwidth]{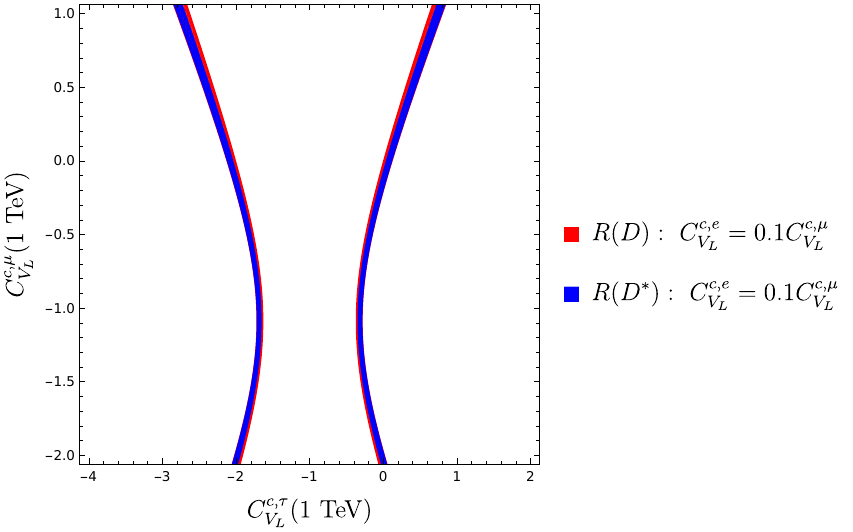}
\caption{Constraints in the $C^{c, \tau}_{V_L}$--$C^{c, \mu}_{V_L}$ plane considering different correlations 
between the Wilson coefficients $C^{c,  e}_{V_L}$ and $C^{c, \mu}_{V_L}$.}
\label{fig:CVL}
\end{center}
\end{figure}
\end{center}

\begin{center}
\begin{figure}[H]
\begin{center}
\includegraphics[width=0.6\textwidth]{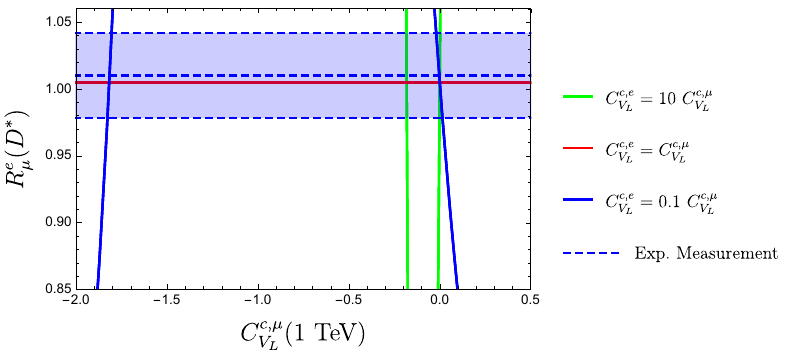}
\caption{Constraints on the Wilson coefficient $C^{c, \mu}_{V_L}$, considering different NP scenarios through its correlation with $C^{c, e}_{V_L}$. The observable taken into account is $R^e_{\mu}(D^*)$.}
\label{fig:CVLLight}
\end{center}
\end{figure}
\end{center}

At first sight, the constraints on  the left-handed Wilson coefficients can be derived using  $\mathcal{R}(D)$, $\mathcal{R}(D^*)$, $F_{L}(D^*)$ and $\mathcal{R}^e_{\mu}(D^*)$. 
The observables $\mathcal{R}(D)$ and $\mathcal{R}(D^*)$ are sensitive to the effects of 
$\tau$ as well as $e$ and $\mu$. On the other hand, 
$F_L(D^*)$ is only sensitive to NP in $\tau$ leptons 
in general. Interestingly, if the vector-left coefficient
is the only effect entering in $F_{L}(D^*)$, then they 
cancel exactly since they have equal contributions in
numerator and denominator, thereby rendering $F_L(D^*)$ insensitive
to $C^{c, \tau}_{V_L}$. This can be verified using Eq.~(\ref{eq:RDstarTau}).
Therefore, the limits on $C^{c, \tau}_{V_L}$ can only be computed through
$\mathcal{R}(D)$ and $\mathcal{R}(D^*)$, which as shown in Figure~\ref{fig:CVL} cannot yield bounded regions in the $C^{c, \tau}_{V_L}$--$C^{c, \mu}_{V_L}$ plane.  Nevertheless, the inclusion of $R^e_{\mu}(D^*)$, which is sensitive to $C^{c,  e}_{V_L}$
and $C^{c, \mu}_{V_L}$, improves this situation. Indeed, by providing bounds on $C^{c,  e}_{V_L}$
and $C^{c, \mu}_{V_L}$ as shown in Fig.~\ref{fig:CVLLight}, the observable $R^e_{\mu}(D^*)$ is also able to constrain $C^{c, \tau}_{V_L}$ because the NP Wilson coefficients are not independent from each other but are correlated through $\mathcal{R}(D)$ and $\mathcal{R}(D^*)$, as can be seen in Fig.~\ref{fig:CVL}. Our findings are summarized in Table~\ref{tab:CVLvalues}. As seen in Fig.~\ref{fig:CVLLight}, when $C^{c,e}_{V_L}=C^{c,\mu}_{V_L}$ the observable $R^{e}_{\mu}(D^*)$ does not provide constrains on $C^{c, e}_{V_L}$ and $C^{c, \mu}_{V_L}$.

\begin{table}
\begin{center}
\begin{tabular}{ | c | c | c |}
\hline
Scenario& $C^{c, \mu}_{V_L}(1~\rm{TeV})$ & $C^{c, \tau}_{V_L}(1~\rm{TeV})$\\
\hline
& $R^e_{\mu}(D^*)$&  $\mathcal{R}(D^*)$, $\mathcal{R}(D)$  and $R^e_{\mu}(D^*)$ \\
\hline
&$[-0.183,-0.181]$&$[-1.913,-1.857]\cup [-0.143,-0.087]$\\
$C^{c,  e}_{V_L}=10C^{c, \mu}_{V_L}$&&\\
&$[-0.001,0.002]$&$[-2.120,-2.045]\cup [0.045,0.120]$\\
\hline
&$[-1.828,-1.805]$& $[-1.911,-1.855]\cup [-0.144,-0.089]$\\
$C^{c,  e}_{V_L}=0.1 C^{c, \mu}_{V_L}$&&\\
&$[-0.020,0.015]$&$[-2.116,-2.042]\cup [0.042,0.116]$\\
\hline
\end{tabular}
\end{center}
\caption{Bounds for $C^{c, \mu}_{V_L}$ and $C^{c, \tau}_{V_L}$ in
different scenarios correlating $C^{c,  e}_{V_L}$ and $C^{c, \mu}_{V_L}$. For the scenario $C^{c,  e}_{V_L}=C^{c,  \mu}_{V_L}$ the allowed NP regions are not bounded and thus not presented.}
\label{tab:CVLvalues}
\end{table}

\begin{center}
\begin{figure}[H]
\begin{center}
\includegraphics[width=0.40\textwidth]{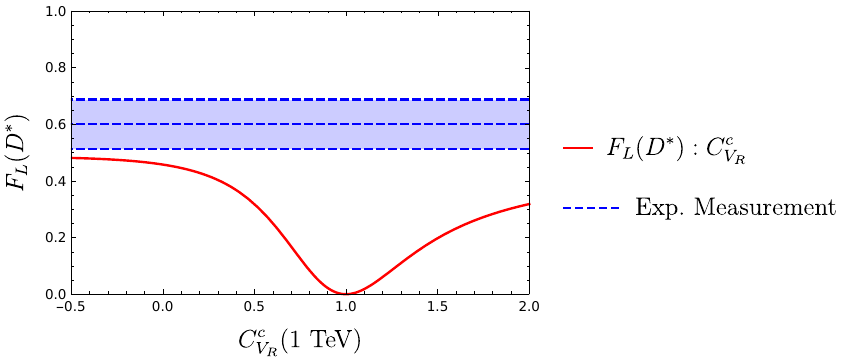}
\includegraphics[width=0.40\textwidth]{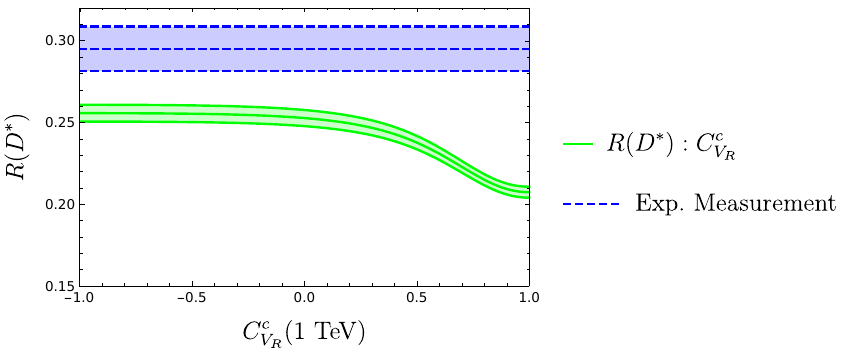}
\caption{Constraints on the Wilson coefficient $C_{V_R}$ following from $F_L(D^\ast)$ (left) and $\mathcal{R}(D^\ast)$ (right).}
\label{fig:CVR}
\end{center}
\end{figure}
\end{center}

In order to study the behaviour of  $C_{V_R}$, we consider the universality behaviour with respect to the different lepton 
flavours as established in Eq.~(\ref{eq:universality}). However, as seen in Fig.~\ref{fig:CVR}, the combination of observables included in this
study cannot explain the current experimental data if 
$C_{V_R}$ is the only NP contribution. 

\subsection{Constraints on Tensor Wilson Coefficients}
\label{subsec:tensor}

\begin{center}
\begin{figure}[H]
\begin{center}
\includegraphics[width=0.40\textwidth]{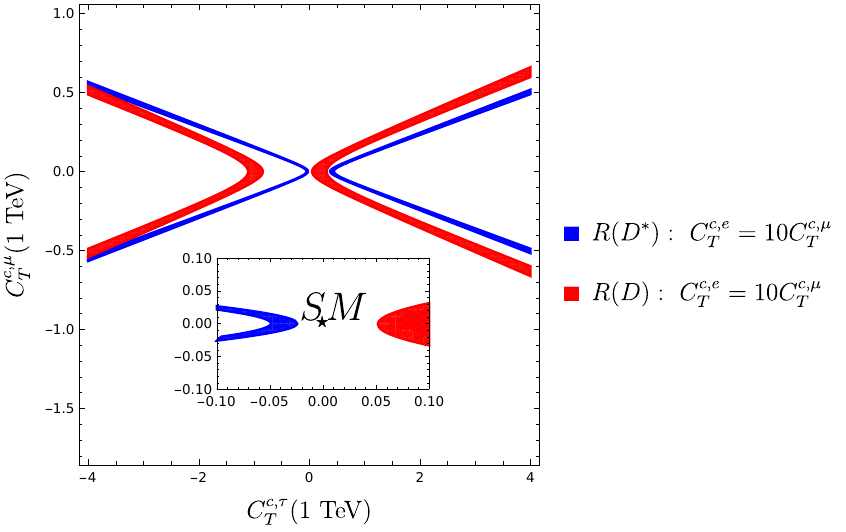}
\includegraphics[width=0.38\textwidth]{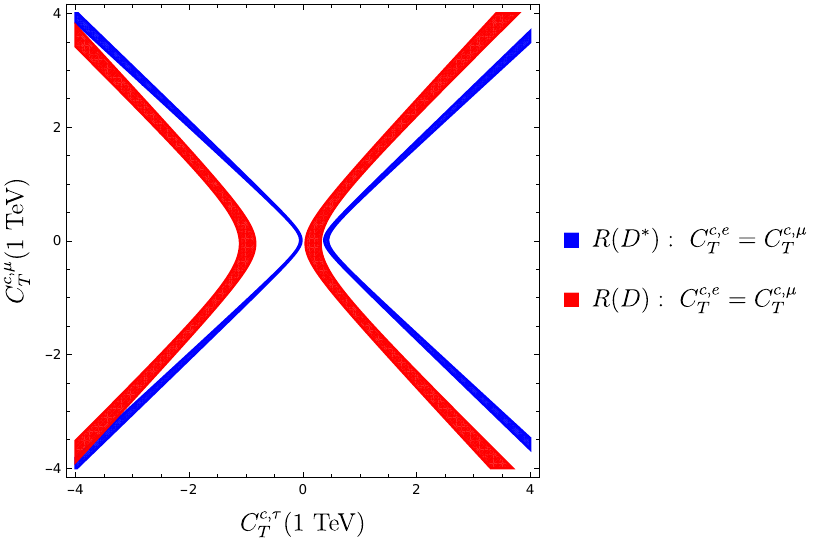}
\includegraphics[width=0.40\textwidth]{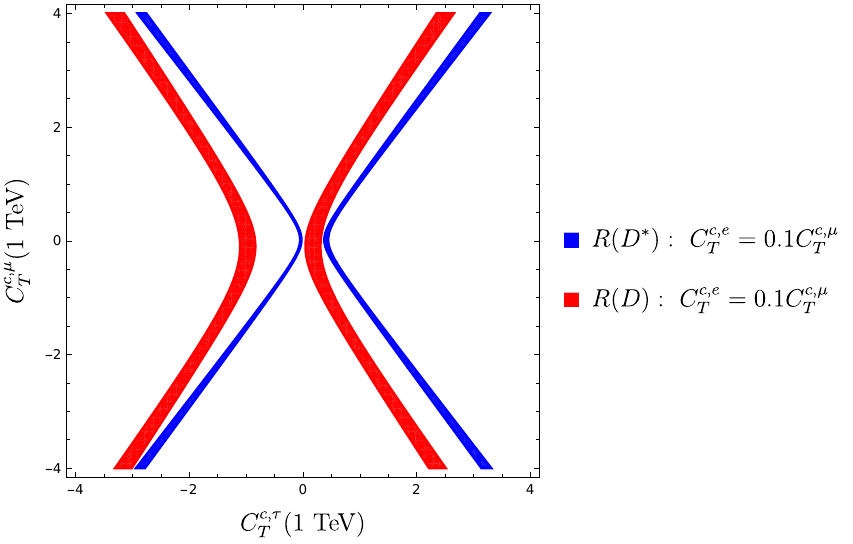}
\caption{Constraints in the $C^{c, \tau}_{T}$--$C^{c, \mu}_{T}$ plane following from $\mathcal{R}(D)$ and $\mathcal{R}(D^\ast)$, considering different correlations 
between the Wilson coefficients $C^{c, e}_{T}$ and $C^{c, \mu}_{T}$.}
\label{fig:CT}
\end{center}
\end{figure}
\end{center}

\begin{center}
\begin{figure}[H]
\begin{center}
\includegraphics[width=0.6\textwidth]{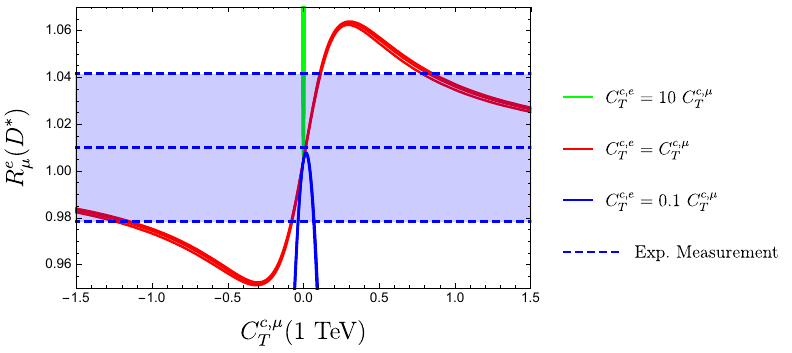}
\caption{Constraints on the Wilson coefficient $C^{c, \mu}_T$, considering different NP scenarios through its correlation with $C^{c, e}_T$. The observable taken into account is $R^e_{\mu}(D^*)$.}
\label{fig:CTLight}
\end{center}
\end{figure}
\end{center}

The tensor Wilson coefficient enters in $\mathcal{R}(D)$, $\mathcal{R}(D^*)$, $F_L(D^*)$ and $R^e_{\mu}(D^*)$. In the case of $F_{L}(D^*)$, the tensor
Wilson coefficient $C^{c, \tau}_T$ does not reproduce the
experimental data within $1~\sigma$. As for $\mathcal{R}(D)$ and 
$\mathcal{R}(D^*)$, the resulting regions are not bounded 
in the $C^{c, \tau}_T$--$C^{c, \mu}_{T}$ plane. This can be
seen in Figure \ref{fig:CT}.  For completeness, we perform the analysis for $R^{e}_{\mu}(D^*)$ finding bounded regions for $C^{c,\mu}_T$ in Figure~\ref{fig:CTLight}, we also present the corresponding numerical intervals in Table~\ref{tab:CTvalues}. However, our results for $\mathcal{R}(D)$ and $\mathcal{R}(D^*)$ already exclude the possibility of explaining data with a tensor interaction at the $1~\sigma$ level.

\begin{table}
\begin{center}
\begin{tabular}{ | c | c |}
\hline
Scenario& $C^{c, \mu}_{T}(1~\rm{TeV})$\\
\hline
&$R^e_{\mu}(D^*)$\\
\hline
&\\
$C^e_T=10C^{\mu}_T$&$[-0.006,0.006]$\\
&\\
\hline
&$(-\infty,-1.242]$\\
$C^e_T=C^{\mu}_T$&$[-0.080,0.115]$\\
&$[0.781,\infty)$\\
\hline	
&\\
$C^e_T=0.1 C^{\mu}_T$&$[-0.037,0.073]$\\
&\\
\hline
\end{tabular}
\end{center}
\caption{Bounds for $C^{c,\mu}_T$ in
different NP scenarios correlating it with $C^{c, e}_T$. The observable considered is $R^e_{\mu}(D^*)$.}
\label{tab:CTvalues}
\end{table}

\boldmath
\section{Extraction of $|V_{cb}|$ and Implications for $\mathcal{B}(B_c\rightarrow \ell^-\bar{\nu}_{\ell})$} \label{sec:Vcb}
\unboldmath

In this section, we proceed with the extraction of the 
values of $|V_{cb}|$ which are compatible with the NP 
contributions determined in the previous subsections. In analogy with the determination of $|V_{ub}|$ presented in Section~\ref{sec:VubPredbu}, we use the bounds for the NP Wilson coefficients  derived from the $|V_{cb}|$-independent observables $\mathcal{R}(D)$, $\mathcal{R}(D^*)$, $F_L(D^*)$ and $R^e_{\mu}(D^*)$ to extract $|V_{cb}|$ through the interplay of an extra  observable which is $|V_{cb}|$-dependent. To accomplish this goal and for the purposes of illustration, we will  focus on the averaged branching fraction $\braket{{\mathcal B}(\bar B\rightarrow D^{*} \tau^- \bar{\nu}_{\tau})}$ which is sensitive to both the NP short-distance contributions and $|V_{cb}|$ itself. Moreover, this particular average is not affected by the ambiguity concerning the leptonic combination in the final state which we encountered while interpreting the experimental measurement of $\braket{{\mathcal B}(\bar B\rightarrow D^{*} \ell^- \bar{\nu}_{\ell})}$, where the contributions from electrons and muons were not transparent.

The possible regions for the short-distance coefficients and $|V_{cb}|$ can be further refined by imposing upper bounds on the leptonic $B_c$ branching ratios as extra constraints, especially in scenarios with non-vanishing pseudoscalar Wilson coefficients at the bottom scale, as they lift the helicity suppression of the leptonic modes. Unfortunately, no direct constraints are available from the LHC. In view of this, an estimate of a bound on $\mathcal{B}(B_c\rightarrow \tau^- \bar{\nu}_{\tau})$ has been derived from LEP data at the $Z$ peak in Ref.~\cite{Akeroyd:2017mhr}. In Refs.~\cite{Blanke:2018yud,Blanke:2019qrx}, it was argued that this bound is too strict and values up to $60 \%$ cannot be excluded. Thus we will take
\begin{eqnarray}
\label{eq:Bcbound}
\mathcal{B}(B_c\rightarrow  \tau^- \bar{\nu}_{\tau})<0.60.
\label{eq:BcUB}
\end{eqnarray}
Information on the lifetime $\tau_{B_c}$ of the $B_c$ mesons can also be converted into bounds for the leptonic branching ratios \cite{Alonso:2016oyd,Watanabe:2017mip}, giving results in the same ballpark \cite{Blanke:2018yud}. However, as this observable depends on any possible $B_c$ decay channel, the interpretation is more complex since other NP contributions may enter.

In order to incorporate Eq.~(\ref{eq:BcUB}) in our studies, we use the theoretical expression
\begin{eqnarray}
\label{eq:BrBcTaunu}	
\mathcal{B}(B_c\rightarrow \tau \nu_{\tau})/
\tilde{\mathcal{B}}r(B_c\rightarrow \tau \nu_{\tau})^{\rm SM}
&=& |V_{cb}|^2 |1 +C^{c, \tau}_{V_L}+ 4.06 C^{c, \tau}_P|^2,
\end{eqnarray}	
with 
\begin{eqnarray}
\tilde{\mathcal{B}}r(B_c\rightarrow \tau \nu_{\tau})^{\rm SM}
&=&
\frac{G^2_F}{8\pi}M_{B_c}m^2_{\tau}
\Bigl(1-\frac{m^2_{\tau}}{M_{B_c}^2}\Bigl)^2f^2_{B_c}
\tau_{B_c},
\end{eqnarray}
and the Wilson coefficients are evaluated at $\mu=m_b$.

For the  purposes of numerical comparison, we make a small digression from our strategy and evaluate the leptonic branching fractions for the $B_c$ meson using the value of $|V_{cb}|$ obtained in \cite{Amhis:2019ckw} from exclusive $\bar{B}\rightarrow D \ell^-\bar{\nu}_{\ell}$ decays:
\begin{eqnarray}
|V_{cb}|&=&0.03958\pm 0.00117.
\label{eq:VcbHFAG}
\end{eqnarray}
Assuming this value of $|V_{cb}|$, we may calculate the following ``SM'' leptonic branching fractions:
\begin{eqnarray}
\mathcal{B}(B_c\rightarrow e^-\bar{\nu}_{e})\Bigl|_{\rm SM}&=&(2.03\pm 0.19)\times 10^{-9},\label{eq:Bcellnuell1}\\
\mathcal{B}(B_c\rightarrow \mu^-\bar{\nu}_{\mu})\Bigl|_{\rm SM}&=&(8.67\pm 0.80)\times 10^{-5},\label{eq:Bcellnuell2}\\
\mathcal{B}(B_c\rightarrow \tau^-\bar{\nu}_{\tau})\Bigl|_{\rm SM}&=&(2.25\pm 0.21)\times 10^{-2}.
\label{eq:Bcellnuell3}
\end{eqnarray}
We would like to stress that since the results in Eq.~(\ref{eq:Bcellnuell1}) to Eq.~(\ref{eq:Bcellnuell3})  use an external determination for $|V_{cb}|$, they were not derived from our strategy and presented only as reference values for our future discussion. For completeness and for the purposes of comparison we also provide the inclusive value of $|V_{cb}|$ as presented in \cite{Charles:2004jd}

\begin{eqnarray}
|V_{cb}|&=&0.04162^{+0.00026}_{-0.00080}.
\label{eq:VcbIncl}
\end{eqnarray}

We are now ready to discuss the extraction of $|V_{cb}|$ in the presence of NP contributions. In view of the results obtained in Section~\ref{sec:btoc}, we can only
fulfill this task for the Wilson coefficients $C^{c, \tau}_P$ and $C^{c, \tau}_{V_L}$ where the resulting NP regions are bounded. We describe both cases in the following subsections.

\boldmath
\subsection{$|V_{cb}|$ Compatible with NP Pseudoscalar Coefficients}
\unboldmath

\begin{center}
\begin{figure}[H]
\begin{center}
    \includegraphics[width=0.6\textwidth]{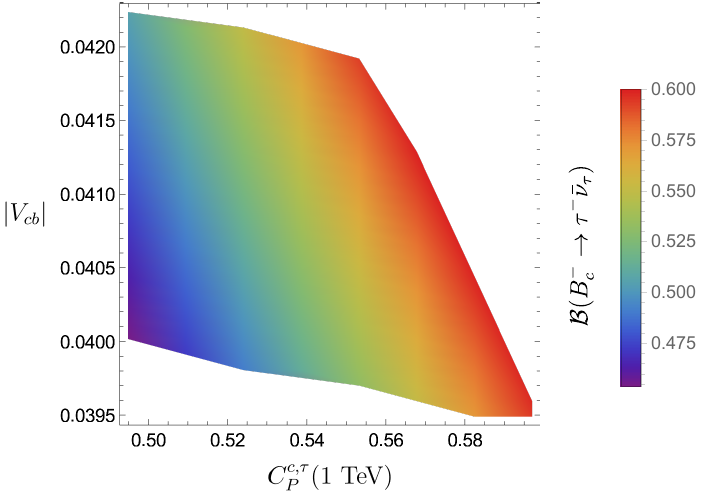}
\caption{Values for $|V_{cb}|$ and $C^{c, \tau}_P$ compatible with the $\mathcal{B}(B_c\rightarrow \tau \bar{\nu}_{\tau})$ bound in Eq.~(\ref{eq:Bcbound}).}
\label{fig:VcbCPTau}
\end{center}
\end{figure}
\end{center}

Using the values for $C^{c, \tau}_{P}$ given in Table~\ref{tab:CPvalues} and the branching fraction $\braket{{\mathcal B}(\bar B\rightarrow D^{*} \tau^- \bar{\nu}_{\tau})}$, we obtain the bounds for $|V_{cb}|$ shown in the second column of Table~\ref{tab:VcbvaluesCP}. Notice that we present our results for each of the possible correlations between $C^{c, e}_{P}$ and $C^{c, \mu}_P$.
We observe that values for $|V_{cb}|$ in the range $0.032\leq|V_{cb}|\leq 0.042$ are compatible with the experimental data considered so far. This seemingly large interval can be further reduced once the constraint  in Eq.~(\ref{eq:Bcbound})
is included. Indeed, this extra restriction has two effects. Firstly, it  restricts the values of $|V_{cb}|$ to the subinterval $0.039\leq|V_{cb}|\leq 0.042$. Secondly, it leads to the following region for the Wilson coefficient $C^{c, \tau}_P$: $0.49\leq C^{c, \tau}_P\leq 0.60$. The interplay between $C^{c, \tau}_P$, $|V_{cb}|$ and $\braket{{\mathcal B}(\bar B\rightarrow D^{*} \tau^- \bar{\nu}_{\tau})}$ is shown in Fig.~\ref{fig:VcbCPTau}. We would like to stress that our values for $|V_{cb}|$ are compatible with the exclusive determination shown in Eq.~(\ref{eq:VcbHFAG}) as well as with the inclusive value shown in Eq.~(\ref{eq:VcbIncl}).

Interestingly, since the Wilson coefficients for tau leptons and muons are correlated through the observable $\mathcal{R}(D^*)$, 
the constraint imposed by the bound in Eq.~(\ref{eq:Bcbound}) over $C^{c, \tau}_P$ has effects on the light generations as well.  In the 5th column of Table~\ref{tab:VcbvaluesCP} we show the new intervals for $C^{c, \tau}_P$ after the inclusion of $\mathcal{B}(B_c\rightarrow \tau^-\bar{\nu}_{\tau})$ in our analysis. In addition we recalculated the new intervals for $C^{c, \mu}_P$ correspondingly and present them in the 4th-column of the same table.

\begin{table}
\begin{center}
\begin{tabular}{ | c | c | c | c | c |}
\hline
Scenario & \multicolumn{2}{| c |}{$|V_{cb}|$}&$C^{c, \mu}_{P}(1~\rm{TeV})$& $C^{c, \tau}_{P}(1~\rm{TeV})$\\
\hline
&\multicolumn{4}{| c |}{$\mathcal{R}(D^*)$, $F_L(D^*)$, $R^e_{\mu}(D^*)$ }\\
\hline
&without &\multicolumn{3}{| c |}{with }\\
&$\mathcal{B}(B_c\rightarrow%
\tau^-\bar{\nu}_{\tau})$&\multicolumn{3}{| c |}{$\mathcal{B}(B_c\rightarrow%
\tau^-\bar{\nu}_{\tau})$}\\
\hline
$C^{c, e}_P=10C^{c, \mu}_P$& $[0.032, 0.042]$ &$[0.039,0.042]$&$[-0.05,0.05]$&$[0.49,0.60]$\\
\hline
$C^{c, e}_P=C^{c, \mu}_P$& $[0.032,0.042]$&$[0.039,0.042]$&$[-0.51,0.41]$&$[0.49,0.60]$\\
\hline	
$C^{c, e}_P=0.1 C^{c, \mu}_P$&$[0.034,0.042]$&$[0.039,0.042]$&$[-0.55,0.34]$&$[0.49,0.60]$\\
\hline
\end{tabular}
\end{center}
\caption{Values of $|V_{cb}|$ obtained in
different NP scenarios correlating $C^{c, e}_P$ and $C^{c, \mu}_P$. For completeness we present the effect of including (excluding) the bound given in Eq.~(\ref{eq:BcUB}) on $\mathcal{B}(B_c\rightarrow \tau^- \bar{\nu}_{\tau})$.}
\label{tab:VcbvaluesCP}
\end{table}

We would like to highlight that, as can be seen in Table~\ref{tab:VcbvaluesCP}, the interval for $C^{c, \tau}_{P}$ does not include the SM value $C^{c, \tau}_{P}=0$, which indicates that the experimental measurements involving $\tau$ leptons require NP effects to be addressed theoretically. In contrast, for muons  $C^{c, \mu}_P=0$ is compatible with the current experimental results which in our case also implies $C^{c, e}_P=0$ in view of the different scenarios that we are studying.

\begin{center}
\begin{figure}[H]
\begin{center}
\includegraphics[width=0.4\textwidth]{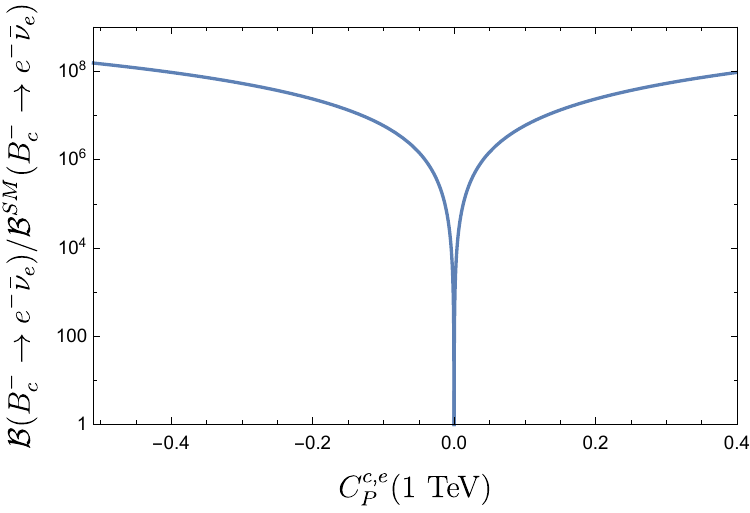}
\includegraphics[width=0.4\textwidth]{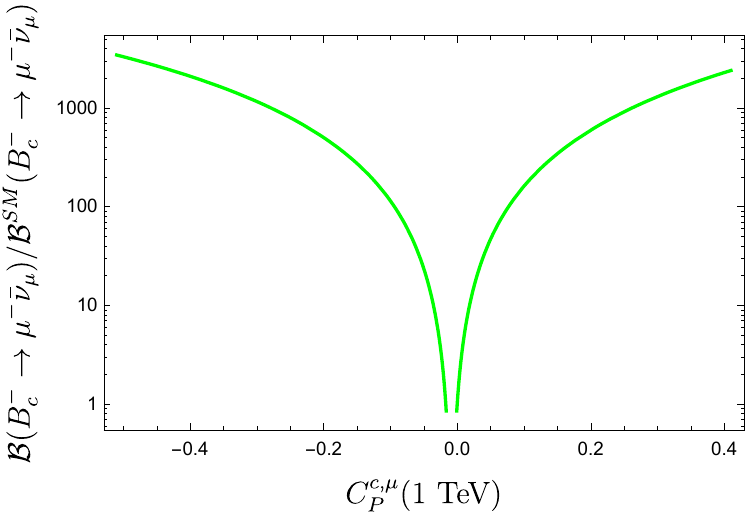}
\caption{Dependence of $\mathcal{B}(B_c\rightarrow \ell' \bar{\nu}_{\ell'})$ normalized to its SM value on $C^{c, e}_P$ and $C^{c, \mu}_P$ for $\ell'=e,\mu$.}
\label{fig:EnhancementmuP}
\end{center}
\end{figure}
\end{center}

\begin{center}
\begin{figure}[H]
\begin{center}
\includegraphics[width=0.4\textwidth]{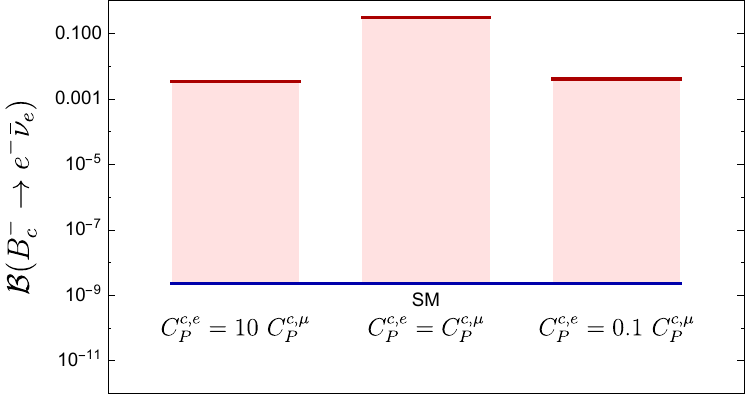}
\includegraphics[width=0.4\textwidth]{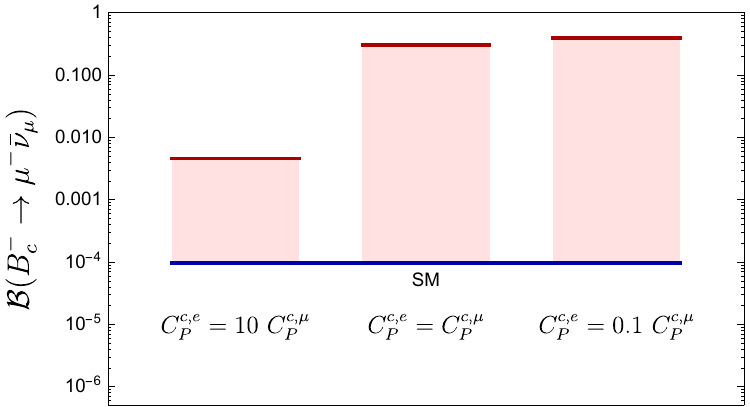}
\caption{Possible enhancements of $\mathcal{B}(B_c\rightarrow e^-\bar{\nu}_{e})$ (left) and $\mathcal{B}(B_c\rightarrow \mu^-\bar{\nu}_{\mu})$ (right) for the different NP scenarios correlating the Wilson coefficients of electrons and muons.}
\label{fig:Prop}
\end{center}
\end{figure}
\end{center}

Another interesting feature that distinguishes $\tau$ leptons from its lighter counterparts is the fact that our final intervals for $C^{c, \mu}_P$ and $C^{c, e}_P$ translate into values for $\mathcal{B}(B_c\rightarrow \mu \bar{\nu}_{\mu})$ and $\mathcal{B}(B_c\rightarrow e \bar{\nu}_{e})$ that are at most about $30\%$. To the best of our knowledge, this does not break any constraint analogous to the one in Eq.~(\ref{eq:BcUB}). Therefore, we can consider the values for  $\mathcal{B}(B_c\rightarrow \mu \bar{\nu}_{\mu})$ and $\mathcal{B}(B_c\rightarrow e \bar{\nu}_{e})$ obtained from our regions for $C^{c, \mu}_P$ and $C^{c, e}_P$ as predictions, which we present in Table~\ref{tab:BcCP}. Notice that the NP contributions can potentially lead possible cancellations which imply  branching fractions below the corresponding SM values.
We find that the masses of the light leptons can enhance the branching fractions $\mathcal{B}(B_c\rightarrow \mu \bar{\nu}_{\mu})$ and $\mathcal{B}(B_c\rightarrow e \bar{\nu}_{e})$ by several orders of magnitude through NP effects entering in pseudoscalar couplings. This effect is most dramatic for electrons, where the enhancement can reach up to eight orders of magnitude. In the case of muons, the largest enhancement is about four orders of magnitude. This  situation is analogous to the one explored in Ref.~\cite{Fleischer:2017ltw} for $B$ leptonic decays. A previous model-dependent study, which accounts for enhancements in the branching fractions of the processes $B_c\rightarrow \ell' \bar{\nu}_{\ell'}$, can be found in Ref.~\cite{Akeroyd:2002cs}. The dependence of 
$\mathcal{B}(B_c\rightarrow e\bar{\nu}_{e})$ and $\mathcal{B}(B_c\rightarrow \mu\bar{\nu}_{\mu})$  on $C^{c,e}_{P}$ and $C^{c,\mu}_{P}$, respectively, is presented in Fig.~\ref{fig:EnhancementmuP} showing the maximal effects achievable. Moreover, in Fig.~\ref{fig:Prop} we illustrate how the enhancements for $\mathcal{B}(B_c\rightarrow \ell'\bar{\nu}_{\ell'})$, with $\ell'=e,\mu$, compare to each other for  each one of our NP scenarios.

\begin{table}
\begin{center}
\begin{tabular}{ | c | c | c |}
\hline
&\multicolumn{2}{| c |}{Predictions}\\
\hline
Scenario &  $\mathcal{B}(B_c\rightarrow%
e^-\bar{\nu}_{e})$ & $\mathcal{B}(B_c\rightarrow%
\mu^-\bar{\nu}_{\mu})$   \\
\hline
$C^{c, e}_{P}=10C^{c, \mu}_{P}$&$[0, 3.33\times 10^{-3}]$&$[0, 4.55\times 10^{-3}]$\\
\hline
$C^{c, e}_{P}=C^{c, \mu}_{P}$&$[0, 0.31]$&$[0, 0.30]$\\
\hline
$C^{c, e}_{P}=0.1 C^{c, \mu}_{P}$&$[0 , 4.04\times 10^{-3}]$&$[0, 0.39]$\\
\hline
\end{tabular}
\end{center}
\caption{Predictions of $\mathcal{B}(B_c\rightarrow \ell^{'-}\bar{\nu}_{\ell'} )$
in the presence of a NP Wilson coefficient $C^{c, \ell'}_{P}$ for different scenarios correlating $C^{c, e}_{P}$ and $C^{c, \mu}_{P}$.  The zero  minima predicted for the different leptonic branching fractions is the 
consequence of a perfect cancellation between the NP contributions and the purely SM ones in Eq.~(\ref{eq:leptonic}). This implies that the effect of a pseudoscalar NP contributions can lead to values of the branching fractions of leptonic decays below their corresponding SM values.}
\label{tab:BcCP}
\end{table}

\boldmath
\subsection{$|V_{cb}|$ Compatible with NP Left-Handed Vector Coefficients}
\unboldmath

As discussed in Sec.~\ref{subsec:vector}, should 
NP enter through a left-handed vector interaction at $1~\rm{TeV}$, the available constraints given by the observables $\mathcal{R}(D)$, $\mathcal{R}(D^*)$ and $R^e_{\mu}(D^*)$ lead to four independent bounded regions for $C^{c, e}_{V_L}$, $C^{c, \mu}_{V_L}$ and $C^{c, \tau}_{V_L}$ only when either $C^{c, e}_{V_L}=10C^{c, \mu}_{V_L}$  or $C^{c, e}_{V_L}=0.1C^{c, \mu}_{V_L}$. 
However, once we include the bound for $\mathcal{B}(B_c\rightarrow \tau^- \bar{\nu}_{\tau})$ given in Eq.~(\ref{eq:Bcbound}), we find that there are actually only two regions in the $C^{c, \tau}_{V_L}$--$C^{c, \mu}_{V_L}$ plane compatible with data, which can be read in the $3$rd and $4$th columns of Table~\ref{tab:VcbvaluesCVL}.

\begin{table}
\begin{center}
\begin{tabular}{ | c | c | c | c | }
\hline
Scenario & $|V_{cb}|$& $C^{c, \mu}_{V_L}(1~\rm{TeV})$ & $C^{c, \tau}_{V_L}(1~\rm{TeV})$\\
\hline
&\multicolumn{3}{| c |}{$\mathcal{R}(D)$, $\mathcal{R}(D^*)$, $R^e_{\mu}(D^*)$ and $\mathcal{B}(B_c\rightarrow%
\tau^-\bar{\nu}_{\tau})$}\\
\hline
& & &$[-2.120, -2.045]$\\
$C^{c,  e}_{V_L}=10C^{c, \mu}_{V_L}$& $[0.038, 0.043]$ & $[-0.001, 0.002]$ & \\
&&&$[0.045,0.120]$\\
\hline
&&&$[-2.116, -2.042]$\\
$C^{c,  e}_{V_L}=0.1 C^{c, \mu}_{V_L}$&$[0.038, 0.043]$&$[-0.020, 0.01]$&\\
&&&$[0.042,0.116]$\\
\hline
\end{tabular}
\end{center}
\caption{Values of $|V_{cb}|$ in
different NP scenarios correlating $C^{c,  e}_{V_L}$ and $C^{c, \mu}_{V_L}$. We do not present the scenario $C^{c,  e}_{V_L}=C^{c,  \mu}_{V_L}$ since in this case there are not bounded regions.}
\label{tab:VcbvaluesCVL}
\end{table}

\begin{center}
\begin{figure}[H]
\begin{center}
\includegraphics[width=0.35\textwidth]{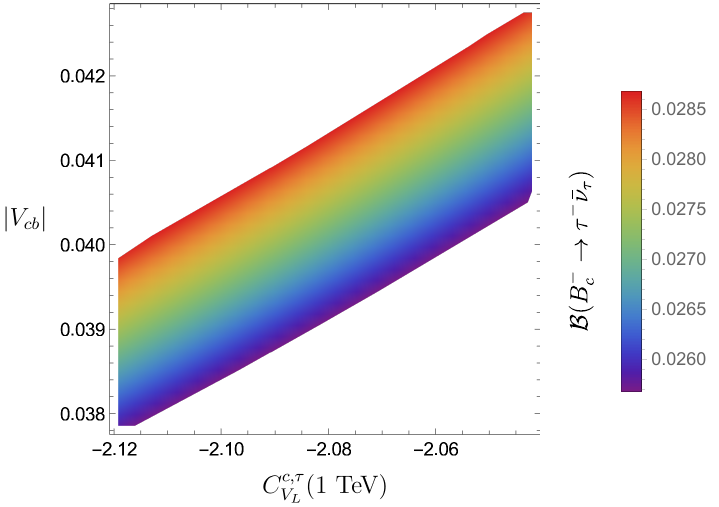}
\includegraphics[width=0.35\textwidth]{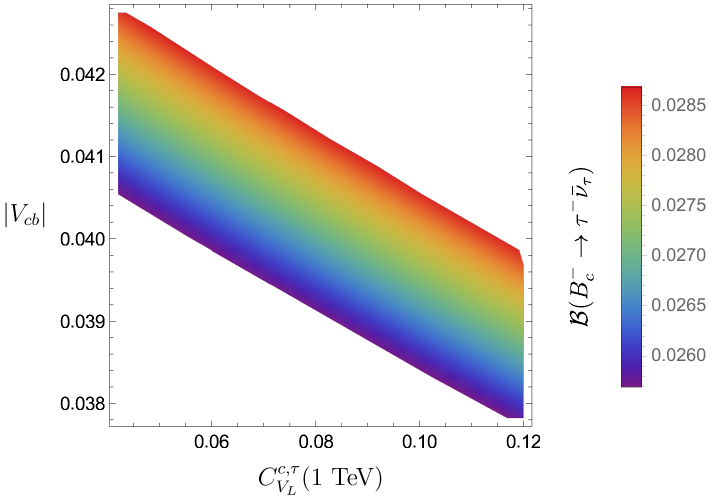}
\caption{Values for $|V_{cb}|$ and $C^{c, \tau}_{V_L}$ compatible with the bound for $\mathcal{B}(B_c\rightarrow \tau^- \bar{\nu}_{\tau})$ presented in Eq.~(\ref{eq:BcUB}).}
\label{fig:VcbCVLTau}
\end{center}
\end{figure}
\end{center}

From our maximally constrained regions for $C^{c, \mu}_{V_L}$ and $C^{c, \tau}_{V_L}$ we proceed with the extraction of $|V_{cb}|$ using the CKM-dependent observable  $\braket{{\mathcal B}(\bar B\rightarrow D^{*} \tau^- \bar{\nu}_{\tau})}$. Just as for the pseudoscalar NP interaction, we do this  by assessing the combinations of values for $C^{c, \tau}_{V_L}$  and $|V_{cb}|$ compatible with the measurement in Eq.~(\ref{eq:BDTauAv}). Our final result for $|V_{cb}|$ is then $0.038\leq |V_{cb}|\leq 0.043$, whose correlation with $C^{c, \tau}_{V_L}$ is shown in Fig.~\ref{fig:VcbCVLTau}. Once more, we find that our determination for $|V_{cb}|$ is compatible with the exclusive and inclusive values presented in Eq.~(\ref{eq:VcbHFAG})
and Eq.~(\ref{eq:VcbIncl}) respectively.

To conclude our study, we analyze the interplay between $C^{\ell}_{V_L}$ and $\mathcal{B}(B_c\rightarrow \ell^- \bar{\nu}_{\ell})$. As we have discussed at the beginning of this section, $\mathcal{B}(B_c\rightarrow \tau^- \bar{\nu}_{\tau})$ reduces the possible values that $C^{c, \mu}_{V_L}$ and $C^{c, \tau}_{V_L}$  can take by eliminating two out of the four regions allowed by the CKM-independent experimental ratios $\mathcal{R}(D)$, $\mathcal{R}(D^*)$ and $R^e_{\mu}(D^*)$. However, it is interesting to notice that the remaining  solutions have $C^{c, \tau}_{V_L}$ components which imply values for $\mathcal{B}(B_c\rightarrow \tau^-\bar{\nu}_{\tau})$ that are below the experimental bound in Eq.~(\ref{eq:BcUB}). Hence, we can interpret these effects as predictions for  $\mathcal{B}(B_c\rightarrow \tau^-\bar{\nu}_{\tau})$. Something similar occurs with the projections on  $C^{c, \mu}_{V_L}$ with respect to $\mathcal{B}(B_c\rightarrow \mu^-\bar{\nu}_{\mu})$, and with the corresponding transformation to $C^{c, e}_{V_L}$ with respect to 
$\mathcal{B}(B_c\rightarrow e^-\bar{\nu}_{e})$. In contrast to the pseudoscalar case discussed in the previous section,
the helicity enhancement does not occur with the vector interaction. In other words, unlike NP entering through pseudo-scalar interactions, the contribution of NP through vector interactions are suppressed by the mass of the lepton in the final state in the decay processes $\mathcal{B}(B_c\rightarrow \ell^- \bar{\nu}_{\ell})$ for $\ell=e,\mu,\tau$. Consequently, our NP scenarios lead to values for $\mathcal{B}(B_c\rightarrow \ell'^- \bar{\nu}_{\ell'})$
that are close to their SM values, as shown in Table~\ref{tab:BcCVL}. 

\begin{table}
\begin{center}
\begin{tabular}{ | c | c | c | c | }
\hline
&\multicolumn{3}{| c |}{Predictions}\\
\hline
Scenario &  $\mathcal{B}(B_c\rightarrow%
e^-\bar{\nu}_{e})$ & $\mathcal{B}(B_c\rightarrow%
\mu^-\bar{\nu}_{\mu})$  & $\mathcal{B}(B_c\rightarrow%
\tau^-\bar{\nu}_{\tau})$  \\
\hline
$C^{c,  e}_{V_L}=10C^{c, \mu}_{V_L}$&$[1.91\times 10^{-9},2.36\times 10^{-9}]$&$[8.39\times 10^{-5},1.10\times 10^{-4} ]$&$[0.026,0.029]$\\
\hline
$C^{c,  e}_{V_L}=0.1 C^{c, \mu}_{V_L}$&$[1.96 \times 10^{-9}, 2.36\times 10^{-9}]$&$[8.08\times 10^{-5},1.10\times 10^{-4} ]$&$[0.026,0.029]$\\
\hline
\end{tabular}
\end{center}
\caption{Predictions for $\mathcal{B}(B_c\rightarrow \ell^{'-}\bar{\nu}_{\ell} )$
in the presence of a NP Wilson coefficient $C^{\ell'}_{V_L}$ for different scenarios correlating $C^{c, e}_{V_L}$ and $C^{c, \mu}_{V_L}$. The scenario $C^{c, e}_{V_L}=C^{c, \mu}_{V_L}$ does not yield bounded NP regions and it is thus not presented.}
\label{tab:BcCVL}
\end{table}

\boldmath
\section{Conclusions}\label{sec:concl}
\unboldmath
We have mapped out the space for NP effects in exclusive $B$ decays originating from $b\to u \ell \bar \nu_{\ell}$ 
and $b\to c \ell \bar \nu_{\ell}$ quark-level transitions, developing and implementing strategies which utilise ratios of various decay observables to avoid input on the CKM matrix elements $|V_{ub}|$ and $|V_{cb}|$. In the first case, considering $B^-\to \ell \bar \nu_{\ell}$ as well as $B\to \pi \ell \bar \nu_{\ell}$ and $B\to \rho \ell \bar \nu_{\ell}$ modes, we updated a previous analysis of NP effects arising from (pseudo)-scalar operators and complemented it with a study of vector and tensor operator contributions. The corresponding constraints for the Wilson coefficients are consistent with the SM at the (1--2)~$\sigma$ level. Since our approach attempts to investigate the effect of NP in all the leptonic generations, we have assumed three possible scenarios which correlate the short-distance contributions between electrons and muons. In particular, we explored specific cases which fall within the following categories: $C_X^e \ll C_X^{\mu}$, $C_X^{\mu} \ll C_X^e$ and $C_X^{\mu}\sim C_X^e$. We have concluded our studies of the $b\to u \ell \bar \nu_{\ell}$ modes by determining $|V_{ub}|$ from the branching ratio of the $B \to \rho \ell' \bar{\nu}_{\ell'}$ channel, where $\ell' = e, \mu$, while allowing simultaneously for NP in a variety of scenarios. When we considered NP entering through the pseudoscalar or tensor coefficients, we found quite similar results, obtaining $3.0 \times 10^{-3} \leq |V_{ub}| \leq 3.7 \times 10^{-3}$ and $2.9 \times 10^{-3} \leq |V_{ub}| \leq 3.7 \times 10^{-3}$, respectively. Both these ranges include the exclusive determination from HFLAV, which assumes the SM, of $|V_{ub}|_{\rm excl.} = (3.67 \pm 0.15) \times 10^{-3}$ but do not reach the inclusive value of $|V_{ub}|_{\rm incl.} = (4.32 \pm 0.12_{-0.13}^{+0.12}) \times 10^{-3}$ \cite{Amhis:2019ckw}. Furthermore, we found in both cases that the value of $|V_{ub}|$ only shows a minor variation with respect to the short-distance coefficient on the relevant scales. On the other hand, we found that the variation with respect to the Wilson coefficient was a lot stronger in the case of new contributions to the left-handed vector coefficients, yielding a larger range of $2.8 \times 10^{-3} \leq |V_{ub}| \leq 5.8 \times 10^{-3}$, which includes both the exclusive and inclusive determination from HFLAV. We then applied our results for the short-distance coefficients and $|V_{ub}|$ to make predictions for the branching ratios of the $B^- \to e \bar{\nu}_e$, $B \to \pi \tau \bar{\nu}_\tau$ and $B \to \rho \tau \bar{\nu}_\tau$ decay modes, which have not yet been measured. For the first two, experimental limits are available, which excluded (part of) the space for the Wilson coefficients and $|V_{ub}|$, which was already taken into account in the results for $|V_{ub}|$ presented above. The largest effects were found in the tensor and pseudoscalar NP scenarios. For the former, the branching ratio of $B \to \pi \tau \bar{\nu}_\tau$ severely constrains the allowed regions. As a result, the current data allow for values of this observable up to the experimental limit ${\mathcal B}(\bar{B}^0 \rightarrow \pi^+ \tau^- \bar{\nu}_{\tau}) < 2.5 \times 10^{-4}$ \cite{Hamer:2015jsa}. Additionally, the branching ratio of $B^- \to e \bar{\nu}_e$ may reach values of up to $2.3 \times 10^{-8}$, being enhanced with respect to its SM value by three orders of magnitude. NP entering through the pseudoscalar operator yields even larger effects for the leptonic mode, with a potential enhancement by four orders of magnitude with respect to the SM, corresponding to $\mathcal{B}(B^- \to e^- \bar{\nu}_e) = 1.4 \times 10^{-7}$.

We have extended this framework to $b\to c \ell \bar \nu_{\ell}$ modes, considering $B\to D \ell \bar \nu_{\ell}$ and $B\to D^{(*)} \ell \bar \nu_{\ell}$ transitions. Moreover, we had a closer look at the leptonic $B^-_c\to \ell \bar \nu_{\ell}$ channels which provide very interesting probes for NP effects. We have provided a state-of-the-art discussion of the SM predictions of the key observables, including the $R_{D^{(*)}}$ ratios, and the constraints on contributions from (pseudo)-scalar, vector and tensor operators. Useful constraints arise also from the polarisation observable $F_L(D^\ast)$ and the ratio $R_\mu^e(D^\ast)$, putting an interesting experimental limit on the flavour universality between electrons and muons.  Just as for the $b\rightarrow u$ transitions, we have performed our studies within the context of three scenarios correlating the Wilson coefficients for electrons and muons. We found that the experimental results for our different observables can be explained through pseudoscalar and left-handed vector NP effects. Following a procedure analogous to the one applied in the case of the $b\rightarrow u$ transitions, we have extracted $|V_{cb}|$ using the CKM-dependent branching fraction $\braket{\mathcal{B}(\bar{B}\rightarrow D^* \tau^- \bar{\nu}_{\tau})}$,
finding values within the range $0.038 \leq |V_{cb}|\leq 0.043$ which are compatible with the exclusive \cite{Amhis:2019ckw} $|V_{cb}|=0.03958\pm 0.00117$  and the inclusive \cite{Charles:2004jd} $|V_{cb}|=0.04162^{+0.00026}_{-0.00080}$  determinations. 

In addition, we have determined potential NP enhancements of the branching fractions for the leptonic decays $B^-_c\rightarrow \mu^-\bar{\nu}_{\mu}$ and $B^-_c\rightarrow e^-\bar{\nu}_{e}$. Interestingly, if NP is present in pseudo-scalar Wilson coefficients,  the branching fraction $\mathcal{B}(B^-_c\rightarrow e^-\bar{\nu}_{e})$ can be enhanced by up to eight orders of magnitude in our scenarios reaching $\mathcal{B}(B^-_c\rightarrow e^-\bar{\nu}_{e})=0.31$. In the case of $\mathcal{B}(B^-_c\rightarrow \mu^-\bar{\nu}_{\mu})$, we may obtain an increase as large as four orders of magnitude leading to $\mathcal{B}(B^-_c\rightarrow \mu^-\bar{\nu}_{\mu})=0.39$. 

In the future, it will be especially useful to have more measurements of the semileptonic decay modes with light leptons in the final state available, separately for electrons and muons, as well as (further) searches for decays that have so far not been measured. We look forward to seeing how the picture for the NP constraints following from leptonic and semileptonic $B_{(c)}$ decays and the methods proposed in this paper will evolve, exploiting the wealth of experimental data to become available.


\section*{Acknowledgements}
This research has been supported by the Netherlands Foundation for Fundamental Research of Matter (FOM) 
programme 156, ``Higgs as Probe and Portal'', and by the National Organisation for Scientific Research (NWO). GTX acknowledges support from the  Deutsche
Forschungsgemeinschaft (DFG, German Research Foundation) under grant  396021762 - TRR 257. We thank Rusa Mandal for useful discussions. 


\appendix

\section{Form Factor Parameterization}
\label{Sec:Appendix1}

The branching ratios of the semileptonic modes $\bar{B} \to \tilde{V} \ell^{-} \bar{\nu}_{\ell}$ and $\bar{B} \to P \ell^{-} \bar{\nu}_{\ell}$, given in Eqs.~(\ref{eq:dBrV})~and~(\ref{eq:dBrP}), respectively, are expressed in terms of hadronic amplitudes. In case of $B \to \tilde{V}$, they are given by \cite{Sakaki:2013bfa}
\begin{eqnarray}
    H_{V, \pm}^{\tilde{V}}(q^2) &=& (M_B + m_{\tilde{V}}) A_1(q^2) \mp \frac{2 M_B |\vec{p}_{\tilde{V}}|}{M_B + m_{\tilde{V}}} V(q^2), \\
    H_{V, 0}^{\tilde{V}}(q^2) &=& \frac{M_B + m_{\tilde{V}}}{2 m_{\tilde{V}} \sqrt{q^2}} \left[ -(M_B^2 - m_{\tilde{V}}^2 -q^2) A_1(q^2) + \frac{4 M_B^2 |\vec{p}_{\tilde{V}}|^2}{(M_B + m_{\tilde{V}})^2} A_2(q^2) \right], \\
    H_{V, t}^{\tilde{V}}(q^2) &=& -\frac{2 M_B |\vec{p}_{\tilde{V}}|}{\sqrt{q^2}} A_0 (q^2), \\
    H_S^{\tilde{V}}(q^2) &\simeq& - \frac{2 M_B |\vec{p}_{\tilde{V}}|}{m_b + m_q} A_0(q^2), \\
    H_{T, \pm}^{\tilde{V}}(q^2) &=& \frac{1}{\sqrt{q^2}} \left[ \pm (M_B^2 - m_{\tilde{V}}^2) T_2(q^2) + 2 M_B |\vec{p}_{\tilde{V}}| T_1(q^2) \right], \\
    H_{T, 0}^{\tilde{V}}(q^2) &=& \frac{1}{2 m_{\tilde{V}}}\left[ -(M_B^2 + 3 m_{\tilde{V}}^2 - q^2) T_2(q^2) + \frac{4 M_B^2 |\vec{p}_{\tilde{V}}|^2}{M_B^2 - m_{\tilde{V}}^2} T_3(q^2) \right]
\end{eqnarray}
and
\begin{eqnarray}
    H_{V,0}^P(q^2) &=& \frac{2 M_B |\vec{p}_P|}{\sqrt{q^2}} F_1(q^2), \\
    H_{V,t}^P(q^2) &=& \frac{M_B^2 - m_P^2}{\sqrt{q^2}} F_0(q^2), \\
    H_S^P(q^2) &\simeq& \frac{M_B^2 - m_P^2}{m_b - m_q} F_0(q^2), \\
    H_T^P(q^2) &=& -\frac{2 M_B |\vec{p}_P|}{M_B + m_P} F_T(q^2)
\end{eqnarray}
for the $B \to P$ transitions. Note that the $|\vec{p}_{\tilde{V}}|$ and $|\vec{p}_P|$ are functions of $q^2$. The seven form factors $V$, $A_0$, $A_1$, $A_2$, $T_1$, $T_2$ and $T_3$ for $B \to \tilde{V}$ and the three form factors $F_0$, $F_1$ and $F_T$ for $B \to P$ have to be determined through non-perturbative techniques. In Appendices~\ref{sec:FFBpr}~and~\ref{sec:FFBD}, we discuss these form factors for $B \to \{\rho, \pi\}$ and $B \to D^{(\ast)}$ modes, respectively.

\boldmath
\subsection{$B \to \{\rho, \pi\}$ Form Factors} \label{sec:FFBpr}
\unboldmath

For the calculation of the branching ratio of the $\bar{B} \to \rho \ell^{-} \bar{\nu}_{\ell}$ decay, we require the form factors $V$, $A_0$, $A_1$, $A_2$, $T_1$, $T_2$ and $T_3$. As we did in Ref.~\cite{Banelli:2018fnx}, we make use of the LCSR calculation from Ref.~\cite{Straub:2015ica}. The form factors are given by
\begin{equation} \label{eq:FFBrhoParam}
F_i (q^2) = \frac{k_{(\rho^0, u)}}{1-q^2/m_{R, i}^2} \sum_{k=0}^2 \alpha_k^i \left[ z_\rho(q^2, t_0^\rho) - z_\rho(0, t_0^\rho) \right]^k,
\end{equation}
where
\begin{equation}
    z_\rho(q^2, t_0^\rho) \equiv \frac{\sqrt{(M_B + m_\rho)^2 - q^2} - \sqrt{(M_B + m_\rho)^2 - t_0^\rho}}{\sqrt{(M_B + m_\rho)^2 - q^2} + \sqrt{(M_B + m_\rho)^2 - t_0^\rho}},
\end{equation}
with
\begin{equation}
t_0^\rho \equiv (M_B + m_\rho) \left[ M_B + m_\rho -2\sqrt{M_B m_\rho} \right].
\end{equation}
The factor $k_{(\rho^0, u)} = f_{\rho^0}^{(u)}/\bar{f}_\rho^{\rho^I}$, where
\begin{eqnarray}
f_{\rho^0}^{(u)} &=& (221.5 \pm 0.3) \times 10^{-3} \text{ GeV}, \\
\bar{f}_\rho^{\rho^I} &=& (213 \pm 5) \times 10^{-3} \text{ GeV},
\end{eqnarray}
takes into account that we have a $b \to u$ instead of a $b \to d$ transition.

The form factors can now be determined using the coefficients $\alpha_k^i$ in Table~\ref{tab:BrhoFFCoeff} and the mass terms $m_{R, i}$ in Table~\ref{tab:BrhoFFmR}. We do not consider the correlations between the $\alpha_k^i$ coefficients in our evaluations. The form factors are valid in the range $0 \leq q^2 \leq 14 \text{ GeV}^2$. We note that Ref.~\cite{Straub:2015ica} considers the form factors $A_1$, $A_{12}$ instead of $A_1$, $A_2$, and $T_2$, $T_{23}$ instead of $T_2$, $T_3$. We can convert the form factors using the following expressions \cite{Straub:2015ica}:
\begin{eqnarray}
    A_2(q^2) &=& \frac{(M_B + m_\rho)^2 (M_B^2 - m_\rho^2 -q^2) A_1(q^2) - 16 M_B m_\rho^2 (M_B + m_\rho) A_{12}(q^2)}{4 M_B^2 |\vec{p}_\rho|^2}, \\
    T_3(q^2) &=& \frac{(M_B^2-m_\rho^2)(M_B^2 + 3 m_\rho^2 - q^2) T_2(q^2) - 8 M_B m_\rho^2 (M_B - m_\rho) T_{23}(q^2)}{4 M_B^2 |\vec{p}_\rho|^2}.
\end{eqnarray}

\begin{table}
    \centering
    \begin{tabular}{|c|c|c|c|}
        \hline
        Form factor $F_i$ & $\alpha_0^i$ & $\alpha_1^i$ & $\alpha_2^i$ \\
        \hline
        $V$ & $0.33 \pm 0.03$ & $-0.86 \pm 0.18$ & $1.80 \pm 0.97$ \\
        $A_0$ & $0.36 \pm 0.04$ & $-0.83 \pm 0.20$ & $1.33 \pm 1.05$  \\
        $A_1$ & $0.26 \pm 0.03$ & $0.39 \pm 0.14$ & $0.16 \pm 0.41$ \\
        $A_{12}$ & $0.30 \pm 0.03$ & $0.76 \pm 0.20$ & $0.46 \pm 0.76$ \\
        $T_1$ & $0.27 \pm 0.03$ & $-0.74 \pm 0.14$ & $1.45 \pm 0.77$ \\
        $T_2$ & $0.27 \pm 0.03$ & $0.47 \pm 0.13$ & $0.58 \pm 0.46$ \\
        $T_{23}$ & $0.75 \pm 0.08$ & $1.90 \pm 0.43$ & $2.93 \pm 1.81$ \\
        \hline
    \end{tabular}
    \caption{Coefficients for the $B \to \rho$ form factors \cite{Straub:2015ica}.}
    \label{tab:BrhoFFCoeff}
\end{table}

\begin{table}
    \centering
    \begin{tabular}{|c|c|}
        \hline
        Form factor $F_i$ & $m_{R, i} / \text{GeV}$ \\
        \hline
        $A_0$ & 5.279 \\
        $V$, $T_1$ & 5.325 \\
        $A_1$, $A_{12}$, $T_2$, $T_{23}$ & 5.724 \\
        \hline
    \end{tabular}
    \caption{Mass terms for the form factor evaluation in Eq.~(\ref{eq:FFBrhoParam}) \cite{Straub:2015ica}.}
    \label{tab:BrhoFFmR}
\end{table}

For the $\bar{B} \to \pi \ell^{-} \bar{\nu}_{\ell}$ form factors, we make use of the LQCD calculations in Refs.~\cite{Lattice:2015tia,Bailey:2015nbd}. Here the form factors are parameterized as
\begin{eqnarray}
    F_0(q^2) &=& \sum_{n=0}^3 b_n^0 z_\pi(q^2, t_0^\pi)^n, \\
    F_1(q^2) &=& \frac{1}{1-q^2/M_{B^\ast}^2} \sum_{n=0}^3 b_n^1 \left[ z_\pi(q^2, t_0^\pi)^n - (-1)^{n-4} \frac{n}{4} z_\pi(q^2, t_0^\pi)^4 \right], \\
    F_T(q^2) &=& \frac{1}{1-q^2/M_{B^\ast}^2} \sum_{n=0}^3 b_n^T \left[ z_\pi(q^2, t_0^\pi)^n - (-1)^{n-4} \frac{n}{4} z_\pi(q^2, t_0^\pi)^4 \right].
\end{eqnarray}
The $z_\pi$ and $t_0^\pi$ are defined in analogy to $z_\rho$ and $t_0^\rho$ with $m_\rho$ replaced by $m_\pi$. The required parameters $b_n^i$ are given in Table~\ref{tab:BpiFFCoeff}. We do not consider correlations between these parameters. These form factors are valid over the full kinematic range.

\begin{table}
    \centering
    \begin{tabular}{|c|c|c|c|c|}
        \hline
        $F_i$ & $b_0^i$ & $b_1^i$ & $b_2^i$ & $b_3^i$ \\
        \hline
        $F_0$ & $0.507 \pm 0.022$ & $-1.77 \pm 0.18$ & $1.27 \pm 0.81$ & $4.2 \pm 1.4$ \\
        $F_1$ & $0.407 \pm 0.015$ & $-0.65 \pm 0.16$ & $-0.46 \pm 0.88$ & $0.4 \pm 1.3$ \\
        $F_T$ & $0.393 \pm 0.017$ & $-0.65 \pm 0.23$ & $-0.6 \pm 1.5$ & $0.1 \pm 2.8$ \\
        \hline
    \end{tabular}
    \caption{Coefficients for the $B \to \pi$ form factors \cite{Lattice:2015tia,Bailey:2015nbd}.}
    \label{tab:BpiFFCoeff}
\end{table}

\boldmath
\subsection{$B \to D^{(\ast)}$ Form Factors} \label{sec:FFBD}
\unboldmath

In order to determine the different hadronic form factors required for the calculation of the $\mathcal{B}( \bar{B}\rightarrow D \ell \bar{\nu})$ 
and  $\mathcal{B}(B\rightarrow D^{*}  \ell \bar{\nu})$ branching ratios, we use the parameterization provided in Ref.~\cite{Bernlochner:2017jka} that 
includes $\mathcal{O}(\Lambda_{\rm QCD}/m_{c,b})$ and $\mathcal{O}(\alpha_s)$ contributions. 

In the case of the $B\rightarrow D$ transition, the connection between the current form 
factors $F_{1}$,  $F_{0}$, 
$F_{T}$ and the HQET hadronic form factors $h_+$, $h_-$ and $h_T$  is given by
\begin{eqnarray}
F_{1}(q^2)&=&\frac{1}{2\sqrt{m_B m_D}}\Bigl[(m_B + m_D) h_+(w) - (m_B - m_D) h_-(w) \Bigl],\nonumber\\
F_{0}(q^2)&=&\frac{1}{2\sqrt{m_B m_D}}\Bigl[\frac{(m_B + m_D)^2-q^2}{m_B + m_D}  h_+(w) - 
\frac{(m_B - m_D)^2-q^2}{m_B - m_D} h_-(w) \Bigl],\nonumber\\
F_{T}(q^2)&=&\frac{m_B + m_D}{2\sqrt{m_B m_D}} h_T(w),
\end{eqnarray}
where $w$ depends on the four momentum transfer $q^2$ to the system composed by the $\ell$ and $\bar{\nu}_{\ell}$, so that it can 
recoil against the $D$ mesons according to
\begin{eqnarray}
\label{eq:w}
w(q^2)&=&\frac{m^2_B + m^2_{D} -q^2}{2 m_B m_{D}}.
\end{eqnarray}

Of particular interest is the zero-recoil point, corresponding to the maximum four-momentum transferred to the lepton pair,
\begin{eqnarray}
q_0^2&=&(M_B-M_{D})^2,
\end{eqnarray}
for which Eq.~(\ref{eq:w}) trivially reduces to
\begin{eqnarray}
\label{eq:wzc}
w_0&\equiv&w(q_0^2)=1.
\end{eqnarray}

Analogously, for the $B\rightarrow D^*$ processes, there are six form factors: 
$V(q^2)$, $A_1(q^2)$, $A_2(q^2)$, $A_0(q^2)$, $T_1(q^2)$, $T_2(q^2)$, $T_3(q^2)$. They can be expressed in terms 
of the HQET form factors $h_V(w)$, $h_{A_i}(w)$, $h_{T_i}(w)$ (for $i=1,2,3$) through
\begin{eqnarray}
V(q^2)&=&\frac{1}{2\sqrt{m_B m_{D^*}}}h_V(w),\nonumber\\
A_1(q^2)&=&\frac{(m_B +  m_{D^*})^2 - q^2}{2\sqrt{m_B m_{D^*}(m_B + m_{D^*})}}h_{A_1}(w),\nonumber\\
A_2(q^2)&=&\frac{m_B +  m_{D^*} }{2\sqrt{m_B m_{D^*}}}\Biggl[h_{A_3}(w) + \frac{m_{D^*}}{m_B}h_{A_2}(w) \Biggl],\nonumber\\
A_0(q^2)&=&\frac{1}{2\sqrt{m_B m_{D^*}}}\Biggl[\frac{(m_B +  m_{D^*})^2 - q^2}{2m_{D^*}}h_{A_1}(w)-
\frac{m^2_B -  m^2_{D^*} + q^2}{2m_{B}}h_{A_2}(w)\nonumber\\
&&-\frac{m^2_B -  m^2_{D^*} - q^2}{2m_{D^*}}h_{A_3}(w)\Biggl],\nonumber\\
T_1(q^2)&=&\frac{1}{2\sqrt{m_B m_{D^*}}}\Biggl[(m_B + m_{D^*})h_{T_1}(w) - (m_B - m_{D^*})h_{T_2}(w) \Biggl],\nonumber\\
T_2(q^2)&=&\frac{1}{2\sqrt{m_B m_{D^*}}}\Biggl[\frac{(m_B +  m_{D^*})^2 - q^2}{m_B+ m_{D^*}}h_{T_1}(w) 
-\frac{(m_B -  m_{D^*})^2 - q^2}{m_B- m_{D^*}}h_{T_2}(w)\Biggl],\nonumber\\
T_3(q^2)&=&\frac{1}{2\sqrt{m_B m_{D^*}}}\Biggl[(m_B - m_{D^*})h_{T_1}(w)-(m_B + m_{D^*})h_{T_2}(w)\nonumber\\
&&-2\frac{m^2_B-m^2_{D^*}}{m_B}h_{T_3}(w)  \Biggl].
\end{eqnarray}

Due to the Heavy Quark Symmetry,  the HQET form factors depend on a single form factor, the Isgur--Wise function, in the heavy-quark limit $\Lambda_{\rm QCD}/m_{b,c} \ll 1$:
\begin{eqnarray}
h(w)&=&\xi(w)\hat{h}(w).
\end{eqnarray}
In the case of the $B\rightarrow D$ transitions, we have
\begin{eqnarray}
\hat{h}_+&=&1 + \frac{\alpha_s}{\pi}\Bigl[ \mathcal{C}_{V_1} + \frac{w+1}{2}(\mathcal{C}_{V_2} + \mathcal{C}_{V_3} )\Bigl]
+ \Bigl(\varepsilon_c + \varepsilon_b\Bigl)\hat{L}_1,\nonumber\\
\hat{h}_-&=&\frac{\alpha_s}{\pi}\frac{w+1}{2}(\mathcal{C}_{V_2} - \mathcal{C}_{V_3} ) + \Bigl(\varepsilon_c - \varepsilon_b\Bigl)\hat{L}_4,
\nonumber\\
\hat{h}_T&=&1 + \frac{\alpha_s}{\pi}  \Bigl(\mathcal{C}_{T_1}-\mathcal{C}_{T_2} + \mathcal{C}_{T_3}\Bigl)+ \Bigl( \varepsilon_c + \varepsilon_b\Bigl)
\Bigl(\hat{L}_1-\hat{L}_4\Bigl).
\label{eq:HQETBD}
\end{eqnarray}
On the other hand, the corresponding expressions for the $B\rightarrow D^*$ mode read
\begin{eqnarray}
\hat{h}_V&=& 1 + \frac{\alpha_s}{\pi}\mathcal{C}_{V_1} + \varepsilon_c\Bigl(\hat{L}_2 - \hat{L}_5\Bigl) + 
\varepsilon_b\Bigl(\hat{L}_1 - \hat{L}_4\Bigl),\nonumber\\
\hat{h}_{A_1}&=& 1 +\frac{\alpha_s}{\pi} \mathcal{C}_{A_1} + \varepsilon_c\Bigl(\hat{L}_2 - \hat{L}_5\frac{w-1}{w+1}\Bigl) 
+ \varepsilon_b\Bigl(\hat{L}_1 - \hat{L}_4\frac{w-1}{w+1}\Bigl),\nonumber\\
\hat{h}_{A_2}&=&\frac{\alpha_s}{\pi}\mathcal{C}_{A_2} + \varepsilon_c\Bigl(\hat{L}_3 + \hat{L}_6\Bigl),\nonumber\\
\hat{h}_{A_3}&=&1 + \frac{\alpha_s}{\pi}\Bigl(\mathcal{C}_{A_1} + \mathcal{C}_{A_3}\Bigl) + 
\varepsilon_c\Bigl(\hat{L}_2 - \hat{L}_3 + \hat{L}_6 - \hat{L}_5 \Bigl) + \varepsilon_b \Bigl(\hat{L}_1- \hat{L}_4\Bigl),\nonumber\\
\hat{h}_{T_1}&=&1 + \frac{\alpha_s}{\pi}\Bigl(\mathcal{C}_{T_1} + \frac{w-1}{2}(\mathcal{C}_{T_2}-\mathcal{C}_{T_3})\Bigl)
+  \varepsilon_c \hat{L}_2 +  \varepsilon_b  \hat{L}_1,\nonumber\\
\hat{h}_{T_2}&=& \frac{\alpha_s}{\pi}\frac{w+1}{2}\Bigl(\mathcal{C}_{T_2} + \mathcal{C}_{T_3}\Bigl) + \varepsilon_c \hat{L}_5 - 
\varepsilon_b \hat{L}_4,\nonumber\\
\hat{h}_{T_3}&=&\frac{\alpha_s}{\pi}\mathcal{C}_{T_2} +  \varepsilon_c\Bigl(\hat{L}_6 - \hat{L}_3\Bigl).
\label{eq:HQETBDstar}
\end{eqnarray}

The $\mathcal{\alpha}_s$ corrections in Eqs.~(\ref{eq:HQETBD})~and~(\ref{eq:HQETBDstar}) are included in the functions 
$\mathcal{C}_{V_i}, \mathcal{C}_{A_i}, \mathcal{C}_{T_i}$ which depend on $w$, defined in Eq.~(\ref{eq:w}), and on the ratio $z=m_c/m_b$. 
On the other hand, the $\mathcal{O}(\Lambda_{\rm QCD}/m_{b,c})$ power corrections are contained in
$\hat{L}_{1,2,3,4,5,6}$, which depend on the subleading Isgur--Wise functions whose linearized version around the
zero recoil point $w=w_0=1$ read
\begin{eqnarray}
\hat{\chi}_2(w)&=&\hat{\chi}_2(1) + \hat{\chi}'_2(1)(w-1) \nonumber \\
\hat{\chi}_3(w)&=& \hat{\chi}'_3(1)(w-1) \nonumber \\
\eta(w)&=&\eta(1) + \eta'(1)(w-1).
\end{eqnarray}
For the leading Isgur--Wise function, we use
\begin{eqnarray}
\xi(w)&=&\xi(w_0)\Biggl[1 - 8 a \bar{\rho}^2_*z_* + z^2_*\Bigl(V_{21} \bar{\rho}^2_* - V_{20} + 
(\varepsilon_b - \varepsilon_c)  \Bigl) \Bigl[2\Xi  \eta'(1)\frac{1-r_{D^{(*)}}}{1+r_{D^{(*)}}}\Bigl] \nonumber\\
&&+(\varepsilon_b + \varepsilon_c)\Bigl[\Xi\Bigl(12 \hat{\chi}'_3(1) - 4\hat{\chi}_2(1) \Bigl) -
16 \Bigl([a^2-1]\Xi-16 a^4\Bigl) \hat{\chi}'_2(1) \Bigl] \nonumber\\
&&+\frac{\alpha_s}{\pi}\Bigl[\Xi\Bigl(C'_{V_1}(w_0) + \frac{C_{V_3}(w_0)+r_{D^{(*)}} C_{V_2}(w_0)}{1+r_{D^{(*)}}}\Bigl) \nonumber\\
&&+2 a^2(\Xi - 32 a^2) \frac{C'_{V_3}(w_0)+r_{D^{(*)}} C'_{V_2}(w_0)}{1+r_{D^{(*)}}} \nonumber\\
&&-64 a^6  \frac{C''_{V_3}(w_0)+r_{D^{(*)}} C''_{V_2}(w_0)}{1+r_{D^{(*)}}} - 32 a^4 C''_{V_1}(w_0)\Bigl]
\Biggl],
\end{eqnarray}
with
\begin{eqnarray}
\Xi&=& 64 a^4  \bar{\rho}^2_* - 16 a^2 -V_{21}, \nonumber\\
z_*(w)&=&\frac{\sqrt{w+1}-\sqrt{2}a}{\sqrt{w+1}+\sqrt{2}a},\nonumber\\
a&=&\Bigl(\frac{1 + r_{D^{(*)}}}{2\sqrt{2}}\Bigl),\nonumber\\
r_{D^{(*)}}&=&\frac{m_{D^{(*)}}}{m_B}.
\end{eqnarray}
We can determine $\xi(w_0)$ by demanding the zero recoil point normalization condition $\xi(1)=1$, our result is
\begin{eqnarray}
\xi(w_0)=0.70\pm 0.015.
\end{eqnarray}

Notice that in Eqs.~(\ref{eq:HQETBD}) and  (\ref{eq:HQETBDstar}), the power corrections in $m_b$ and $m_c$ are included through 
the terms multiplying
\begin{eqnarray}
\varepsilon_{b, c}=\frac{\bar{\Lambda}}{2m_{b, c}},
\end{eqnarray}
where $\bar{\Lambda}$ is of $\mathcal{O}(\Lambda_{\rm QCD})$:
\begin{eqnarray}
\bar{\Lambda}&=&\overline{m}_B - m_b(m^{1S}_b)  + \lambda_1/(2m^{1S}_b).
\label{eq:Lambdabar}
\end{eqnarray}
The required inputs in Eq.~(\ref{eq:Lambdabar}) are
\begin{eqnarray}
\overline{m}_B=5.313~\hbox{GeV},&~~&\lambda_1=-0.3~\hbox{GeV}^2~\hbox{ \cite{Ligeti:2014kia}},
\end{eqnarray}
and as indicated in \cite{Bernlochner:2017jka}, the prescription
\begin{eqnarray}
m_b(m^{1S}_b)\rightarrow m^{1S}_b,
\end{eqnarray}
has to be used everywhere except in those terms  where the $\bar{\Lambda}/m_{b,c}$ factors do not multiply subleading Isgur--Wise functions
in Eq.~(\ref{eq:HQETBD})~and Eq.~(\ref{eq:HQETBDstar}) to ensure the cancellation of leading renormalons.

It can be seen how at leading order in $\alpha_s$ and $\Lambda_{\rm QCD}/m_{b,c}$ the different 
HQET form factors either reduce to a common Isgur--Wise function $\xi$ or vanish.

Since we want to explore the possibility
of having NP effects in light leptons, we consider the set of input parameters shown in Table  \ref{eq:InputsHQET} and present the corresponding correlation matrix in Table~\ref{eq:InputsHQETCorr}.  
They were obtained from a  fit to QCDSR and LQCD calculations as presented in \cite{Ligeti:2014kia}. 
This set of inputs avoids the usage
of the differential distributions for the processes $B\rightarrow D^{(*)} \ell \bar{\nu}_{\ell}$, for $\ell=e, \mu$,
to fit the data.

\begin{table}
\begin{center}
  \begin{tabular}{ | c | c | }
    \hline
    Input &  Value\\
    \hline
    \hline
    $\bar{\rho}^2_*$ & $1.24\pm 0.08$\\ 
    \hline
    $\hat{\chi}_2(1)$ & $-0.06\pm 0.02$  \\ 
    \hline
    $\hat{\chi}'_2(1)$ & $0.00\pm 0.02$ \\ 
    \hline
    $\hat{\chi}'_3(1)$ & $0.04\pm 0.02$\\
    \hline
    $\eta(1)$ & $0.31\pm 0.04$\\
    \hline
    $\eta'(1)$ & $0.05\pm 0.10$\\
    \hline
    $m^{1s}_b~\hbox{(GeV)}$& $4.71\pm 0.05$\\
    \hline
    $\delta m_{bc}~\hbox{(GeV)}$& $3.40\pm 0.02$ \\
    \hline
  \end{tabular}
\end{center}
\caption{Inputs used in the evaluation of the  HQET form factors.}
\label{eq:InputsHQET}
\end{table}

\begin{table}
\begin{center}
  \begin{tabular}{|c|c|c|c|c|c|c|c|c|c|}
    \hline
    &$\bar{\rho}^2_*$  & $\hat{\chi}_2(1)$ & $\hat{\chi}'_2(1)$ & $\hat{\chi}'_3(1)$ & $\eta(1)$ & $\eta'(1)$ & $m^{1s}_b$ & $\delta m_{bc}$\\
    \hline
    $\bar{\rho}^2_*$  &$1.00$ & $-0.27$ & $-0.13$ & $0.81$ & $0.08$ & $-0.07$& $0.24$ & $0.02$\\
    \hline 
    $\hat{\chi}_2(1)$ & $-0.27$ & $1.00$ & $0.00$ & $0.01$ & $0.01$ & $0.03$& $-0.01$ & $0.00$\\
    \hline
    $\hat{\chi}'_2(1)$ & $-0.13$ & $0.00$ & $1.00$ & $-0.01$ & $-0.01$ & $0.01$& $0.01$& $0.00$\\ 
    \hline
    $\hat{\chi}'_3(1)$ & $0.81$ & $0.01$ & $-0.01$ & $1.00$ & $-0.02$ & $-0.09$& $0.04$&  $0.00$\\
    \hline
    $\eta(1)$ & $0.08$ & $0.01$ & $-0.01$ & $-0.02$ & $1.00$ & $0.11$& $-0.48$& $0.04$\\
    \hline
    $\eta'(1)$ & $-0.07$ & $0.03$ & $0.01$ & $-0.09$ & $0.11$ & $1.00$& $0.07$& $-0.01$\\
    \hline 
    $m^{1s}_b$ &$-0.24$ & $-0.01$ & $0.01$ & $0.04$ & $-0.48$ & $0.07$& $1.00$& $0.00$\\
    \hline
    $\delta m_{bc}$ &$0.02$ & $0.00$ & $0.00$ & $0.00$ & $0.04$ & $-0.01$& $0.00$&$1.00$\\
    \hline
  \end{tabular}
\end{center}
\caption{Correlation matrix for the input parameters used to calculate the  HQET form factors for the $B\rightarrow D^{(*)}\ell\bar{\nu}_{\ell}$ decays.}
\label{eq:InputsHQETCorr}
\end{table}

\section{Input Parameters} \label{Sec:Appendix2Input}

For convenience, we summarize the numerical values and sources of the input parameters used in this paper in Table~\ref{tab:inputParam}.

\begin{table}
    \centering
    \begin{tabular}{|c|c|c|c|}
        \hline
        Parameter & Value & Unit & Reference \\
        \hline \hline
        $m_e$ & $0.5109989461(31) \times 10^{-3}$ & GeV & \cite{Tanabashi:2018oca} \\
        \hline
        $m_\mu$ & $105.6583745(24) \times 10^{-3}$ & GeV & \cite{Tanabashi:2018oca} \\
        \hline
        $m_\tau$ & $(1776.86 \pm 0.12) \times 10^{-3}$ & GeV & \cite{Tanabashi:2018oca} \\
        \hline \hline
        $m_u$ & $(2.2_{-0.4}^{+0.5}) \times 10^{-3}$ & GeV & \cite{Tanabashi:2018oca} \\
        \hline
        $m_c$ &
        $1.275^{+0.025}_{-0.035}$ & GeV & \cite{Tanabashi:2018oca}\\
        \hline
        $m_b$ & $4.18_{-0.03}^{+0.04}$ & GeV & \cite{Tanabashi:2018oca} \\
        \hline
         $m^{1S}_b$ & $4.71 \pm 0.05$ & GeV & \cite{Ligeti:2014kia} \\
        \hline \hline
        $M_{B^\pm}$ & $(5279.32 \pm 0.14) \times 10^{-3}$ & GeV & \cite{Tanabashi:2018oca} \\
        \hline
        $M_{B_d}$ & $(5279.63 \pm 0.15) \times 10^{-3}$ & GeV & \cite{Tanabashi:2018oca} \\
        \hline
        $M_{B_c}$ & $6274.9(8) \times 10^{-3}$ & GeV & \cite{Tanabashi:2018oca} \\
        \hline
        $M_{B^\ast}$ & $(5324.65 \pm 0.25) \times 10^{-3}$ & GeV & \cite{Tanabashi:2018oca} \\
        \hline \hline
        $m_{\pi^\pm}$ & $139.57061(24) \times 10^{-3}$ & GeV & \cite{Tanabashi:2018oca} \\
        \hline
        $m_{\pi^0}$ & $134.9770(5) \times 10^{-3}$ & GeV & \cite{Tanabashi:2018oca} \\
        \hline \hline
        $m_{\rho^\pm}$ & $(775.11 \pm 0.34) \times 10^{-3}$ & GeV & \cite{Tanabashi:2018oca} \\
        \hline
        $m_{\rho^0}$ & $(775.26 \pm 0.25) \times 10^{-3}$ & GeV & \cite{Tanabashi:2018oca} \\
        \hline
        \hline
        $M_{D^0}$ & $1864.83(5)\times 10^{-3} $ & GeV & \cite{Tanabashi:2018oca} \\
        \hline
        $M_{D^{+}}$ & $1869.65(5)\times 10^{-3} $ & GeV & \cite{Tanabashi:2018oca} \\
        \hline
        \hline
        $M_{D^{*0}}$ & $2.00685(5)\times 10^{-3} $ & GeV & \cite{Tanabashi:2018oca} \\
        \hline
        $M_{D^{*+}}$ & $2.01026(5)\times 10^{-3} $ & GeV & \cite{Tanabashi:2018oca} \\
        \hline
        \hline
        $\tau_{B^\pm}$ & $(1.638 \pm 0.004) \times 10^{-12}$ & s & \cite{Tanabashi:2018oca} \\
        \hline
        $\tau_{B_d}$ & $(1.520 \pm 0.004) \times 10^{-12}$ & s & \cite{Tanabashi:2018oca} \\
        \hline
        $\tau_{B_c}$ & $(0.510 \pm 0.009) \times 10^{-12}$ & s & \cite{Tanabashi:2018oca} \\
        \hline \hline
        $|V_{ub}|$ & $(3.67 \pm 0.15) \times 10^{-3}$ & & \cite{Amhis:2019ckw} \\
        \hline
        $|V_{cb}|$ & $(3.958 \pm 0.117) \times 10^{-2}$ & & \cite{Amhis:2019ckw} \\
        \hline \hline
        $f_{B^\pm}$ & $(190.0 \pm 1.3) \times 10^{-3}$ & GeV & \cite{Aoki:2019cca} \\
        \hline
         $f_{B_c}$ & $(434 \pm 15) \times 10^{-3}$ & GeV & \cite{Colquhoun:2015oha} \\
        \hline \hline
        $G_{\rm F}$ & $1.1663787(6) \times 10^{-5}$ & $\text{GeV}^{-2}$ & \cite{Tanabashi:2018oca} \\
        \hline
    \end{tabular}
    \caption{Collection of input parameters used in this paper.}
    \label{tab:inputParam}
\end{table}

\bibliographystyle{utphys}
\bibliography{biblio}
\end{document}